\pdfoutput=1

\documentclass[11pt,twoside,a4paper,cmspaper,final,collab]{cms-tdr}
\def\svnVersion{183634}\def\cmsCernNo{2012-322}\def\cmsCernDate{\today}\def\cmsMessage{Submitted to the European Physical Journal C}

\begin{document}\cmsNoteHeader{TOP-11-013}

\hyphenation{had-ron-i-za-tion}
\hyphenation{cal-or-i-me-ter}
\hyphenation{de-vices}

\RCS$Revision: 167978 $
\RCS$HeadURL: svn+ssh://alverson@svn.cern.ch/reps/tdr2/papers/TOP-11-013/trunk/TOP-11-013.tex $
\RCS$Id: TOP-11-013.tex 167978 2013-01-28 15:21:57Z alverson $

\ifthenelse{\boolean{cms@external}}{\providecommand{\suppMaterial}{the supplemental material [URL will be inserted by publisher]}}{\providecommand{\suppMaterial}{App.~\ref{app:suppMat}}}
\ifthenelse{\boolean{cms@external}}{\providecommand{\breakhere}{\linebreak[4]}}{\providecommand{\breakhere}{\relax}}
\ifthenelse{\boolean{cms@external}}{\providecommand{\cmsLeft}{top}}{\providecommand{\cmsLeft}{left}}
\ifthenelse{\boolean{cms@external}}{\providecommand{\cmsRight}{bottom}}{\providecommand{\cmsRight}{right}}
\newcommand{\pT} {\pt}
\newcommand{\mtt}{$m_{\ttbar}$\xspace}
\newcommand{\Wjets}{W+jets\xspace}
\newcommand{\Zjets}{Z+jets\xspace}
\newcommand{\ejets}{e+jets\xspace}
\newcommand{\mujets}{$\mu$+jets\xspace}
\newcommand{\ljets}{$\ell$+jets\xspace}
\newcommand{\mumu}{$\mu^+\mu^-$\xspace}
\newcommand{\ee}{$\mathrm{e^+e^-}$\xspace}
\newcommand{\mue}{$\mu^{\pm}\mathrm{e^{\mp}}$\xspace}
\newcommand{\pb}{\mbox{\ensuremath{\,\text{pb}}}\xspace}
\newlength{\fiveup}\setlength{\fiveup}{0.4\textwidth}

\cmsNoteHeader{TOP-11-013} 

\title{Measurement of differential top-quark-pair production cross sections in pp collisions at \texorpdfstring{$\sqrt{s} = 7\TeV$}{sqrt(s) = 7 TeV}}

\author{The CMS Collaboration}

\date{\today}

\abstract{
Normalised differential top-quark-pair production cross sections are measured in pp~collisions at a centre-of-mass energy of 7\TeV at the LHC with the CMS detector using data recorded in 2011 corresponding to an integrated luminosity of 5.0\fbinv. The measurements are performed in the lepton+jets decay channels (e+jets and $\mu$+jets) and the dilepton decay channels ($\mathrm{e^+e^-}$, $\mu^+\mu^-$, and $\mu^{\pm}\Pe^{\mp}$). The $\ttbar$ differential cross section is measured as a function of kinematic properties of the final-state charged leptons and jets associated to b quarks, as well as those of the top quarks and the $\ttbar$ system. The data are compared with several predictions from perturbative QCD calculations up to approximate next-to-next-to-leading-order precision. No significant deviations from the standard model are observed.}

\hypersetup{%
pdfauthor={CMS Collaboration},%
pdftitle={Measurement of differential top-quark-pair production cross sections in pp colisions at sqrt(s) = 7 TeV},%
pdfsubject={CMS},%
pdfkeywords={CMS, physics, top physics, cross section}}

\maketitle 

\section{Introduction}

Measurements of top-quark production cross section and properties have played a major role in testing the standard model (SM) and in searches for new physics beyond it. The large top-quark production rates at the Large Hadron Collider (LHC) give access to a new realm of precision measurements. For the first time, the \ttbar pair production rate is sufficiently high to perform a detailed and precise measurement of the \ttbar\ production cross section differentially as a function of various kinematic observables in \ttbar\ events~\cite{bib:ATLAS}. These measurements are crucial to verify the top-quark production mechanism at the LHC energy scale in the context of SM predictions with various levels of perturbative quantum chromodynamics (QCD) approximations. Furthermore, scenarios beyond the SM, for example decays of massive Z-like bosons into top-quark pairs, could be revealed in such measurements, most prominently as resonances in the invariant \ttbar mass spectrum~\cite{bib:laneref31, bib:laneref32, bib:technicolor}.

Here, measurements of the normalised differential \ttbar production cross section in proton-proton (pp) collisions at a centre-of-mass energy $\sqrt{s}$~of 7\TeV with the Compact Muon Solenoid (CMS) detector are presented. These results complement the recent CMS measurements of the \ttbar production cross section~\cite{bib:TOP-10-002_paper, bib:TOP-10-003_paper, bib:TOP-11-003_paper, bib:TOP-11-006_paper, bib:TOP-11-005_paper}. The analysis makes use of the full set of data recorded in 2011, corresponding to an integrated luminosity of $5.0\pm0.1\fbinv$. The cross section is determined as a function of the kinematic properties of the leptons and of the jets associated to b quarks or antiquarks (b jets) from top-quark decays, of the top quarks themselves, as well as of the \ttbar system. The results are compared to several theoretical predictions obtained with \MADGRAPH~\cite{bib:madgraph}, \MCATNLO~\cite{bib:mcatnlo}, \POWHEG~\cite{bib:powheg1,bib:powheg2,bib:powheg3}, and to the latest next-to-leading-order (NLO) plus next-to-next-to-leading-logarithm (NNLL)~\cite{bib:ahrens_mttbar} and approximate next-to-next-to-leading-order (NNLO)~\cite{bib:kidonakis_pt, bib:kidonakis_y} calculations.

The measurements are performed in several decay channels of the \ttbar\ system, both in the \textit{\ljets} channels ($\ell = \Pe \text{ or }\mu$), with a single isolated lepton and at least four jets in the final state, and in the \textit{dilepton} channels, with two oppositely charged leptons (\ee, \mumu, \mue) and at least two jets. The top-quark-pair candidate events are selected by requiring isolated leptons and jets with high transverse momenta. Backgrounds to \ttbar\ production are suppressed by use of b-tagging techniques. The top-quark kinematic properties are obtained through kinematic fitting and reconstruction algorithms. The normalised differential \ttbar\ production cross section is determined by counting the number of \ttbar\ signal events in each bin of the measurement, correcting for the detector effects and dividing by the measured total cross section. Correlations between the bins of the measurement are taken into account by using regularised unfolding techniques.

The measurement performed here refers to kinematical distributions. To remove systematic uncertainties on the normalisation, the absolute differential cross section is normalised to the in-situ measured inclusive cross section. The inclusive cross section, as obtained in this analysis, is consistent with the results from dedicated CMS measurements~\cite{bib:TOP-10-002_paper, bib:TOP-10-003_paper, bib:TOP-11-003_paper, bib:TOP-11-006_paper, bib:TOP-11-005_paper}. To avoid additional model uncertainties due to the extrapolation of the measured cross section into experimentally inaccessible phase space regions, the results for directly measurable quantities, such as the kinematic properties of leptons and b jets, are reported in a visible phase space. This phase space is defined as the kinematic region in which all selected final state objects are produced within the detector acceptance and are thus measurable experimentally. For top-quark and \ttbar\ distributions, the measurements are performed in the full phase space, allowing for comparison with calculations up to the approximate NNLO precision.

This document is structured as follows. A brief description of the CMS detector is provided in Section~\ref{sec:detector}, followed by details of the event simulation in Section~\ref{sec:simulation}, and the event selection and reconstruction in Section~\ref{sec:selection}. The estimated systematic uncertainties on the measurements of the cross section are described in Section~\ref{sec:errors}. The result of the differential cross section measurements are presented in Section~\ref{sec:diffxsec}, followed by a summary in Section~\ref{sec:concl}.

\section{The CMS Detector}
\label{sec:detector}

The central feature of the CMS apparatus is a superconducting solenoid of 13\,m length and 6\,m internal diameter, which provides an axial magnetic field of 3.8\unit{T}. Within the field volume are the silicon pixel and strip trackers, the crystal electromagnetic calorimeter (ECAL), and the brass/scin\-til\-la\-tor hadron calorimeter (HCAL).
Charged particle trajectories are measured by the inner tracker, covering $0<\phi<2\pi$ in azimuth and $|\eta|<2.5$, where the pseudorapidity $\eta$ is defined as $\eta =-\ln[\tan{\theta/2}]$, and $\theta$ is the polar angle of the trajectory of the particle with respect to the anticlockwise-beam direction.
The ECAL and the HCAL surround the tracking volume, providing high-resolution energy and direction measurements of electrons, photons, and hadronic jets. Muons are measured in gas-ionisation detectors embedded in the steel field return yoke. Extensive forward calorimetry complements the coverage provided by the barrel and endcap detectors. The detector is nearly hermetic, allowing for energy balance measurements in the plane transverse to the beam directions. A two-tier trigger system selects the pp collision events for use in physics analysis. A more detailed description of the CMS detector can be found in Ref.~\cite{bib:JINST}.

\section{Event Simulation and Theoretical Calculations}
\label{sec:simulation}

Event generators, interfaced with detailed detector simulations, are used to model experimental effects, such as reconstruction and selection efficiencies as well as detector resolutions.
For the simulation of the \ttbar\ signal sample, the \MADGRAPH event generator (v.~5.1.1.0) is used, which implements the relevant matrix elements up to three additional partons. The value of the top-quark mass is fixed to $m_{\text{t}}=172.5\GeV$ and the proton structure is described by the parton density functions (PDF) CTEQ6L1~\cite{bib:cteq}. The generated events are subsequently processed with \PYTHIA (v.~6.424)~\cite{bib:pythia} for parton showering and hadronisation, and the MLM prescription~\cite{bib:MLM} is used for the matching of the jets with parton showers.

Standard-model background samples are simulated with \MADGRAPH, \POWHEG (r1380)~\cite{bib:powheg}, or \PYTHIA, depending on the process. For the \ljets channels, W- and Z/$\gamma^{*}$-boson production with additional jets (referred to as \Wjets and \Zjets, respectively, in the following), single-top-quark production ($s$-, $t$-, and tW-channel), diboson (WW, WZ, and ZZ), and QCD multijet events are considered as background processes and listed according to their importance. For the dilepton channels, the main background contributions (in decreasing order of importance) stem from \Zjets, single-top-quark, \Wjets, diboson, and QCD multijet events. The \Wjets and \Zjets samples, including the W/Z+$\rm{c \bar c/b \bar b}$ processes, are simulated with \MADGRAPH with up to four partons in the final state. \POWHEG is used for single-top-quark production, while \PYTHIA is used to simulate diboson and QCD multijet events. Parton showering and hadronisation are also simulated with \PYTHIA in all the background samples. The \PYTHIA Z2 tune~\cite{bib:Z2tune} is used to characterise the underlying event in both the \ttbar signal and the background samples. The CMS detector response is simulated using \GEANTfour (v.~9.4)~\cite{bib:geant}.

For comparison with the measured distributions, the events in the simulated samples are normalised to an integrated luminosity of $5.0\fbinv$ according to their predicted cross sections. The latter are taken from NNLO (\Wjets, \Zjets), NLO+NNLL (single-top-quark $s$-~\cite{bib:schan}, $t$-~\cite{bib:tchan}, and tW-~\cite{bib:twchan} channels), NLO (diboson~\cite{bib:mcfm:diboson}), and leading-order (LO) (QCD multijet~\cite{bib:pythia}) calculations. Correction factors described in Section~\ref{sec:errors} are applied where necessary to improve the description of the data by the simulation. The \ttbar simulation sample is normalised to the data to present expected rates in figures in Section~\ref{sec:selection}.

In addition to the \MADGRAPH prediction, theoretical calculations obtained with the NLO generators \POWHEG and \MCATNLO (v.~3.41), and the latest NLO+NNLL~\cite{bib:ahrens_mttbar} and approximate NNLO~\cite{bib:kidonakis_pt, bib:kidonakis_y} predictions are compared, when available, to the final results presented in Section~\ref{sec:diffxsec}. The proton structure is described by the PDF sets CTEQ6M~\cite{bib:cteq} both for \POWHEG and \MCATNLO, while the NNLO MSTW2008~\cite{bib:MSTW} PDF set is used for the NLO+NNLL and for the approximate NNLO calculations. The events generated with \POWHEG and \MCATNLO are further processed with \PYTHIA and \HERWIG (v.~6.520)~\cite{bib:HERWIG}, respectively, for the subsequent parton showering and hadronisation. While \POWHEG and \MCATNLO are formally equivalent up to the NLO accuracy, they differ in the techniques used to avoid double counting of radiative corrections that may arise from interfacing with the parton showering generators. Furthermore, the parton showering in \PYTHIA is based on a transverse-momentum-ordered evolution scale, whereas in \HERWIG it is angular-ordered.

\section{Event Reconstruction and Selection}
\label{sec:selection}

The event selection is based on the decay topology of the top quark, where each top quark decays into a W boson and a b quark. The \ljets channels refer to events with only one leptonic W-boson decay, whereas in the dilepton channels each of the two W bosons decays leptonically (muon or electron). These signatures imply the identification of isolated leptons with high transverse momentum \pt, large missing transverse momentum due to neutrinos from W-boson decays escaping the detector, and highly energetic jets. The heavy-quark content of the jets is identified through b-tagging techniques.

\subsection{Lepton and Jet Reconstruction}
\label{subsec:lepjetreco}
Events are reconstructed using a particle-flow technique~\cite{bib:pf2009, bib:pf2010}, which combines signals from all sub-detectors to enhance the reconstruction performance by identifying individual particle candidates in pp~collisions. Charged hadrons from pileup events, \ie those originating from a vertex other than the one of the hard interaction, are subtracted event-by-event. Subsequently, the remaining neutral-hadron pileup component is subtracted at the level of jet energy corrections~\cite{bib:PUSubtraction}.

Electron candidates are reconstructed from a combination of the track momentum at the main interaction vertex, the corresponding energy deposition in the ECAL, and the energy sum of all bremsstrahlung photons attached to the track. They are required to have a transverse energy $\et > 30\GeV$ within the pseudorapidity interval $|\eta| < 2.1$ for the \ljets channels, while electrons in the dilepton channels have to fulfil $\et > 20\GeV$ and $|\eta| < 2.4$. As an additional quality criterion, a relative isolation $I_{\text{rel}}$ is computed. It is defined by the sum of the transverse momenta of all neutral and charged reconstructed particle candidates inside a cone around the electron in $\eta-\phi$~space of $\Delta R\equiv\sqrt{(\Delta\eta)^2 + (\Delta\phi)^2} < 0.4$ for the \ljets channels and $\Delta R<0.3$ for the dilepton channels, divided by the \pt of the electron. The transverse momentum associated with the electron is excluded from the sum.
 A relative isolation of the electron $I_{\text{rel}} < 0.125$ is demanded for the \ljets channels and $I_{\text{rel}} < 0.17$ for the dilepton channels. In addition, electrons from photon conversions, identified by missing hits in the silicon tracker, or being close to a second electron track, are rejected.

Muon candidates are reconstructed by matching the track information from the silicon tracker and the muon system. They are required to have $\pt > 30\GeV$ and $|\eta| < 2.1$ for the \ljets channels, while in the dilepton channels the corresponding selections require $\pt>20\GeV$ and $|\eta|<2.4$. Isolated muon candidates are selected if they fulfil $I_{\text{rel}} < 0.125$ for the \ljets channels and $I_{\text{rel}} < 0.20$ for the dilepton channels.
To further increase the purity of muons originating from the primary interaction and to suppress misidentified muons or muons from decay-in-flight processes, additional quality criteria, such as a minimal number of hits associated with the muon track, are required in both the silicon tracker and the muon system.

Jets are reconstructed by clustering the particle-flow candidates~\cite{bib:JME-10-011:JES} using the anti-$k_{\text{t}}$ clustering algorithm with a distance parameter of 0.5~\cite{bib:antikt}. Electrons and muons passing less stringent selections on lepton kinematic quantities and isolation compared to the ones mentioned above have been identified and are excluded from the clustering process. A jet is selected if it has $\pt > 30\GeV$ and $|\eta| < 2.4$ for both the \ljets and dilepton channels. In addition, jets originating from b quarks are identified in each decay channel by a ``combined secondary-vertex'' (CSV) algorithm~\cite{bib:btag004}, which provides a b-tagging discriminant by combining secondary vertices and track-based lifetime information. The chosen working point in the \ljets channels results in an efficiency for tagging a b jet of about 60\%, while the probability to misidentify light-flavour jets as b jets (mistag rate) is only about 1.5\%. In the dilepton channels, the working point is selected such that the b-tagging efficiency and mistag rate are about 80--85\% and around 10\%, respectively~\cite{bib:btag004}.

The missing transverse energy \ETslash is defined as the magnitude of the transverse momentum imbalance $\vec{\PTslash}$, which is the negative of the vectorial sum of the transverse momenta of all the particles reconstructed with the particle-flow algorithm~\cite{bib:MET}.

\subsection{Event Selection}
\label{subsec:evsel}
The event selection in the \ljets channels proceeds as follows. In the \ejets channel, events are triggered by an isolated electron and three or more jets fulfilling transverse momentum thresholds. The trigger efficiency within the acceptance of this analysis is above 96\%. Events in the \mujets channel are triggered by the presence of an isolated muon fulfilling $\pt$ thresholds and geometrical acceptance requirements. In this channel, the trigger efficiency is above 87\%. For the final analysis, only triggered events that have exactly one isolated lepton (leading lepton) according to the lepton identification criteria described in Section~\ref{subsec:lepjetreco} are retained.
Events with additional muons with $\pt > 10\GeV$, $|\eta| < 2.5$, and relative isolation $I_{\text{rel}} < 0.2 $ are rejected. Furthermore, in the \ejets channel, events are rejected if additional electrons have $\ET > 20\GeV$, $|\eta| < 2.5$, and $I_{\text{rel}} < 0.2$, and form a dielectron mass within $15\GeV$ of the mass of the Z boson. In the \mujets channel, events are rejected if they contain electron candidates with $\ET >15\GeV$, $|\eta| < 2.5$, and $I_{\text{rel}} < 0.2$. These lepton vetoes are meant to suppress background events from Z-boson and diboson production.
An event must contain at least four reconstructed jets satisfying the criteria mentioned in Section~\ref{subsec:lepjetreco}. At least two of them are required to be tagged as b jets in order to suppress the background contribution mainly from \Wjets events. After this selection, the remaining backgrounds are dominantly single-top-quark and top-quark-pair events from other decay channels, i.e.\,events with missing transverse energy signature. Therefore, no requirement on missing transverse energy is imposed.

In the dilepton channels, at least two isolated leptons of opposite charge are required. These events are triggered using combinations of two leptons fulfilling transverse momentum thresholds and isolation requirements. The trigger efficiency is greater than 95\% in the \mumu channel and greater than 97\% in the \mue and \ee channels. For the final analysis, only triggered events passing the lepton identification criteria described above are retained. In events with more than two leptons, only the lepton pair with the highest \pt sum is considered. Events with an invariant mass of the lepton pair smaller than $12\GeV$ are removed in order to suppress events from heavy-flavour resonance decays. Dilepton events are required to have at least two jets. At least one of the jets is required to be identified as a b jet to reduce the background contribution. In addition, backgrounds from \Zjets processes in the \mumu and \ee channels are further suppressed by requiring the dilepton invariant mass to be outside a Z-boson mass window of $91 \pm 15\GeV$ and \ETslash to be larger than $30\GeV$.

\begin{figure*}[htbp]
  \begin{center}
    \includegraphics[width=0.48\textwidth]{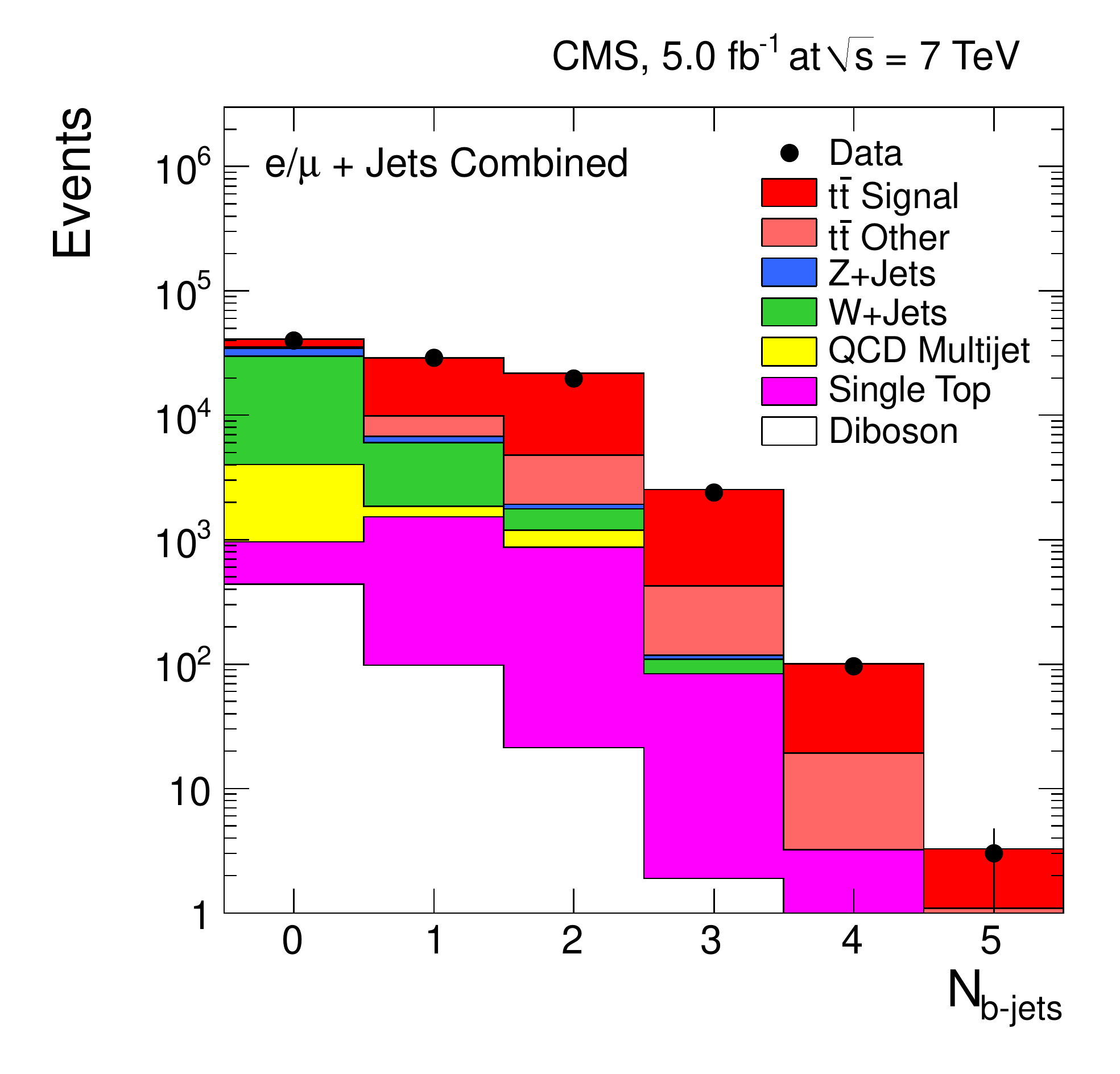}
    \includegraphics[width=0.48\textwidth]{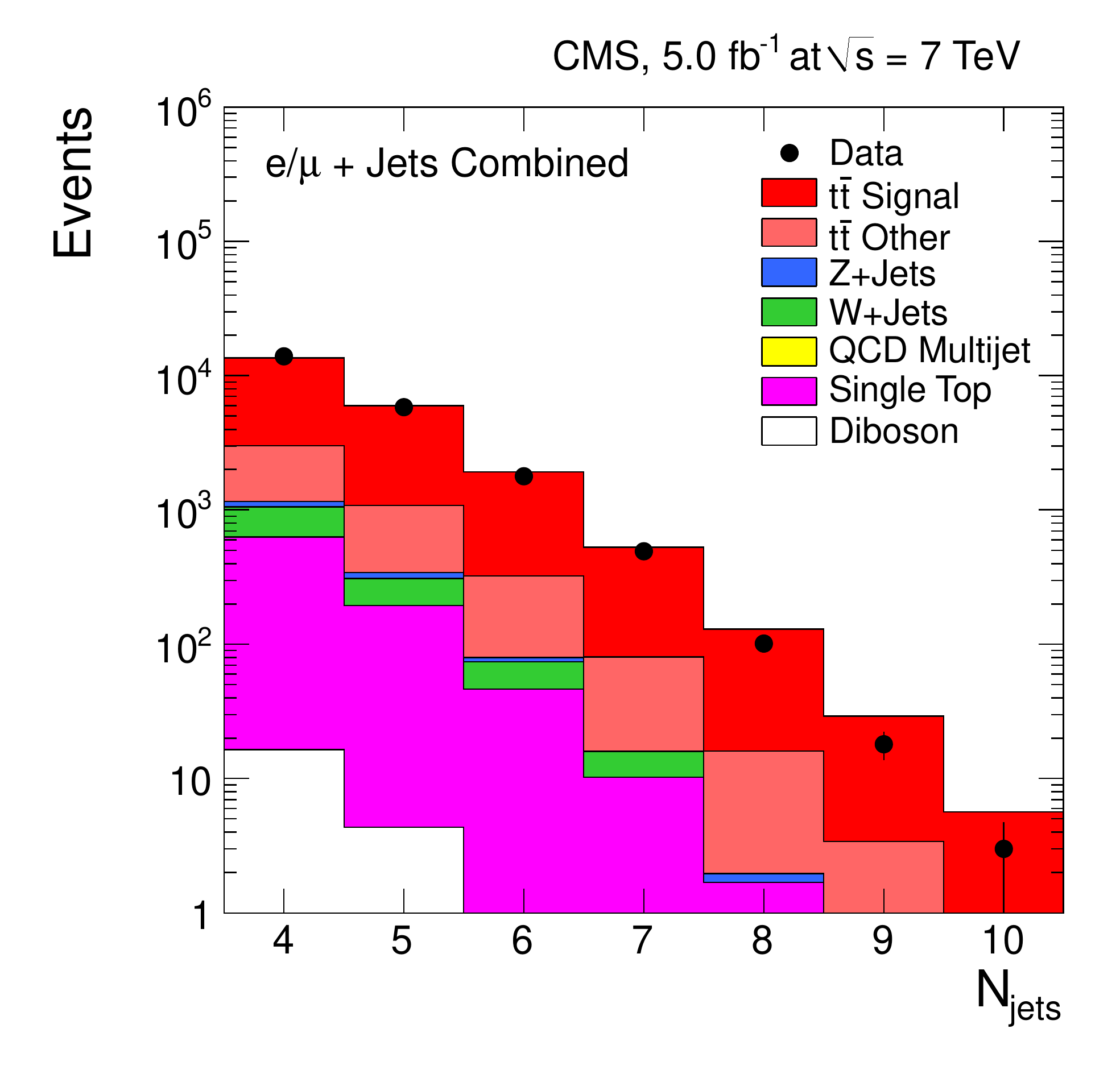}
    \includegraphics[width=0.48\textwidth]{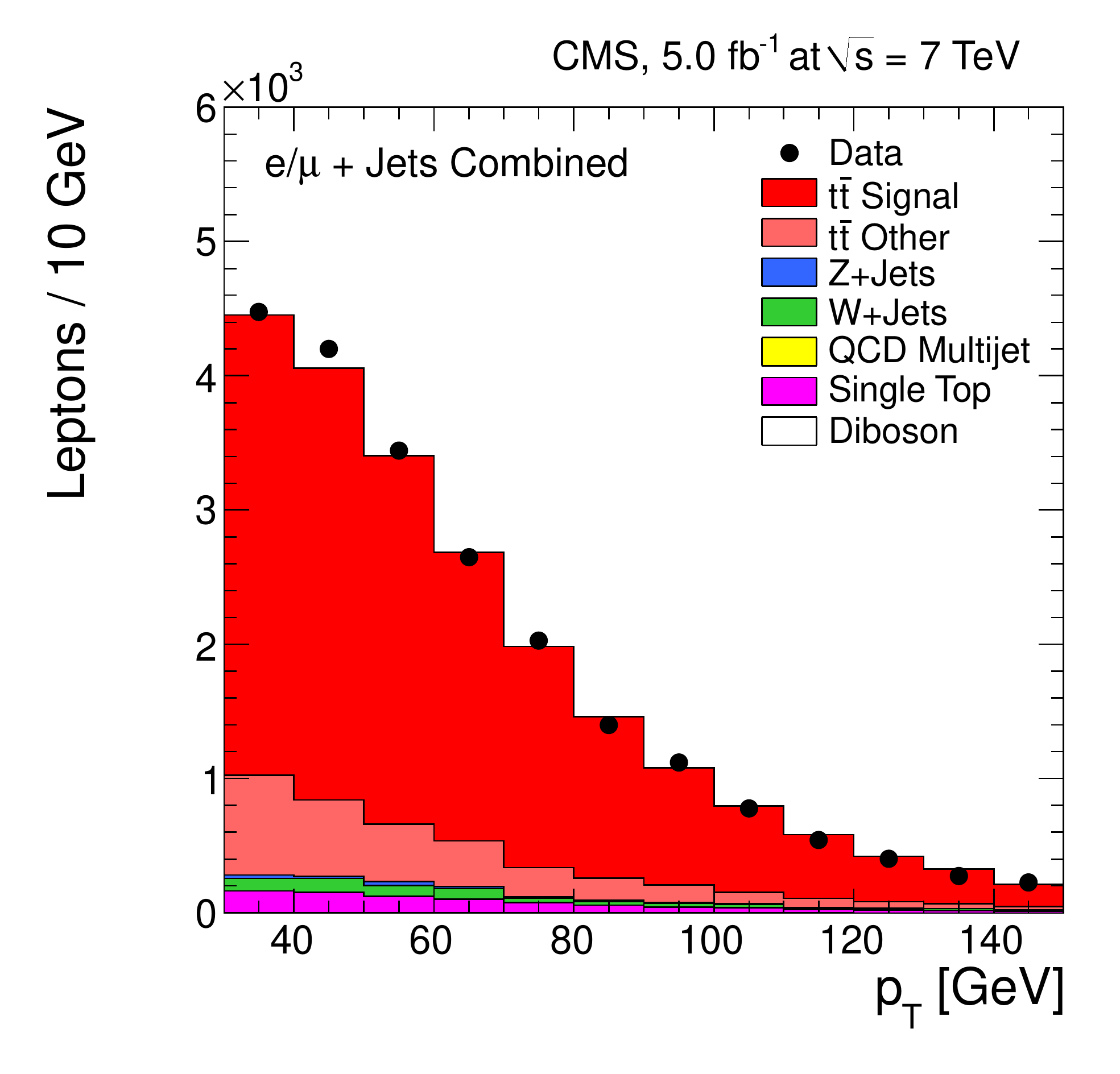}
    \includegraphics[width=0.48\textwidth]{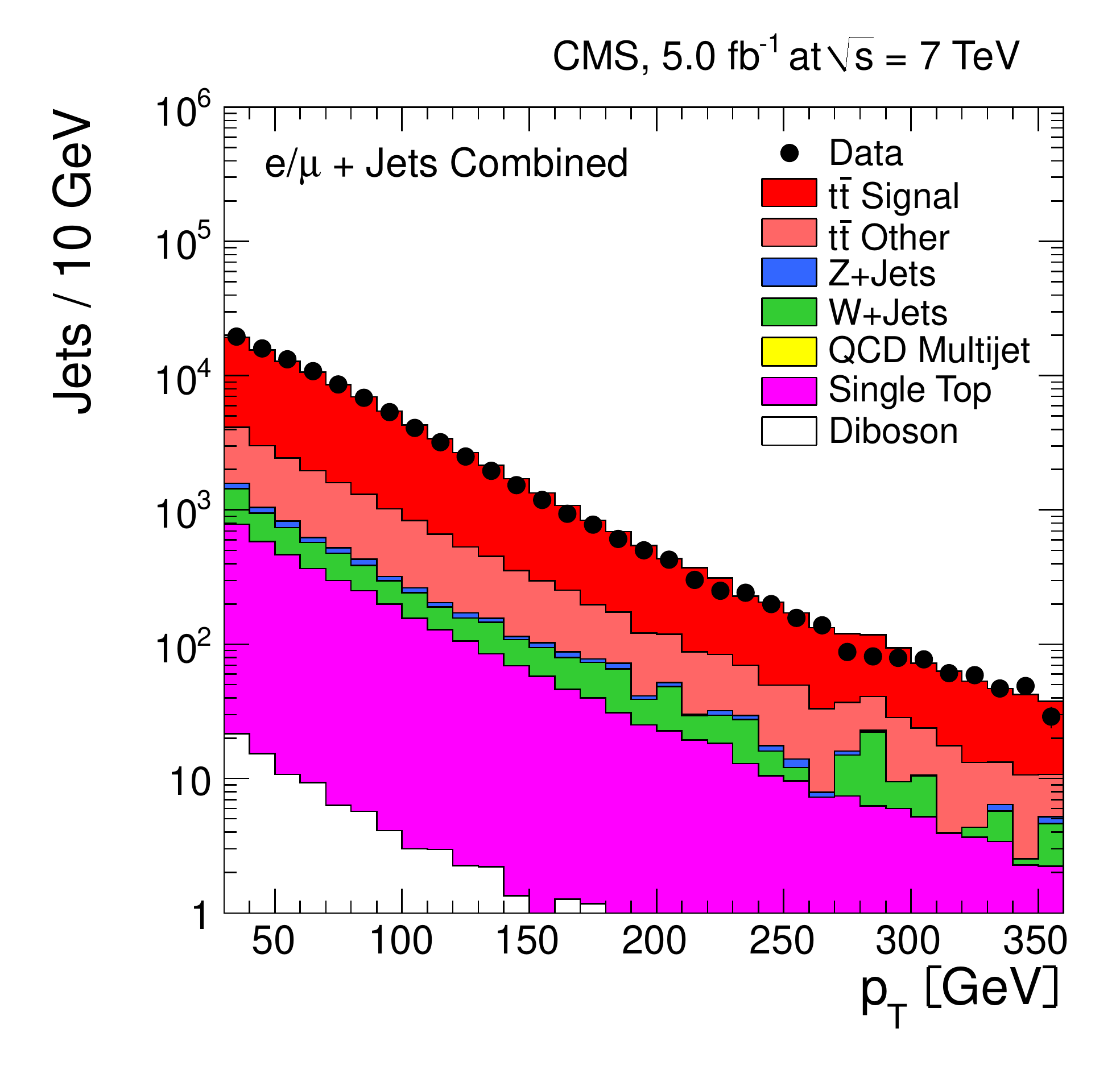}
	\setlength{\unitlength}{\textwidth}
    \caption{Basic kinematic distributions after event selection for the \ljets channels. The top left plot shows the multiplicity of the reconstructed b-tagged jets. The multiplicity of the reconstructed jets (top right), the \pt of the selected isolated leptons (bottom left), and the \pt of the reconstructed jets (bottom right) are shown after the b-tagging requirement.}
    \label{fig:ctrl:ljets}
  \end{center}
\end{figure*}

\begin{figure*}[htbp]
  \begin{center}
	\includegraphics[width=0.48\textwidth]{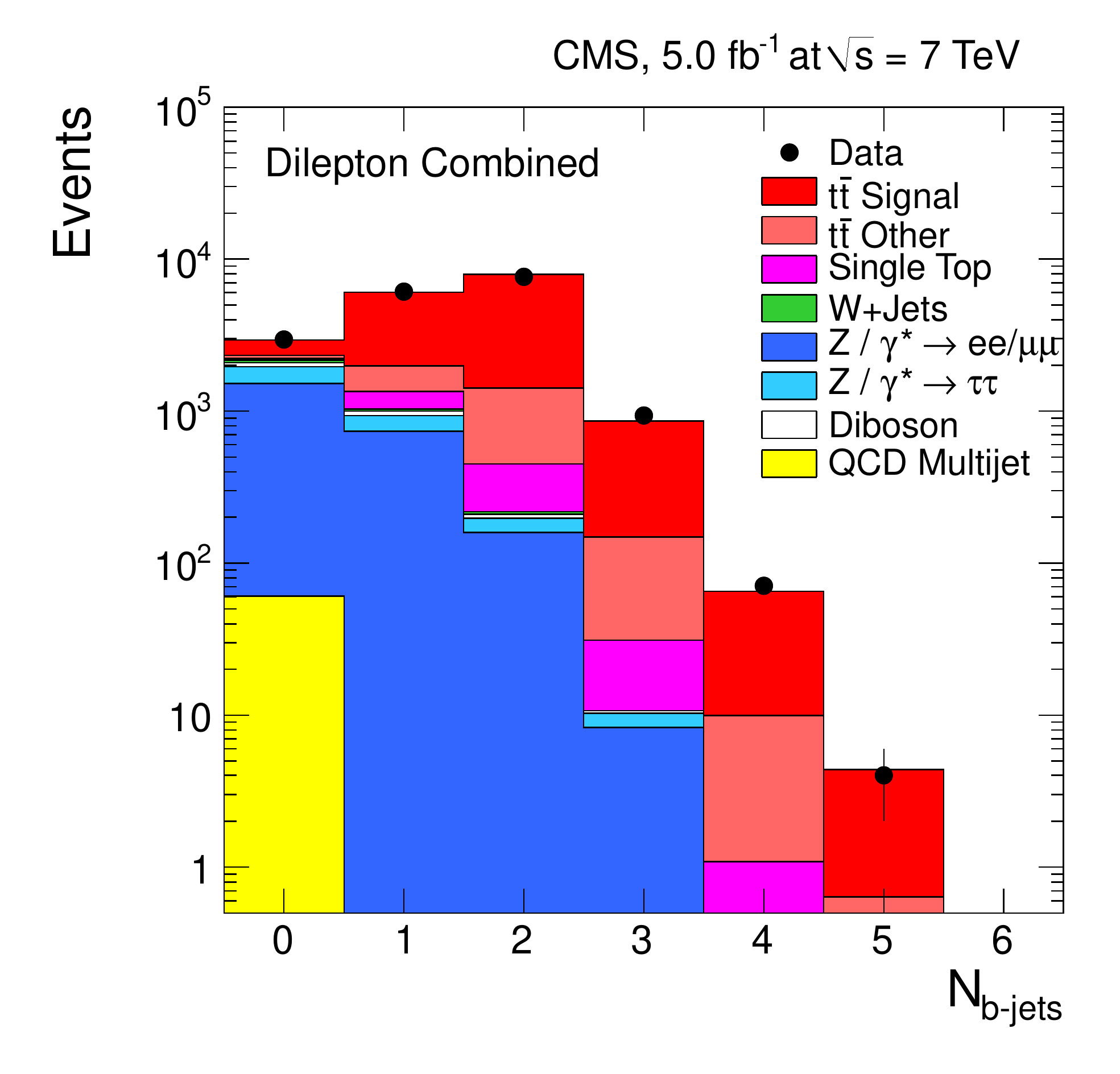}
	\includegraphics[width=0.48\textwidth]{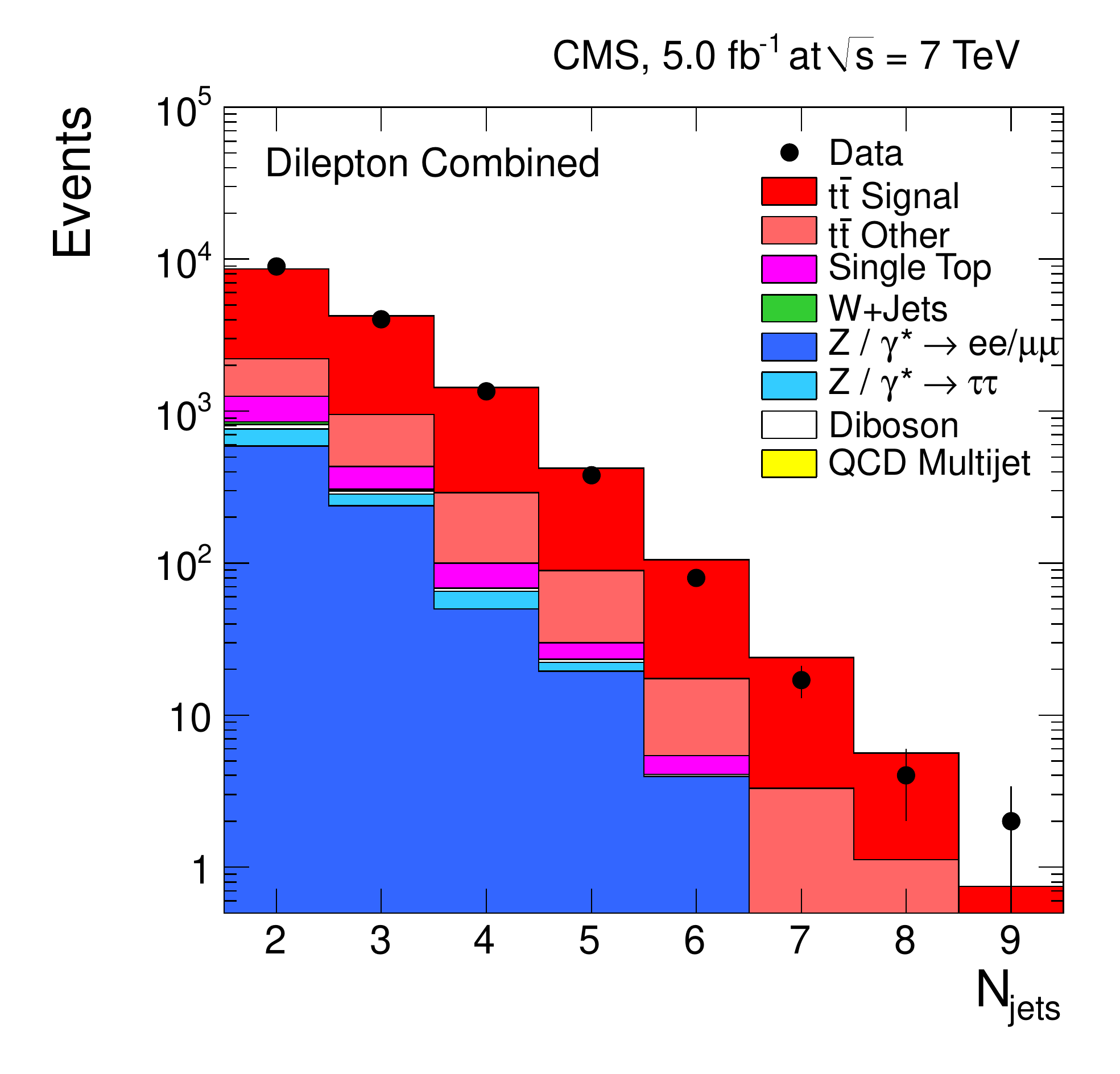}
	\includegraphics[width=0.48\textwidth]{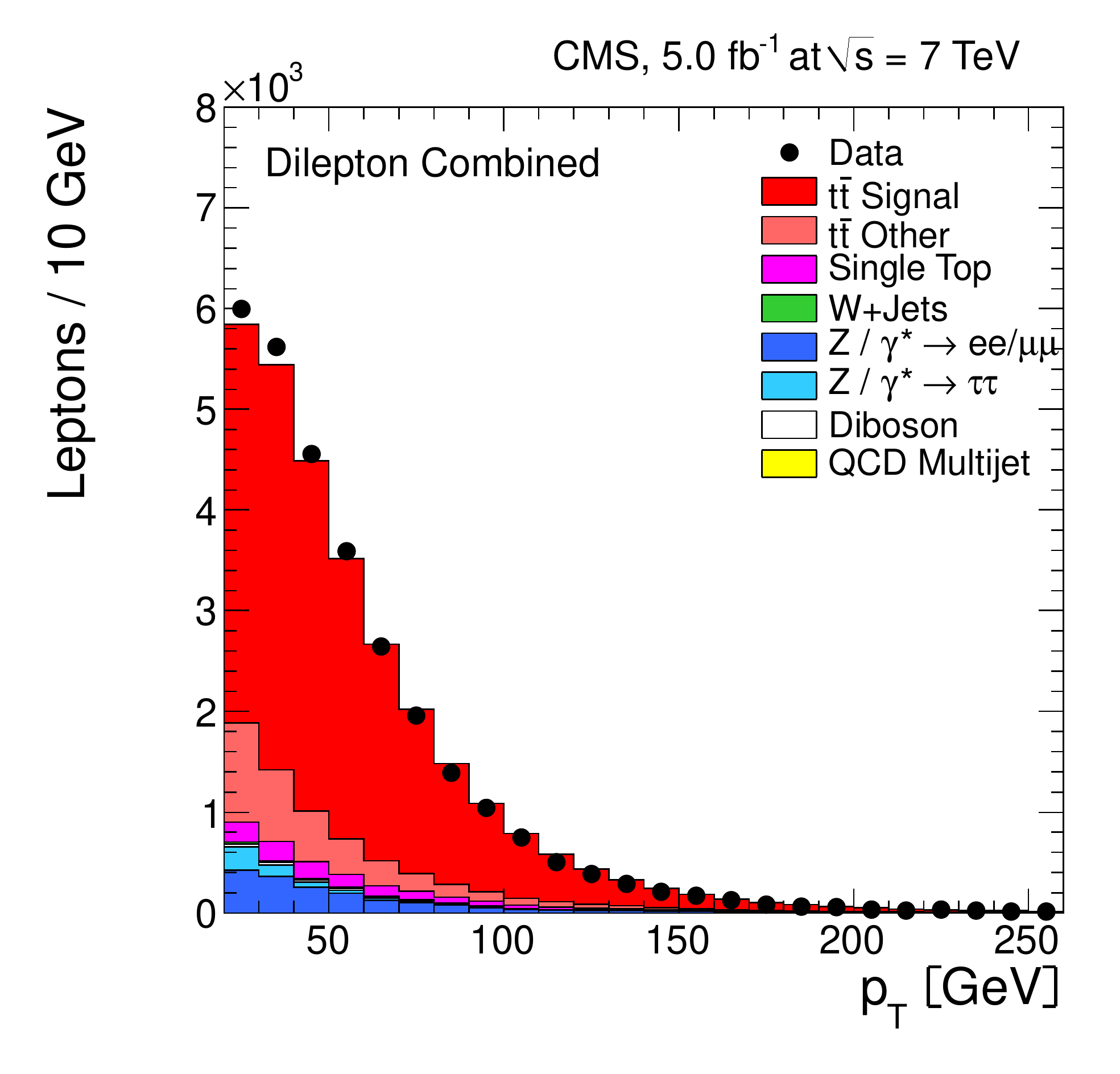}
	\includegraphics[width=0.48\textwidth]{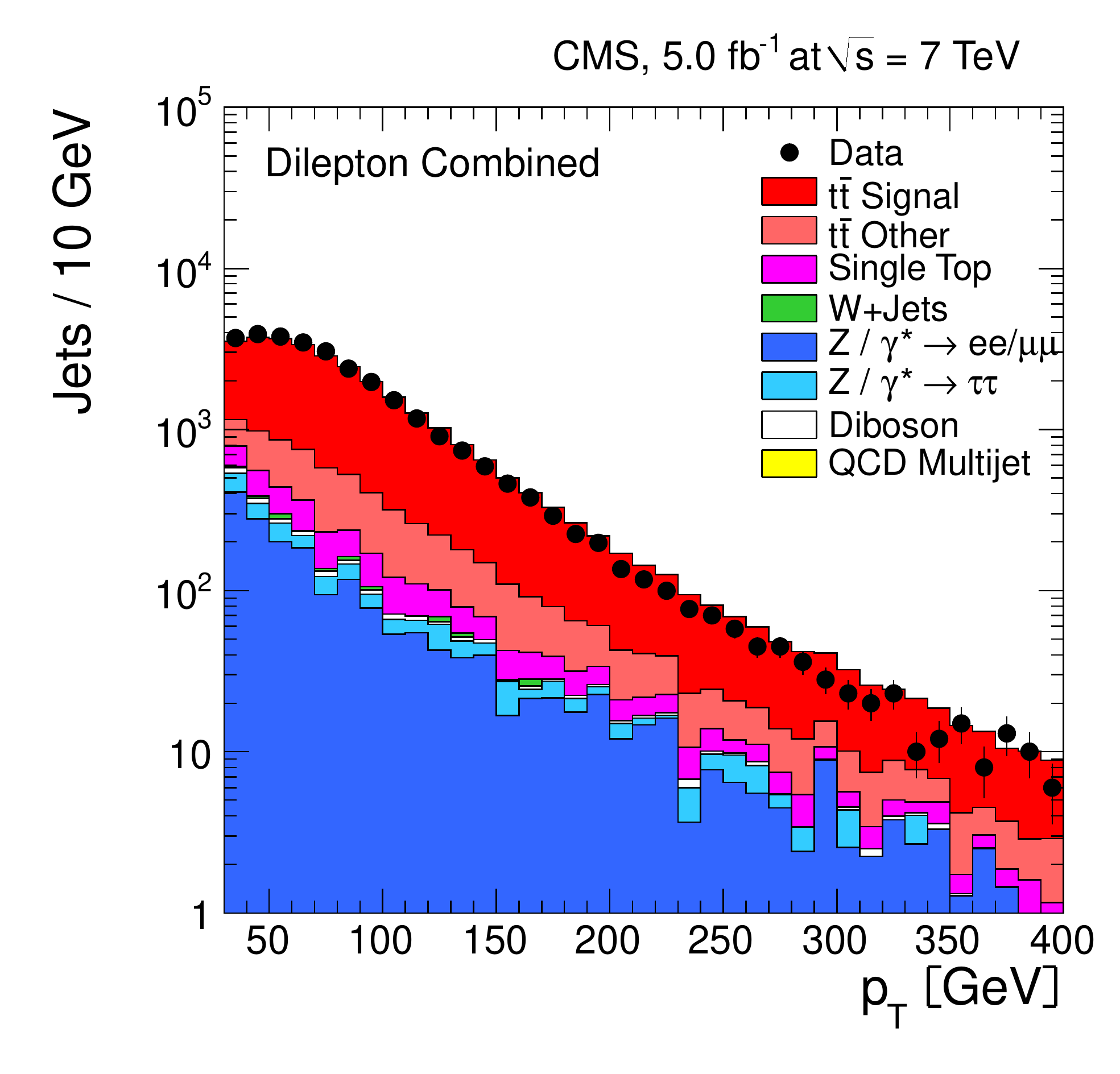}
    \setlength{\unitlength}{\textwidth}
    \caption{Basic kinematic distributions after event selection for the dilepton channels. The top left plot shows the multiplicity of the reconstructed b-tagged jets. The multiplicity of the reconstructed jets (top right), the \pt of the selected isolated leptons (bottom left), and the \pt of the reconstructed jets (bottom right) are shown after the b-tagging requirement. The Z/$\gamma^{*}$+jets background is determined from data (cf. Section~\ref{subsec:evsel}).}
  \label{fig:ctrl:dileptons}
  \end{center}
\end{figure*}

Basic distributions of the \ljets and dilepton event samples are shown in Figs.~\ref{fig:ctrl:ljets} and~\ref{fig:ctrl:dileptons}, respectively, for different steps of the selection. The data are well described by the simulation. It has been verified that the result of the measurement is unaffected by the small remaining differences.

In the \ljets channels, the main contributions to the background arise from \Wjets and QCD multijet events, which are efficiently suppressed after the b-tagging requirement. After performing the full event selection, including the kinematic top-quark-pair reconstruction described in Section~\ref{sec:kinfit}, 9\,076 events are found in the \ejets channel and 10\,766 events in the \mujets channel. In both decay channels, the \ljets signal contribution to the final event sample is about 80\%. The remaining fraction of events contains around 13\% \ttbar decays other than the \ljets channels, including \ttbar decays into $\tau$ leptons originating from the primary interaction, about 4\% single-top-quark events, around 3\% \Wjets events, and negligible fractions of \Zjets, diboson, and QCD multijet events. The background contributions are all estimated from simulation, normalised as described in Section~\ref{sec:simulation} and subtracted from the data in each bin of the measurement.

In the dilepton channels, after performing the full event selection, including the kinematic top-quark-pair reconstruction (cf. Section~\ref{sec:kinfit}), 2\,632 events are found in the \ee channel, 3\,014 in the \mumu channel, and 7\,498 in the \mue channel. Only \ttbar\ events with two leptons (electrons or muons) in the final state are considered as signal and constitute about 70--80\% of the final event sample, depending on the decay channel. All other \ttbar\ events, specifically those originating from decays via $\tau$ leptons, are considered as background and amount to 12--14\% of the final event sample. Dominant backgrounds to the \ee and \mumu channels originate from \Zjets processes. Their contribution is estimated from data following the procedure described in Ref.~\cite{bib:topPAS11_002}. The background normalisation is determined using the number of events inside the Z-peak region (removed from the candidate sample), and a correction needed for non-\Zjets backgrounds in this control region is derived from the \mue channel. The fraction of \Zjets events is found to be around 13\%. Other sources of background, including single-top-quark production and diboson events, are estimated from simulation and found to be about 6\%. The contribution arising from misidentified or genuine leptons within jets is estimated from data using like-sign events in a non-isolated region and found to be smaller than 1\%, consistent with the simulation. The background contributions are subtracted from the data in each bin of the measurement.

\subsection{Kinematic Top-Quark-Pair Reconstruction}
\label{sec:kinfit}

For both the \ljets and dilepton channels, the kinematic properties of the top-quark pair are determined from the four-momenta of all final-state objects by means of kinematic reconstruction algorithms.

In the \ljets channels, a constrained kinematic fitting algorithm is applied~\cite{bib:CMSNOTE2006023}. In the fit, the four-momenta of the selected lepton, up to five leading jets, and the $\vec{\PTslash}$ representing the transverse momentum of the neutrino, are varied according to their resolutions. The longitudinal component of the neutrino is treated as a free parameter. Moreover, the fit is constrained to reconstruct two W bosons, each with a mass of $80.4\GeV$, and top quark and antiquark with identical masses. In events with several combinatorial solutions, only the one with the minimum $\chi^{2}$ of the fit is accepted.

In the dilepton channels, an alternative kinematic reconstruction method is used~\cite{bib:Abbott:1997fv}.
In these channels, due to the presence of two neutrinos, the kinematic reconstruction is underconstrained, even after imposing a transverse-momentum balance of the two neutrinos, a W-boson invariant mass of $80.4\GeV$, and equality of the top-quark and antiquark masses. The top-quark mass can be reconstructed in a broad mass range due to detector resolution effects. To account for this, the top-quark mass for each lepton-jet combination is assumed between $100\GeV$ and $300\GeV$ in steps of $1\GeV$. In the case that an event produces more than one physical solution, those using two b-tagged jets are preferred to the ones using one b-tagged jet, and solutions using one b-tagged jets are preferred to those using no b-tagged jets. After this selection, if an event has more than one solution with the preferred b-tagging, these are ranked according to how the neutrino energies match with a simulated neutrino energy spectrum, and the highest ranked one is chosen.

For both decay channels, the kinematic reconstruction yields no physical solution for about 11\% of the events. These events are excluded from further analysis. The simulation provides a good description of the data before and after this requirement.

\begin{figure*}[htbp]
  \begin{center}
	\includegraphics[width=0.48\textwidth]{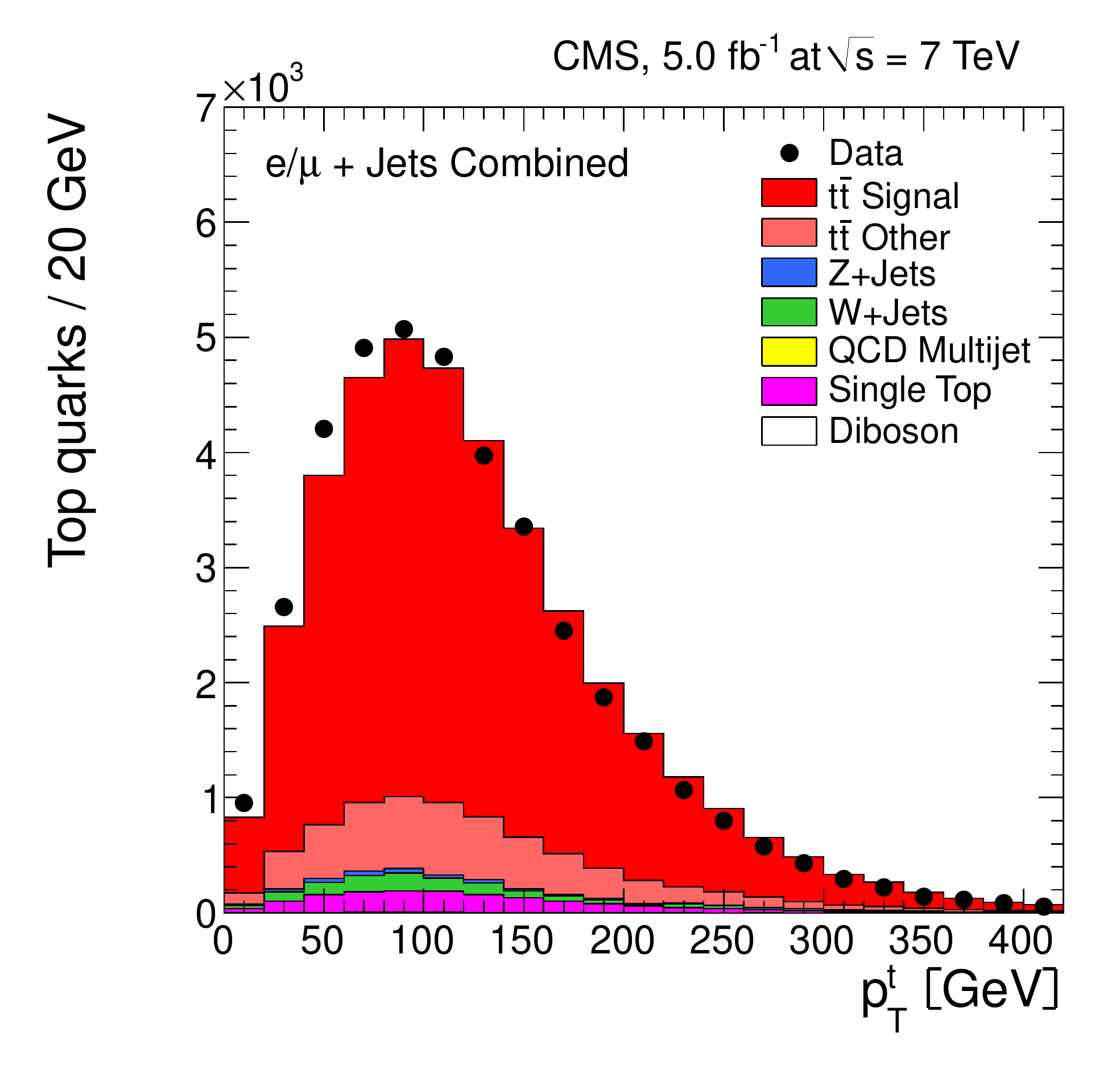}
	\includegraphics[width=0.48\textwidth]{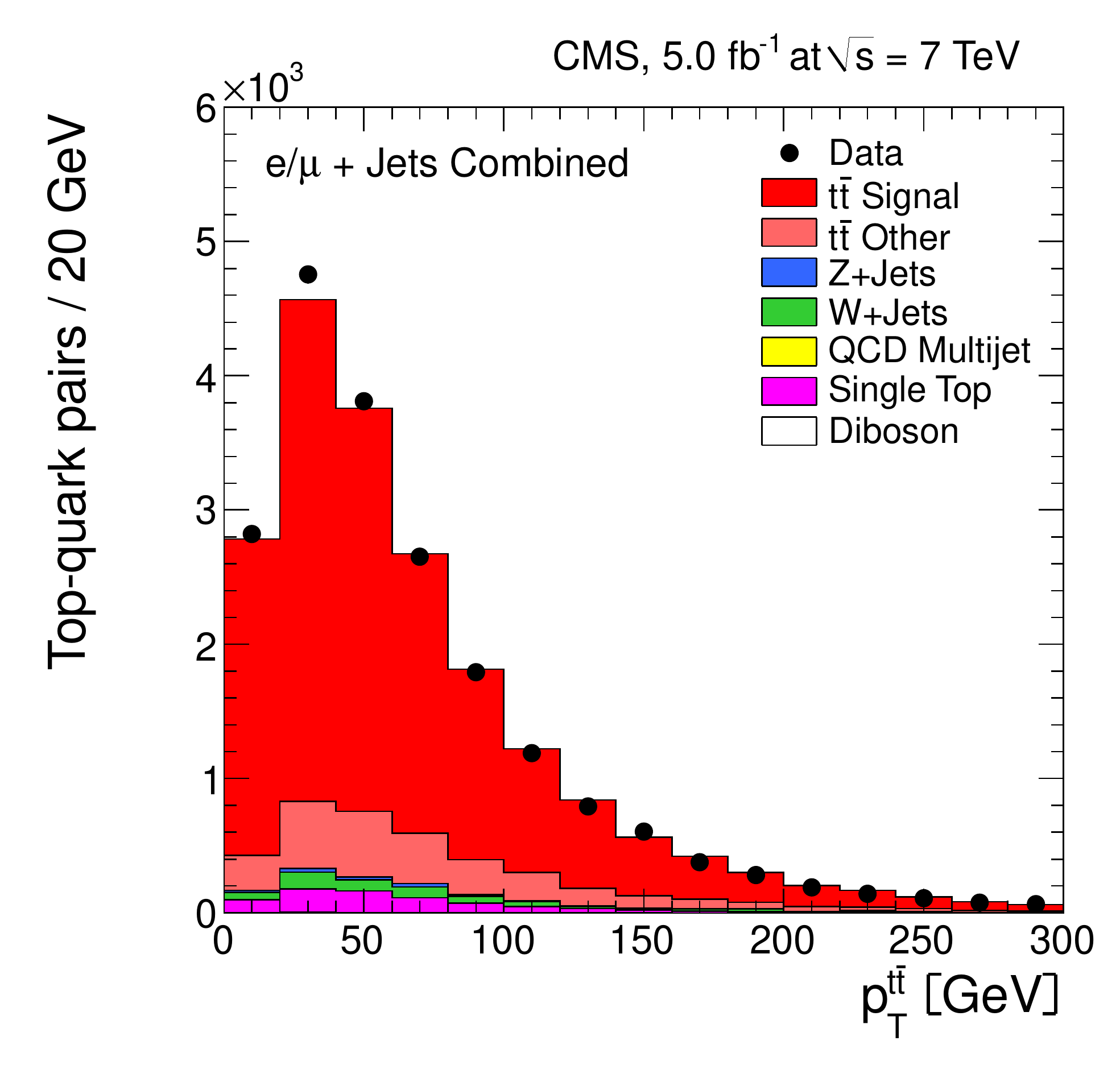}
	\includegraphics[width=0.48\textwidth]{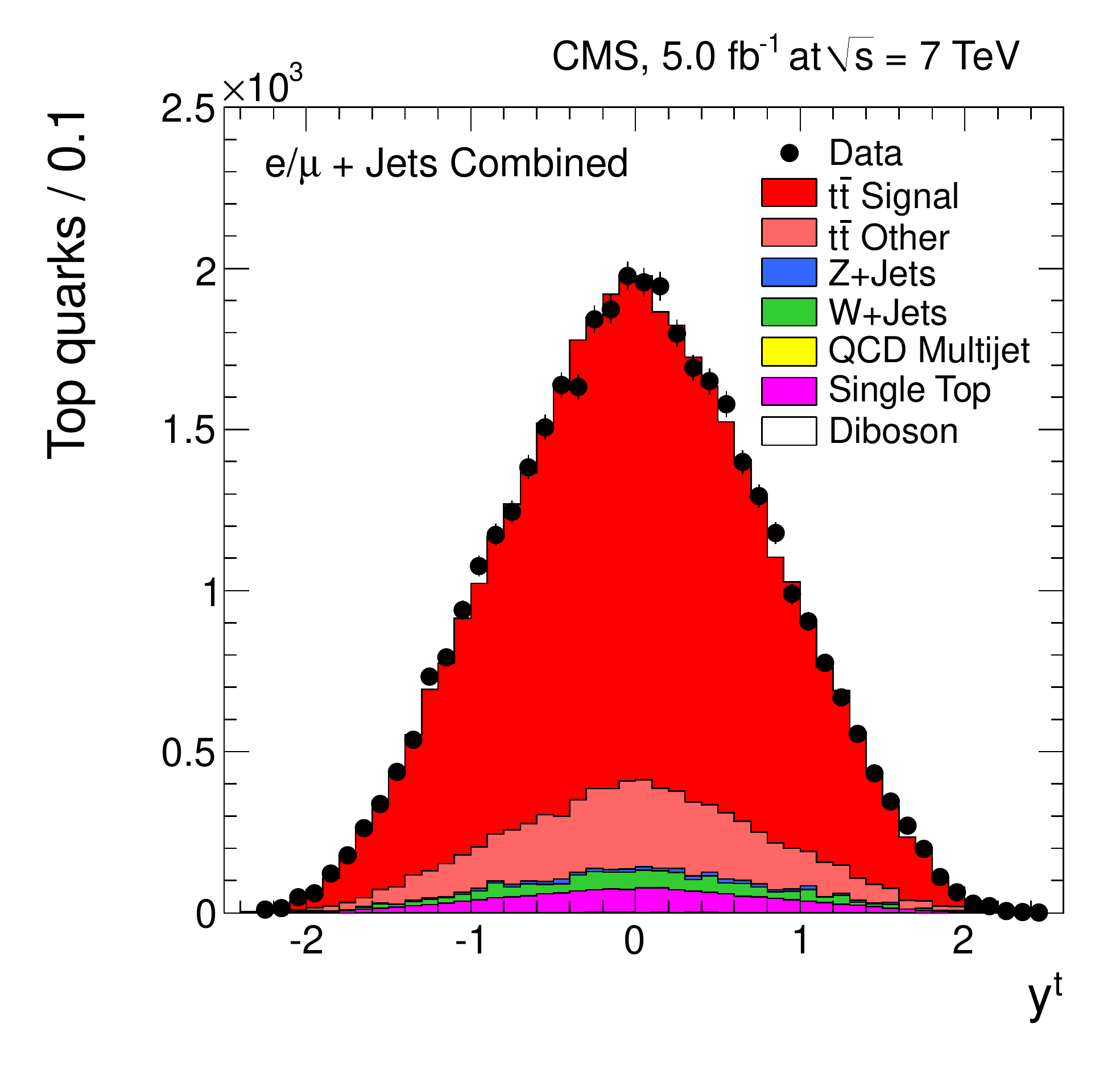}
	\includegraphics[width=0.48\textwidth]{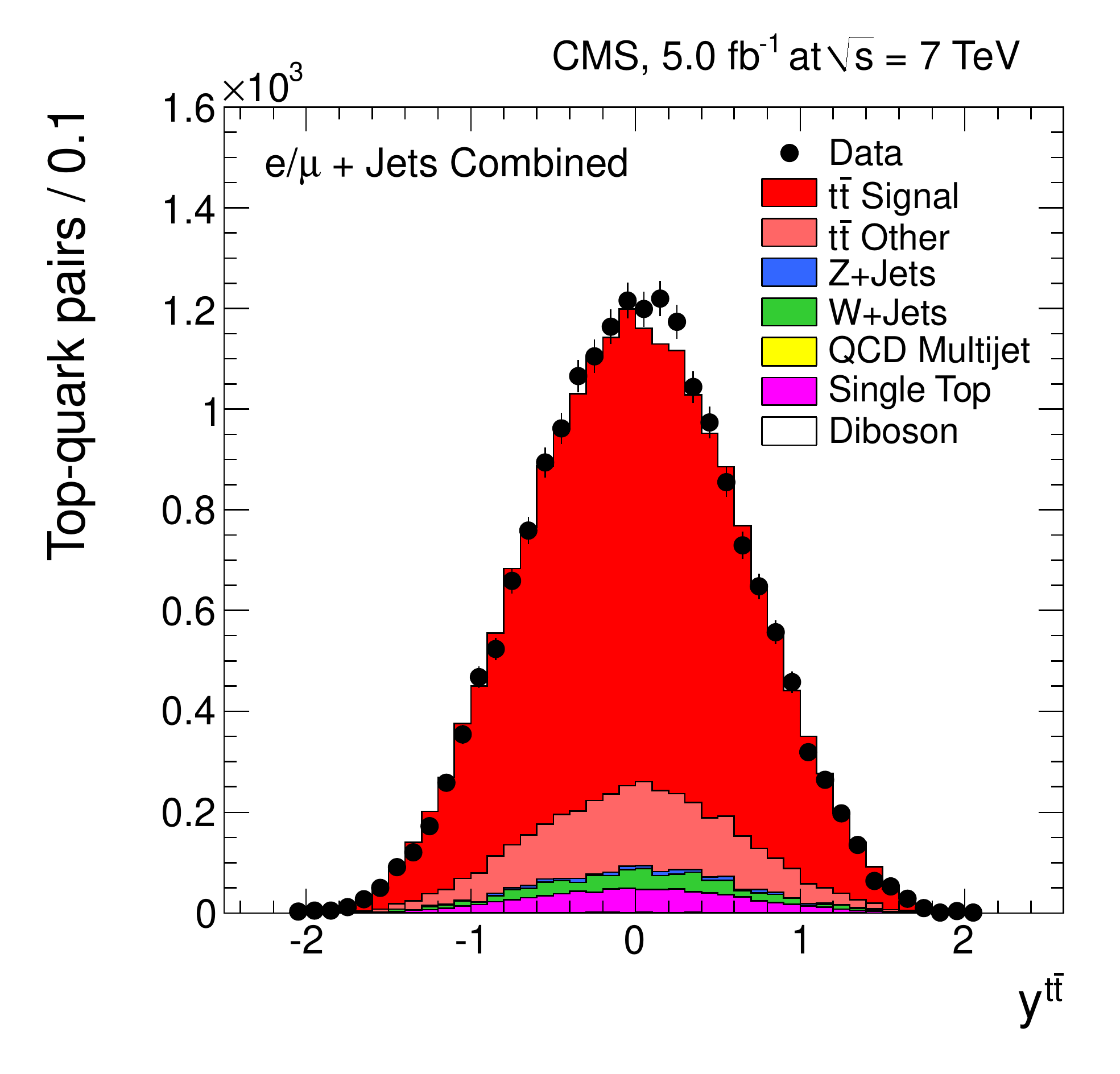}
    \caption{Distribution of top-quark and \ttbar\ quantities as obtained from the kinematic reconstruction in the \ljets channels. The left plots show the distributions for the top quarks or antiquarks; the right plots show the \ttbar\ system. The top row shows the transverse momenta, and the bottom row shows the rapidities.}
    \label{fig:kinreco:ljets}
  \end{center}
\end{figure*}

\begin{figure*}[htbp]
  \begin{center}
	\includegraphics[width=0.48\textwidth]{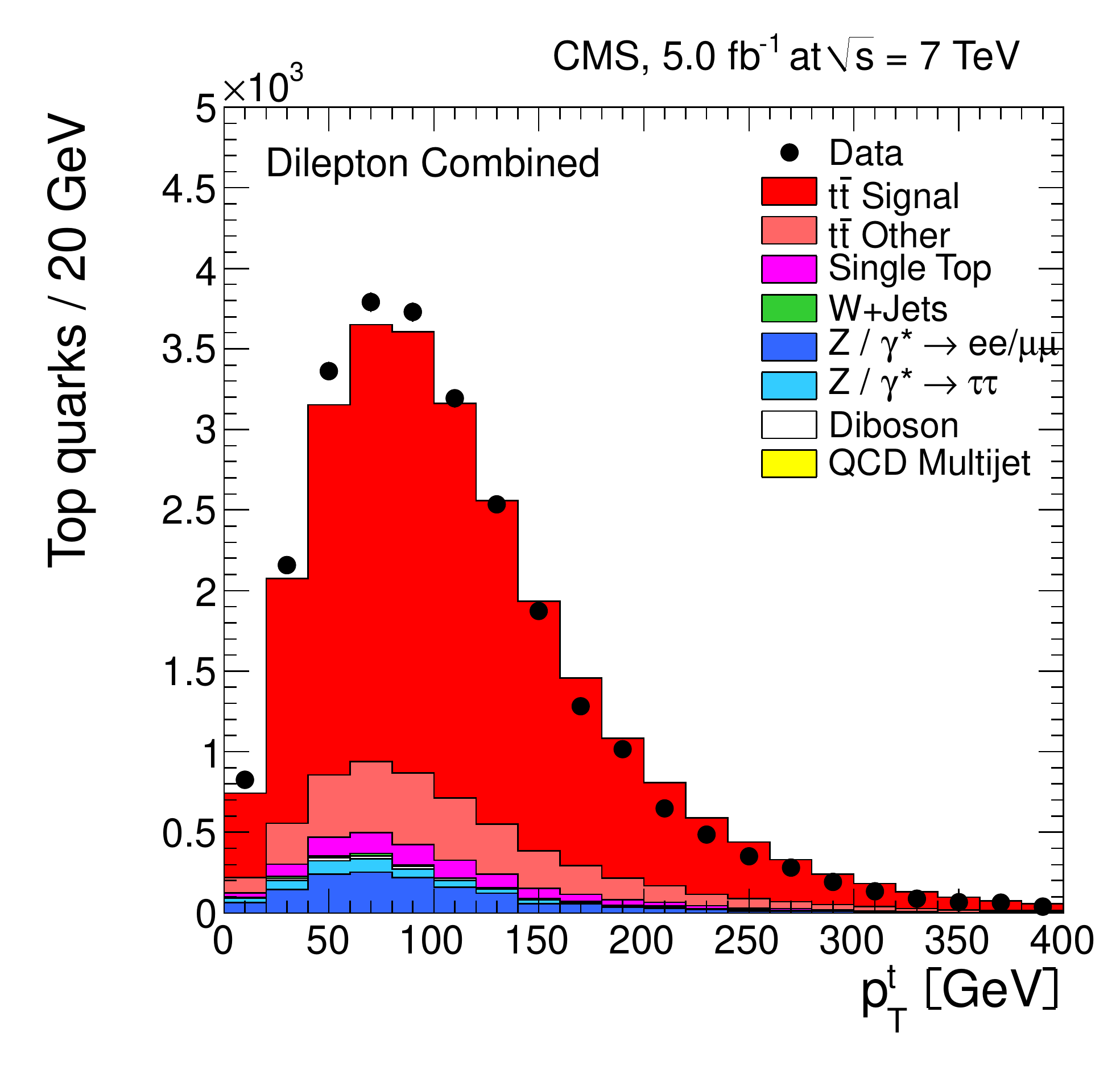}
    	\includegraphics[width=0.48\textwidth]{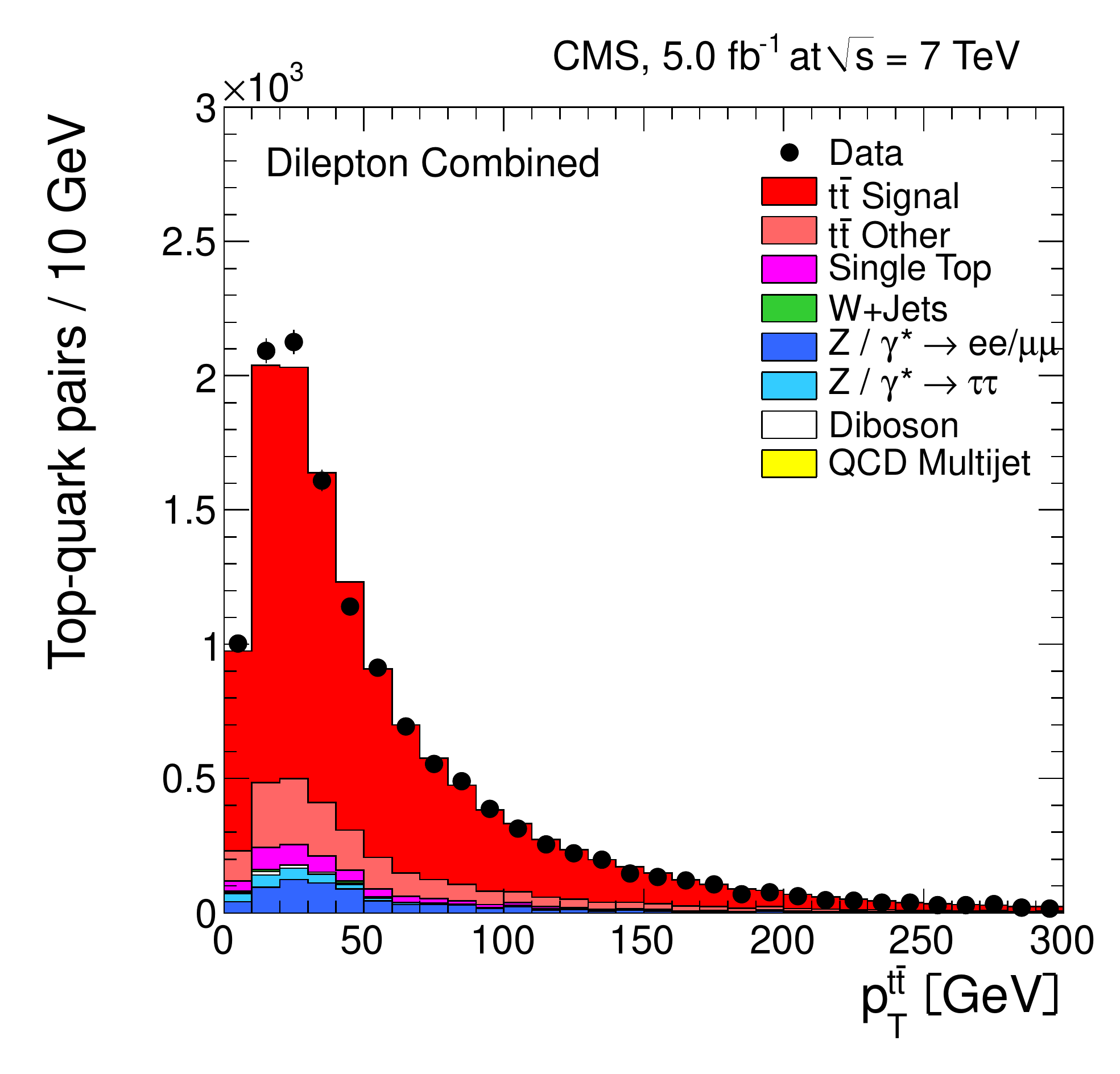}
	\includegraphics[width=0.48\textwidth]{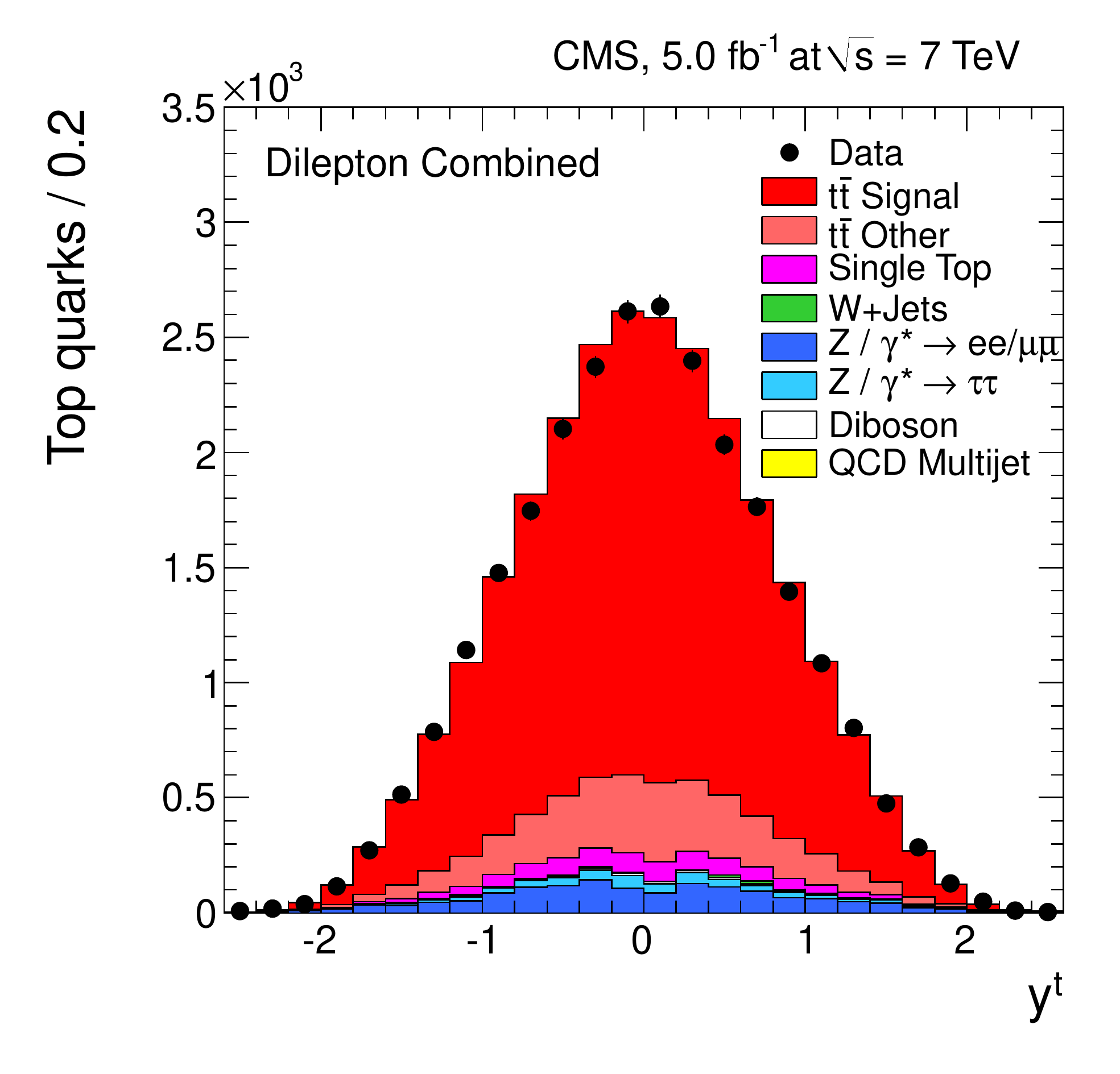}
	\includegraphics[width=0.48\textwidth]{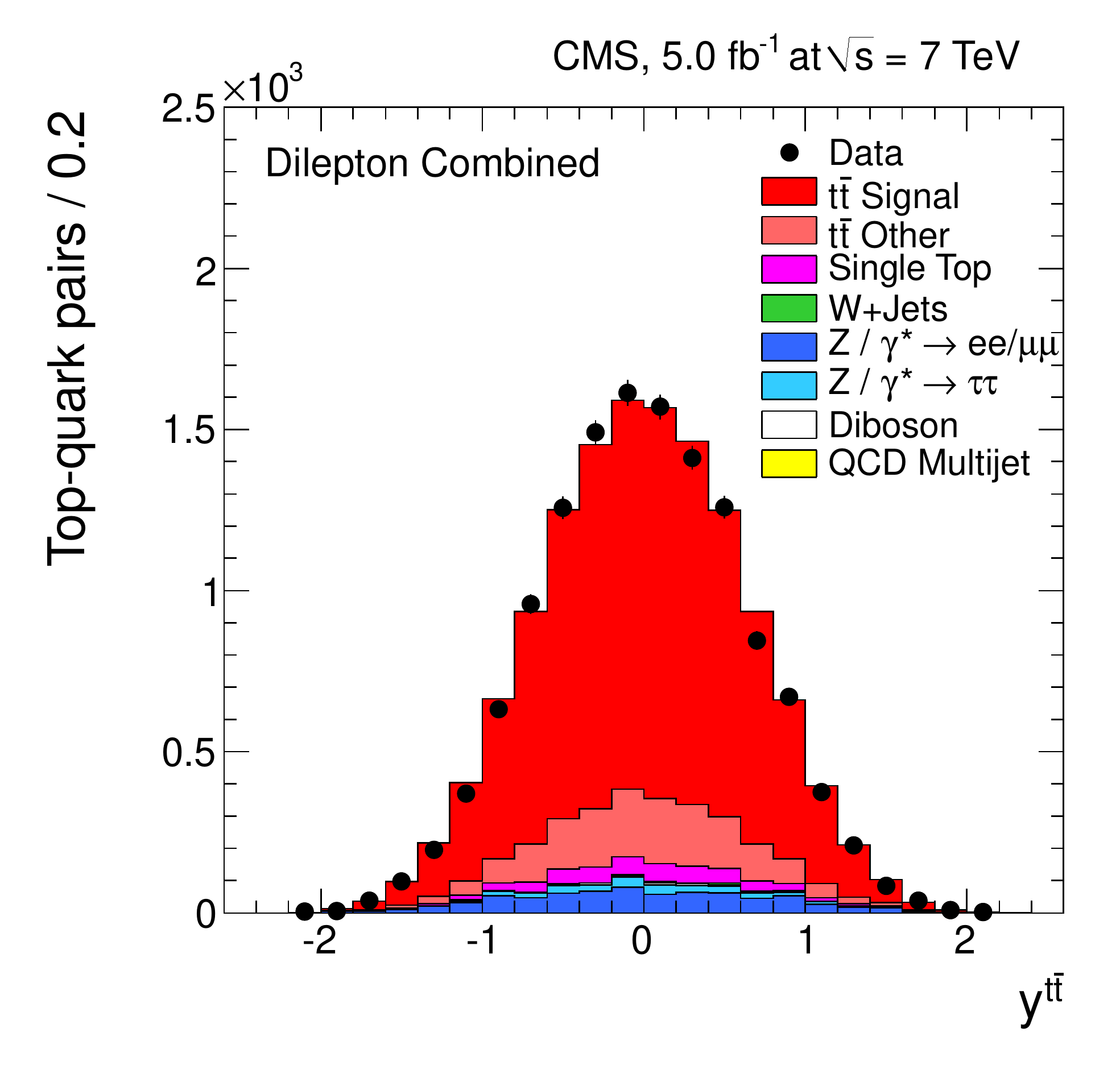}
    \caption{Distribution of top-quark and \ttbar\ quantities as obtained from the kinematic reconstruction in the dilepton channels. The left plots show the distributions for the top quarks or antiquarks; the right plots show the \ttbar\ system. The top row shows the transverse momenta, and the bottom row shows the rapidities. The Z/$\gamma^{*}$+jets background is determined from data (cf. Section~\ref{subsec:evsel}).}  	
    \label{fig:kinreco:dileptons}
  \end{center}
\end{figure*}

Distributions of the top-quark or antiquark and \ttbar\ kinematic observables ($\pt^{\text{t}}$, $y^{\text{t}}$, $\pt^{\ttbar}$, and $y^{\ttbar}$, where $y$ is the rapidity defined as $y =1/2 \cdot \ln[(E+p_z)/(E-p_z)]$, with $E$ and $p_z$ denoting the particle energy and the momentum along the anticlockwise-beam axis, respectively) as obtained from the kinematic reconstruction, are presented in Fig.~\ref{fig:kinreco:ljets} for the \ljets event sample and in Fig.~\ref{fig:kinreco:dileptons} for the dilepton event sample. In general, the data are well described by the simulation within uncertainties. As in Figs.~\ref{fig:ctrl:ljets} and~\ref{fig:ctrl:dileptons}, the final results are not affected by the small remaining differences in normalisation between data and simulation. For both channels, the measured \pt distributions show a trend of being shifted to lower transverse momenta compared to the simulated distributions.

\section{Systematic Uncertainties}
\label{sec:errors}

Systematic uncertainties on the measurement arise from detector effects as well as from theoretical uncertainties. Each systematic uncertainty is investigated separately, and determined individually in each bin of the measurement, by variation of the corresponding efficiency, resolution, or scale within its uncertainty. Correction factors, subsequently referred to as scale factors, are applied where necessary to improve the description of the data by the simulation. For each variation, the measured normalised differential cross section is recalculated, and the difference of the varied result to the nominal result in each bin is taken as the systematic uncertainty. The overall uncertainty on the measurement is then derived by adding the individual contributions in quadrature. The dominant uncertainties on the normalised differential cross section originate from the lepton selection, the b tagging, and from model uncertainties. A summary of the typical systematic uncertainties of the normalised differential cross section, obtained by averaging over all quantities and bins, is given in Table~\ref{tab:systematicsDiff} and a detailed description is given in Sections~\ref{subsec::ExpUncertainties} and \ref{subsec::ModelUncertainties}.

\subsection{Experimental Uncertainties}
\label{subsec::ExpUncertainties}

The efficiency of the single-muon trigger in \mujets events is determined using the ``tag-and-probe'' method~\cite{bib:tp} with Z-boson event samples. A dependence on the pseudorapidity of the muon of a few percent is observed and scale factors are derived. In order to determine the efficiency of the electron-trijet trigger in \ejets events, the tag-and-probe method is also applied to the electron branch, while independent control triggers are used for the hadronic part. Good agreement is observed between data and simulation, and scale factors very close to unity are applied. The lepton identification and isolation efficiencies for the \ljets channels obtained with the tag-and-probe method agree well between data and simulation, so that corrections very close or equal to unity are applied. The systematic uncertainties are determined by shape-dependent variations of trigger and selection efficiencies within their uncertainties. Lepton trigger efficiencies in the dilepton channels are measured using triggers that are only weakly correlated to the dilepton triggers.
The lepton identification and isolation uncertainties in the dilepton channels are also determined using the tag-and-probe method, and are found to be described very well by the simulation for both electrons and muons. The overall difference between data and simulation in bins of pseudorapidity and transverse momentum is estimated to be less than 2\% for electrons, while scale factors for muons are found to be close to unity.

To estimate the uncertainty on the jet energy scale, the reconstructed jet energy is varied as a function of the transverse momentum and the pseudorapidity of the jet (typically by a few percent)~\cite{bib:JME-10-011:JES}.
The uncertainty on the jet energy resolution (JER) is determined by variation of the simulated JER up and down by about $\pm6\%$, $\pm9\%$, and $\pm20\%$, for the pseudorapidity regions $|\eta| < 1.7$, $1.7 < |\eta| < 2.3$, and $|\eta| > 2.3$, respectively~\cite{bib:JME-10-011:JES}.

The uncertainty due to background normalisation is determined by variation of the background yields. For the \ljets channels, the background normalisation is varied by $\pm30\%$ for the single-top-quark and diboson samples, and by $\pm50\%$ for the QCD samples~\cite{bib:TOP-10-002_paper, bib:TOP-10-003_paper}. For the W/Z-boson samples, this uncertainty is covered by variations of the kinematic scales of the event process (renormalisation and factorisation scales and jet-parton matching), as described in ~Section~\ref{subsec::ModelUncertainties}.
In the \ee and \mumu channels, the dominant background from \Zjets processes as determined from data (cf. Section~\ref{sec:selection}) is varied in normalisation by $\pm30\%$. In addition, variations of the background contributions from single-top-quark and diboson events up and down by $\pm30\%$ are performed~\cite{bib:TOP-11-002_paper, bib:TOP-11-005_paper}.

The uncertainty on the b-tagging efficiency is determined by dividing the b-jet distributions for transverse momentum and pseudorapidity into two bins at the median of the respective distributions. These are $\pt=65\GeV$ and $|\eta|=0.7$ for the \ljets and $\pt=65\GeV$ and $|\eta|=0.75$ for the dilepton channels. The b-tagging scale factors for the b jets in the first bin are scaled up by half of the uncertainties quoted in Ref.~\cite{bib:btag004}, while those in the second bin are scaled down and vice versa, so that a maximum variation is assumed and the difference between the scale factors in the two bins amounts to the full uncertainty. The variations are performed separately for the transverse-momentum and pseudorapidity distributions.

The kinematic reconstruction of the top quarks is generally found to be very well described by the simulation, and the resulting uncertainties are small. In the case of the \ljets analysis, the uncertainty of the kinematic fit is included in the variations of jet energy scales and resolutions. In the dilepton analysis, the bin-to-bin uncertainty is determined from the small remaining difference between the simulation and the data.

The pileup model estimates the mean number of additional pp interactions to be about 9.5 events for the analysed data. This estimate is based on the total inelastic proton-proton cross section, which is determined to be 73.5 mb~\cite{bib:ppInelXSec}. The systematic uncertainty is determined by varying this cross section within its uncertainty of $\pm8\%$.

\subsection{Model Uncertainties}
\label{subsec::ModelUncertainties}

The impact of theoretical assumptions on the measurement is determined by repeating the analysis, replacing the standard \MADGRAPH signal simulation by dedicated simulation samples, as described below.

The uncertainty on the modeling of the hard-production process is assessed by varying the renormalisation and factorisation scale in the \MADGRAPH signal samples up and down by a factor of two with respect to its nominal value, equal to the $Q^2$ of the hard process ($Q^2 = m^2_{\text{t}} + \Sigma p^2_{\mathrm{T}}$). Furthermore, the effect of additional jet production in \MADGRAPH is studied by varying the threshold between jet production at the matrix-element level and via parton showering up and down by a factor of two with respect to the nominal value of 20\GeV. In the \ljets channels, variations of the renormalisation and factorisation scale are also applied to single-top-quark events to determine a shape uncertainty for this background contribution. Additionally, both kinematic scales are varied for W- and Z-boson background events to associate a shape and background normalisation uncertainty to these samples. Each type of variation is applied simultaneously for the W- and Z-boson samples.

The uncertainty due to the hadronisation model is determined by comparing samples simulated with \POWHEG and \MCATNLO using \PYTHIA and \HERWIG, respectively, for hadronisation.
The dependence of the measurement on the top-quark mass is estimated from dedicated \MADGRAPH simulation samples in which the top-quark mass is varied with respect to the value used for the default simulation. The resulting variations are scaled linearly according to the present world average uncertainty of $0.9\GeV$.
The effect of the uncertainty from parton density functions on the measurement is assessed by reweighting the sample of simulated \ttbar signal events. For this reweighting, the minimum and maximum variations with respect to the nominal value is obtained by following the PDF4LHC prescription~\cite{bib:PDF4LHC} using the NLO PDF sets CT10~\cite{bib:CT10}, \breakhere MSTW2008NLO, and NNPDF2.1~\cite{bib:NNPDF}.

\begin{table}
	\begin{center}
		\caption{Breakdown of typical systematic uncertainties for the normalised differential cross section in the \ljets and dilepton channels. The background uncertainty for the \ljets channels includes normalisation uncertainties as well as uncertainties due to variations of the kinematic scales in W/Z-boson events.}
		\label{tab:systematicsDiff}
		\begin{tabular}{lcccc}
			\hline
			\hline
			\textbf{Source} & \multicolumn{2}{c}{\textbf{Systematic uncertainty (\%)}} \\
			                             & \ljets &  dileptons       \\
			\hline
			Trigger efficiency              &  0.5  &  1.5  \\
			Lepton selection                &  0.5  &  2.0  \\
			Jet energy scale                &  1.0  &  0.5  \\
			Jet energy resolution           &  0.5  &  0.5  \\
			Background                      &  3.5  &  0.5  \\
			b tagging                       &  1.0  &  0.5  \\
			Kin. reconstruction                   &  --   &  0.5  \\
			Pileup                         &  0.5  &  0.5  \\
			\hline
			Fact./renorm. scale               &  2.0  &  1.0  \\
                        ME/PS threshold                &  2.0  &  1.0  \\
			Hadronisation                   &  2.0  &  2.0  \\
			Top-quark mass                  &  0.5  &  0.5  \\
                        PDF choice                     &  1.5  &  1.0  \\
			\hline
			\hline
		\end{tabular}
	\end{center}
\end{table}

\section{Normalised Differential Cross Section}
\label{sec:diffxsec}

The normalised cross section in each bin $i$ of each observable $X$ is determined through the relation:

\begin{equation}
\label{eq:diffXsec}
\frac{1}{\sigma}\frac{\rd\sigma_{i}}{\rd X}=\frac{1}{\sigma}\frac{x_{i}}{\Delta^{\text{X}}_{i}  \; \mathcal{L}}
\end{equation}

In each bin of the measurement, $x_{i}$ represents the number of signal events in data determined after background subtraction and corrected for detector efficiencies, acceptances, and migrations, as described below. The normalised differential cross section is then derived by scaling to the integrated luminosity $\mathcal{L}$ and by dividing the corrected number of events by the width $\Delta_{i}^{\text{X}}$ of the bin and by the measured total cross section $\sigma$ in the same phase space. Due to the normalisation, those systematic uncertainties that are correlated across all bins of the measurement, and therefore only affect the normalisation, cancel out.

Effects from trigger and detector efficiencies and resolutions, leading to the migration of events across bin boundaries and statistical correlations among neighbouring bins, are corrected by using a regularised unfolding method~\cite{bib:svd, bib:blobel}. For each measured distribution, a response matrix that accounts for migrations and efficiencies is calculated from the simulated \MADGRAPH \ttbar\ signal sample. The generalised inverse of the response matrix is used to obtain the unfolded distribution from the measured distribution by applying a $\chi^2$ technique.
To avoid non-physical fluctuations, a smoothing prescription (regularisation) is applied. The regularisation level is determined individually for each distribution using the averaged global correlation method~\cite{bib:james}.
To keep the bin-to-bin migrations small, the width of the bins of the measurement are chosen according to their purity and stability. For a certain bin~$i$, the number of particles generated and correctly reconstructed~$N^{\text{gen\&rec}}_i$ is determined. The purity~$p_i$ is then this number divided by the total number of reconstructed particles in the same bin $N^{\text{rec}}_i$: $p_i=N^{\text{gen\&rec}}_i/N^{\text{rec}}_i$. Similarly, the stability~$s_i$ is defined as that number scaled to the total number of generated particles in the particular bin $N^{\text{gen}}_i$, yielding $s_i=N^{\text{gen\&rec}}_i/N^{\text{gen}}_i$. In this analysis, the purity and stability of the bins are typically 50\% or larger. The performance of the unfolding procedure is tested for a possible bias due to the choice of the input model (the \ttbar \MADGRAPH signal simulation). It has been verified that, by either reweighting the signal simulation or injecting a resonant \ttbar\ signal into the signal simulation, the unfolding procedure still reproduces the results correctly when using the default \MADGRAPH \ttbar signal simulation to account for migrations and efficiencies.

The analysis proceeds by measuring the normalised differential cross section in the \ljets channels and dilepton channels. For each kinematic distribution, the event yields in the separate channels are added up, the background is subtracted, and the unfolding is performed. As a cross-check, it has been verified that the measurements in the individual channels are in agreement with each other within the uncertainties.

The systematic uncertainties in each bin are assessed from the variations of the combined cross sections. This means that the full analysis is repeated for every systematic variation and the difference with respect to the nominal combined value is taken as the systematic uncertainty for each bin and each measured observable. By using this method, the possible correlations of the systematic uncertainties between the different channels and bins are taken into account.

The normalised differential \ttbar\ cross section ${1}/{\sigma}\cdot{\rd\sigma}/{\rd X}$ is determined as a function of the kinematic properties of the leading leptons, the lepton pair, the b jets, the top quarks, and the top-quark pair, and presented in the following sections. In order to avoid additional model uncertainties due to the extrapolation of the measurement outside experimentally well-described phase space regions, the normalised differential cross sections for the measured leptons and b jets are determined in a visible phase space defined by the kinematic and geometrical acceptance of the final state leptons and jets. In contrast, the top-quark and the top-quark-pair quantities are presented in the full phase space in order to allow for comparisons with recent QCD calculations up to approximate NNLO precision.
To facilitate comparison with theory curves independently of the binning, a horizontal bin-centre correction is applied. In each bin, the measured data points are presented at the horizontal position in the bin where the predicted bin-averaged cross section equals the differential cross section according to the \MADGRAPH calculation (cf.~\cite{bib:BCCs}). The measurement is compared to the predictions from \MADGRAPH, \POWHEG, and \MCATNLO. For the latter, uncertainty bands corresponding to the PDF (following the PDF4LHC prescription~\cite{bib:PDF4LHC}), the top-quark mass, and renormalisation and factorisation scale variations are also given. The top-quark and \ttbar\ results are also compared to the latest approximate NNLO~\cite{bib:kidonakis_pt, bib:kidonakis_y} and NLO+NNLL~\cite{bib:ahrens_mttbar} predictions, respectively. All measured normalised differential cross section values, including bin boundaries and centres, are available in tabular form in \suppMaterial.

\subsection{Lepton and b-Jet Differential Cross Sections}

For the \ljets\ channels, the normalised differential \ttbar\ cross section as a function of the lepton and b-jet kinematic properties is defined at the particle level for the visible phase space where the lepton from the W-boson decay has a pseudorapidity $|\eta^{\ell}|<2.1$ and a transverse momentum $\pt^{\ell} > 30\GeV$, and at least four jets with $|\eta| < 2.4$ and $\pt > 30\GeV$, out of which two are b jets. A jet is defined at the particle level as a b jet if it contains the decay products of a B hadron. For this analysis, the two highest transverse momentum b jets originating from different B hadrons are selected.

\begin{figure}[htbp]
  \begin{center}
       \includegraphics[width=0.48\textwidth]{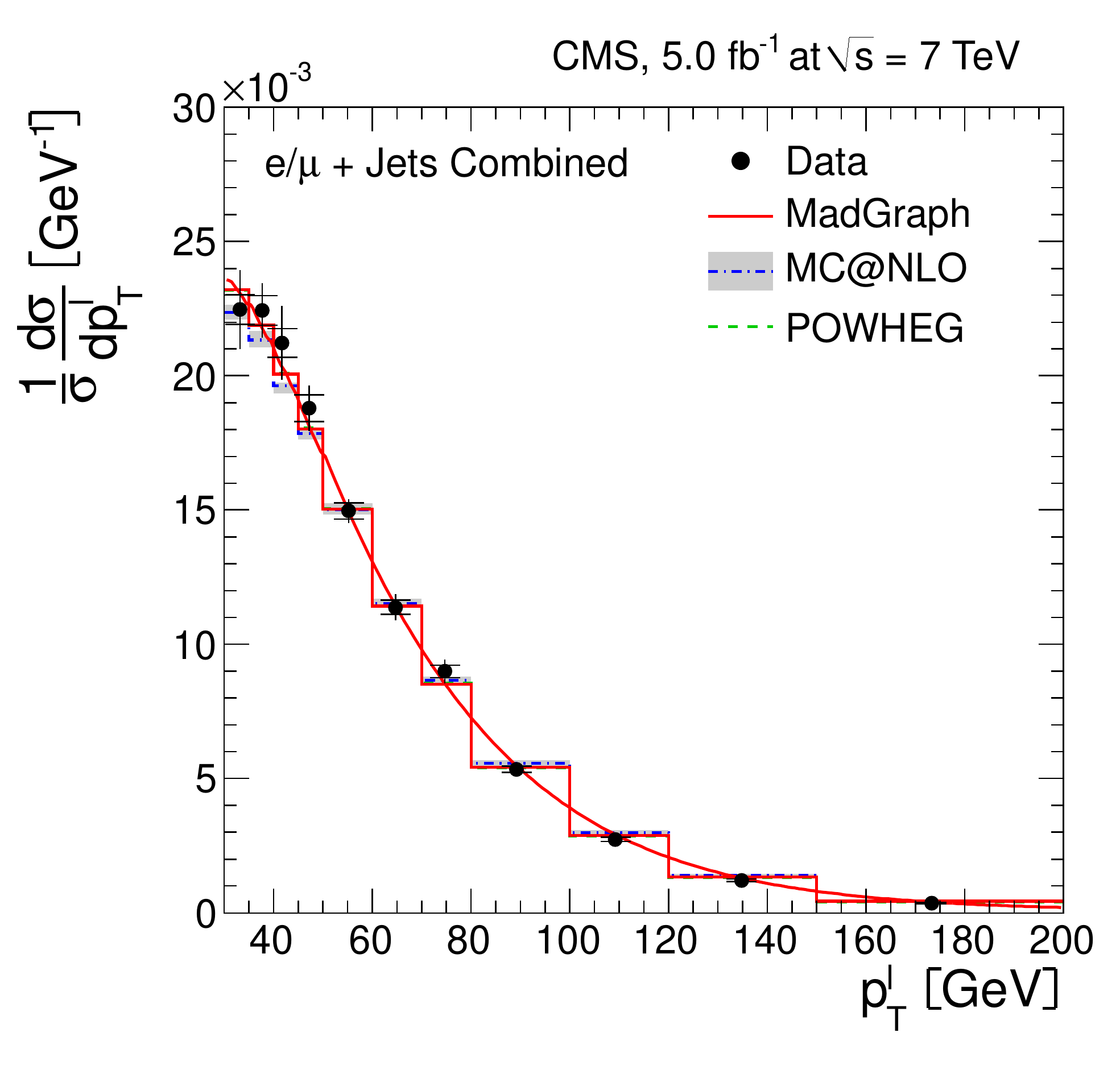}
	\includegraphics[width=0.48\textwidth]{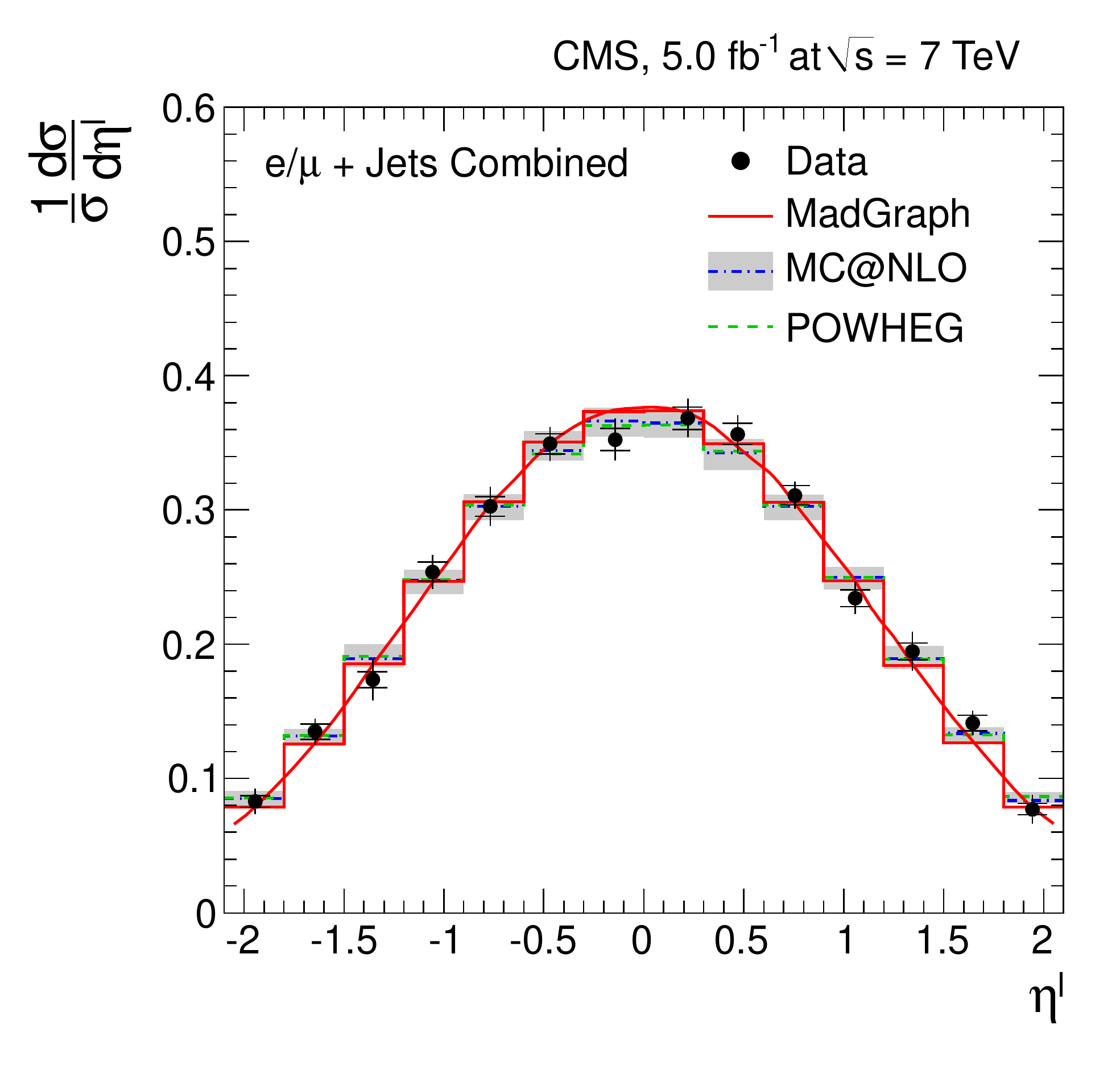}
    \caption{Normalised differential \ttbar\ production cross section in the \ljets channels as a function of the $\pt^{\ell}$ (\cmsLeft) and $\eta^{\ell}$ (\cmsRight) of the lepton. The superscript `$\ell$' refers to both $\ell^{+}$ and $\ell^{-}$. The inner (outer) error bars indicate the statistical (combined statistical and systematic) uncertainty. The measurements are compared to predictions from \MADGRAPH, \POWHEG, and \MCATNLO. The \MADGRAPH prediction is shown both as a curve and as a binned histogram.}
    \label{fig:diffXSec:l:ljets}
  \end{center}
\end{figure}

\begin{figure}[htbp]
  \begin{center}
        \includegraphics[width=0.48\textwidth]{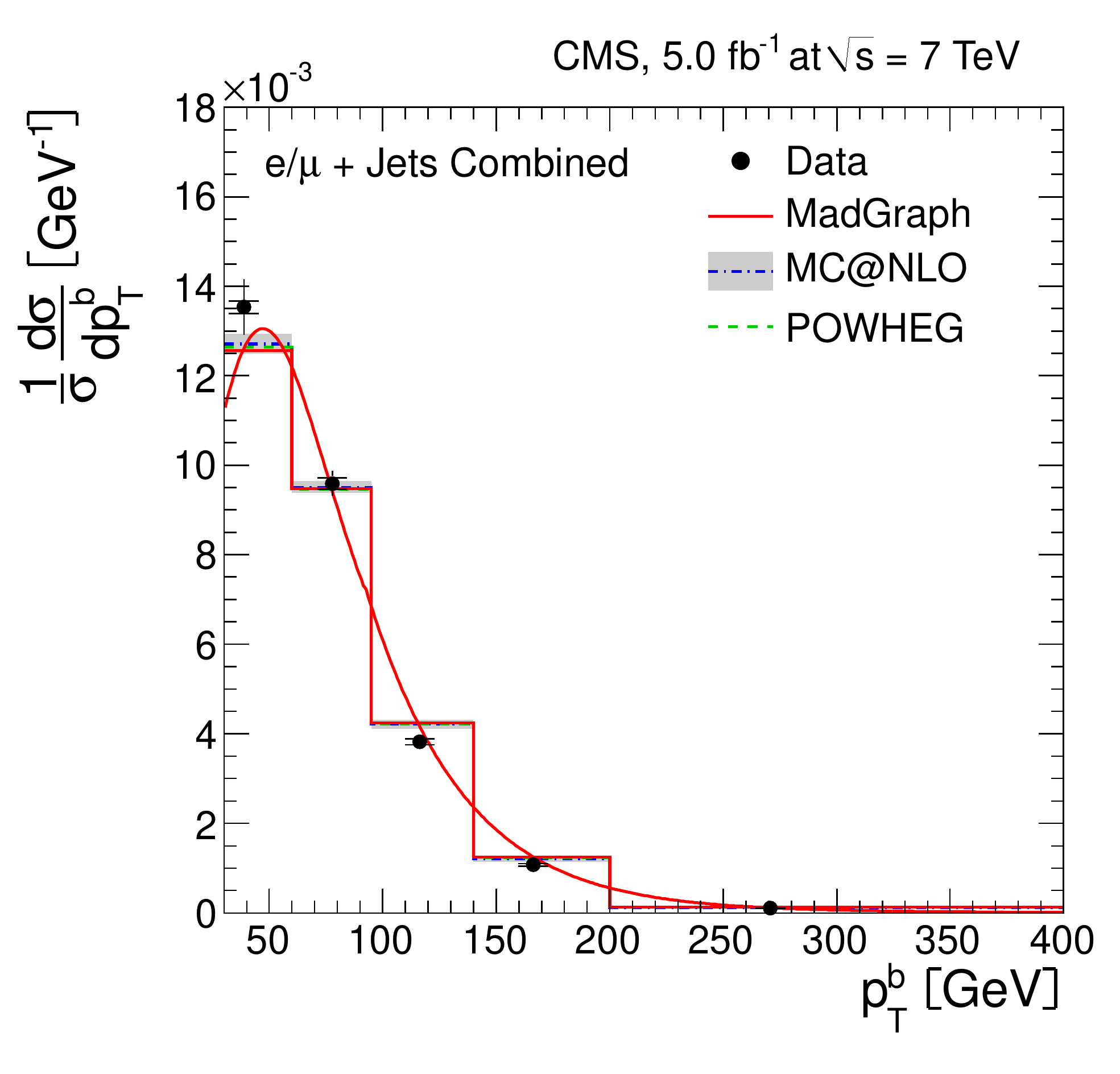}
	\includegraphics[width=0.48\textwidth]{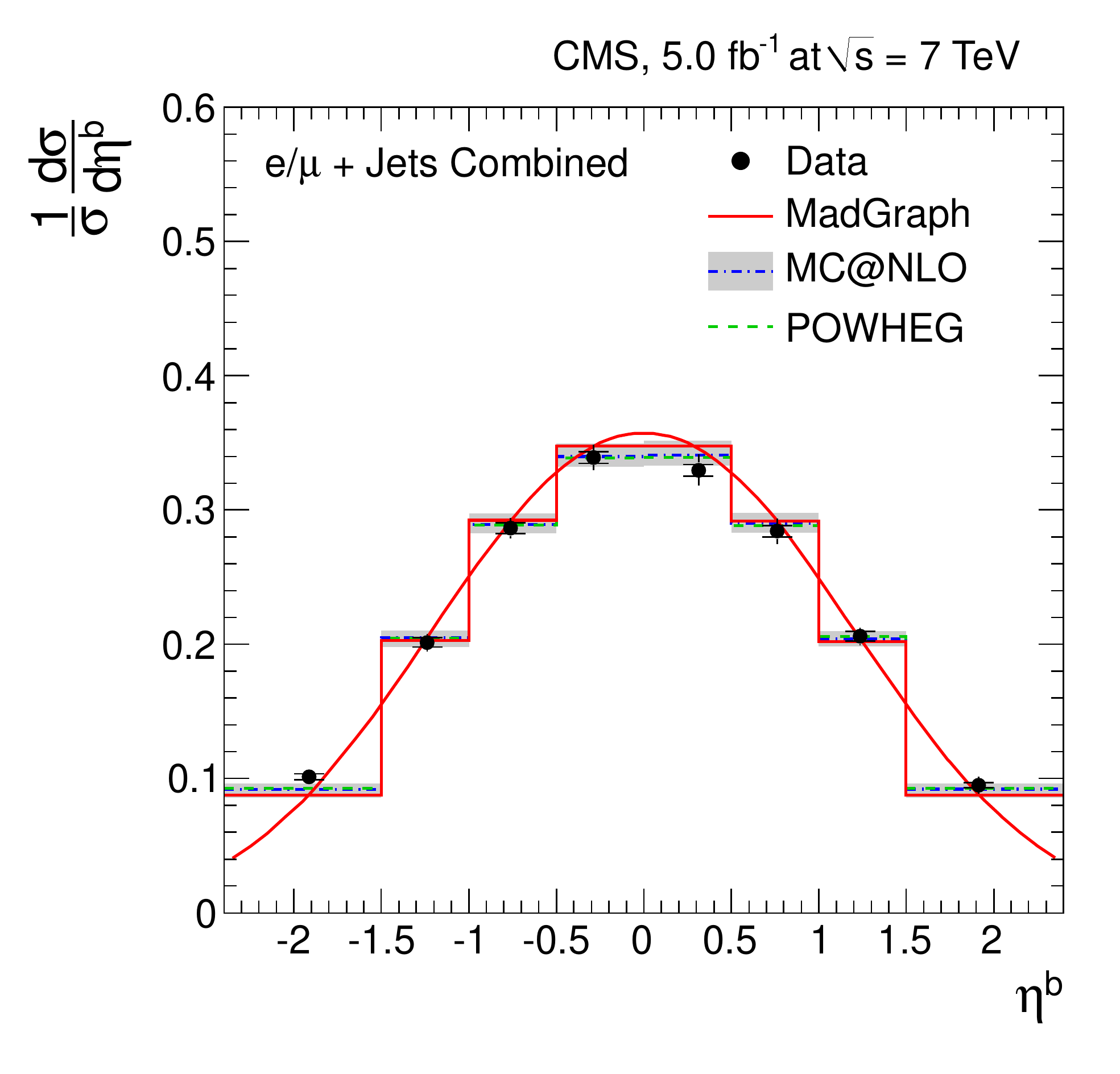}
    \caption{Normalised differential \ttbar\ production cross section in the \ljets channels as a function of the $\pt^{\text{b}}$ (\cmsLeft) and $\eta^{\text{b}}$ (\cmsRight) of the b jets. The superscript `b' refers to both b and $\bbbar$ jets. The inner (outer) error bars indicate the statistical (combined statistical and systematic) uncertainty. The measurements are compared to predictions from \MADGRAPH, \POWHEG, and \MCATNLO. The \MADGRAPH prediction is shown both as a curve and as a binned histogram.}
    \label{fig:diffXSec:bjets:ljets}
  \end{center}
\end{figure}

In Fig.~\ref{fig:diffXSec:l:ljets}, the normalised differential cross section is presented as a function of the lepton transverse momentum $\pt^{\ell}$ and pseudorapidity $\eta^{\ell}$. In Fig.~\ref{fig:diffXSec:bjets:ljets}, the distributions for the transverse momentum of the b jets, $\pt^{\text{b}}$, and their pseudorapidity, $\eta^{\text{b}}$, are shown. Also shown are predictions from \MADGRAPH, \POWHEG, and \MCATNLO. Good agreement is observed between the data and the theoretical predictions within experimental uncertainties.

For the dilepton channels, the normalised \ttbar\ differential cross section as a function of the lepton and b jet kinematic properties is defined at the particle level for the visible phase space where the leptons have $ |\eta^{\ell}| < 2.4$ and $\pt^{\ell} > 20\GeV$, and the b jets from the top-quark decays both lie within the range $|\eta| < 2.4$ and $\pt > 30\GeV$. The b jet at the particle level is defined as described above for the \ljets analysis.

\begin{figure*}[phtb]
  \begin{center}
	\includegraphics[width=0.48\textwidth]{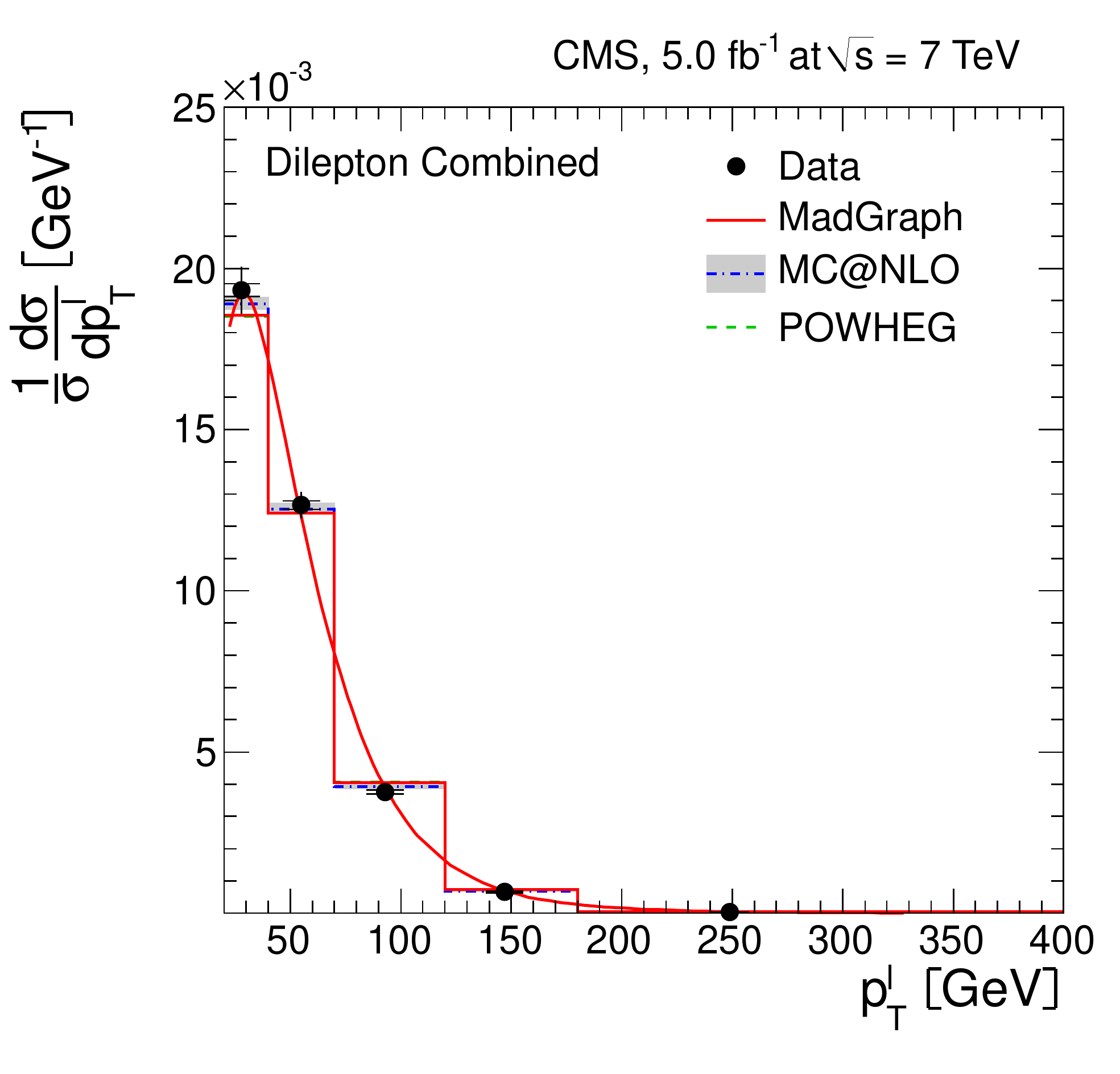}
	\includegraphics[width=0.48\textwidth]{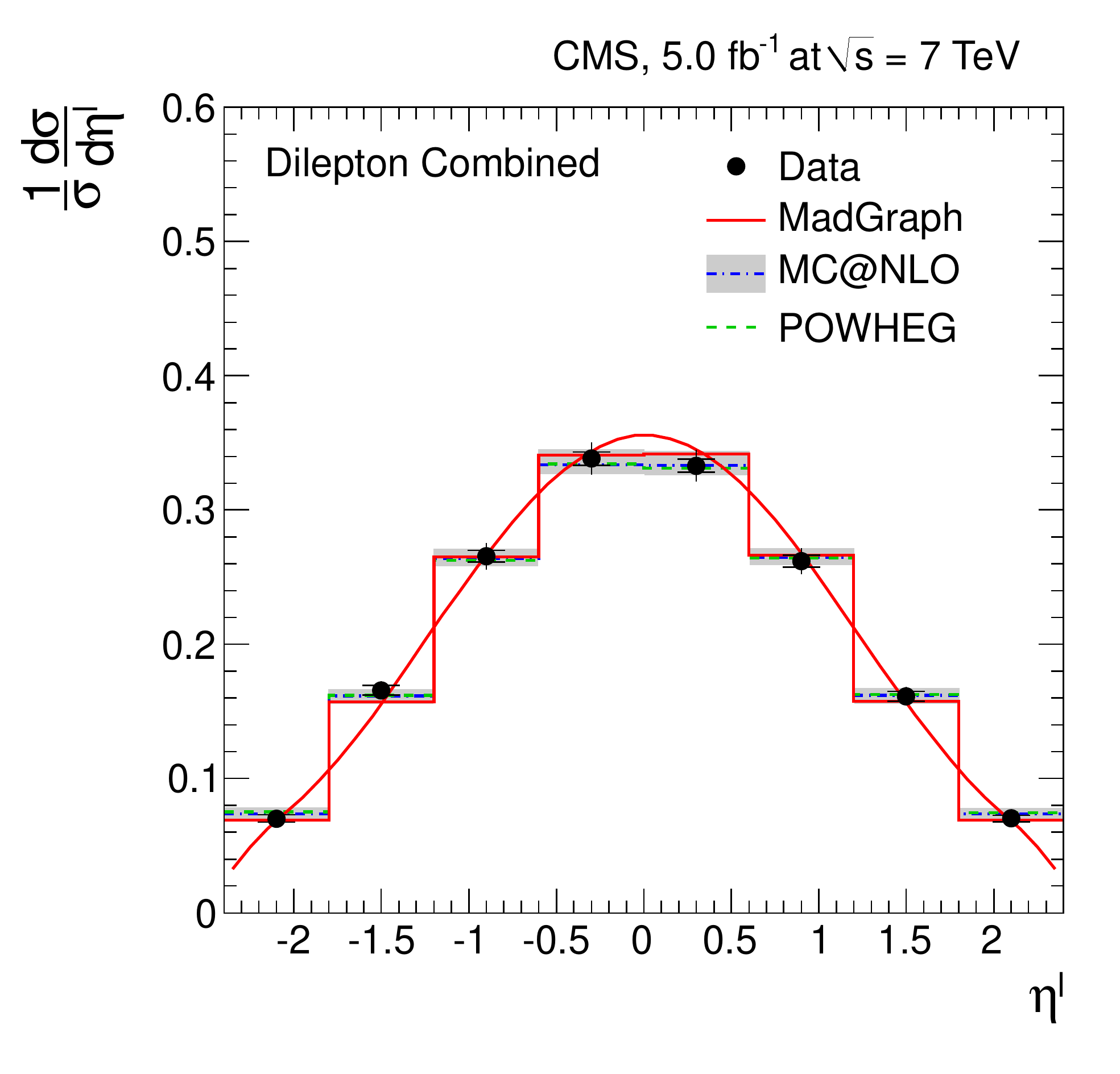}
	\includegraphics[width=0.48\textwidth]{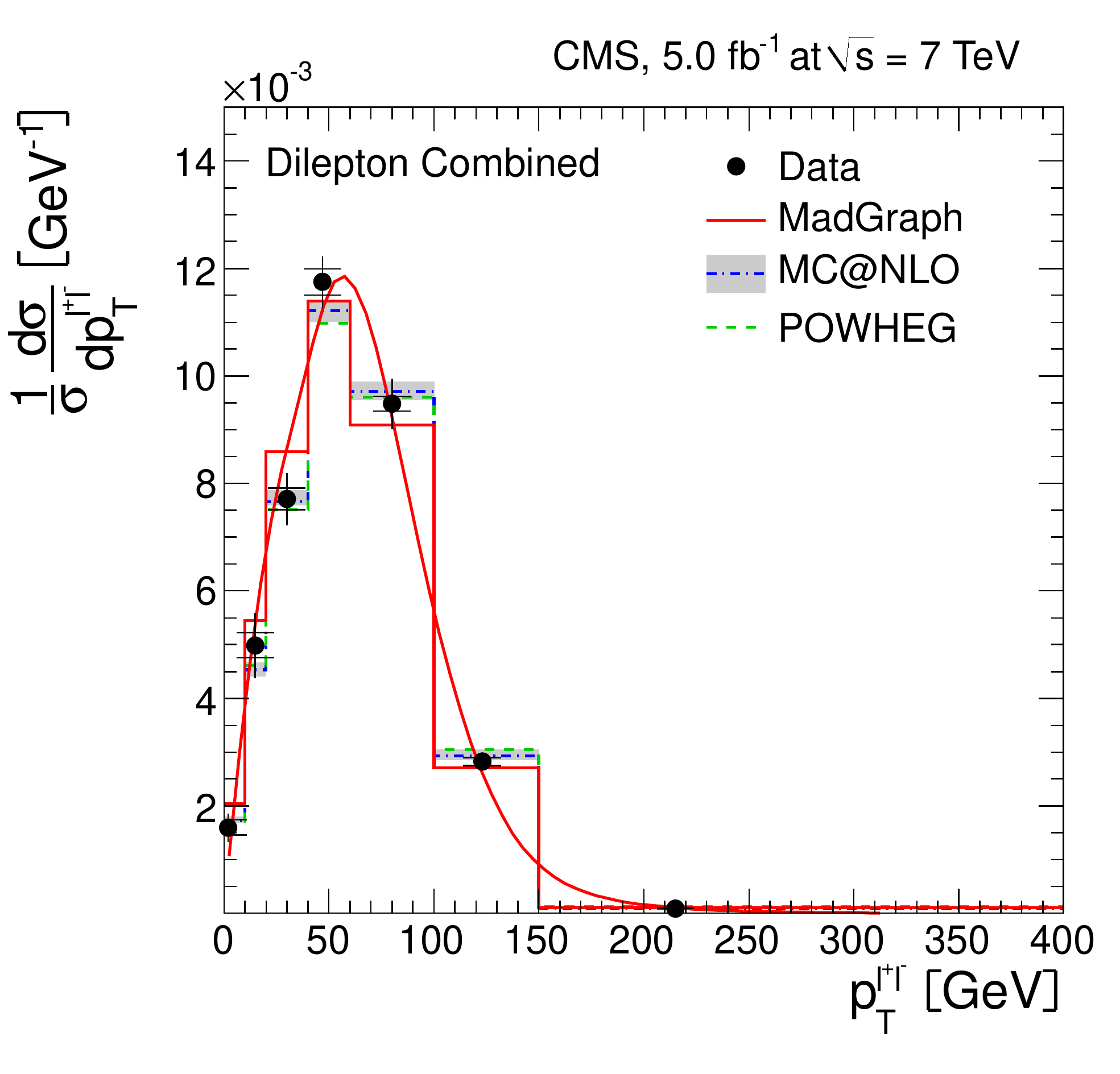}
	\includegraphics[width=0.48\textwidth]{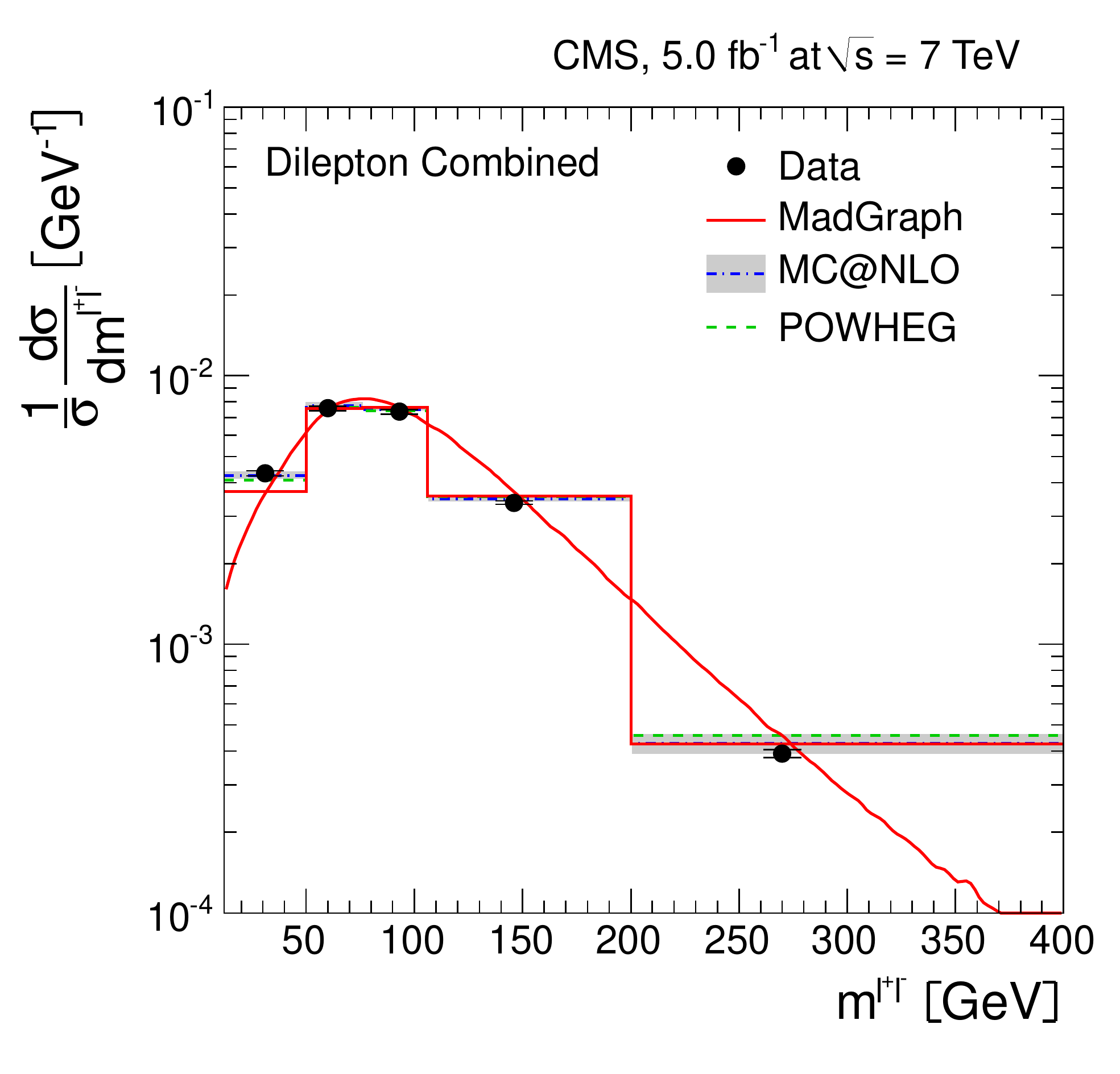}
    \caption{Normalised differential \ttbar\ production cross section in the dilepton channels as a function of the $\pt^{\ell}$ (top left) and $\eta^{\ell}$ (top right) of the leptons, and the $\pt^{\ell^{+}\ell^{-}}$ (bottom left), and $m^{\ell^{+}\ell^{-}}$ (bottom right) of the lepton pair. The superscript `$\ell$' refers to both $\ell^{+}$ and $\ell^{-}$. The inner (outer) error bars indicate the statistical (combined statistical and systematic) uncertainty. The measurements are compared to predictions from \MADGRAPH, \POWHEG, and \MCATNLO. The \MADGRAPH prediction is shown both as a curve and as a binned histogram.}
    \label{fig:diffxsec:ll:dilepton}
  \end{center}
\end{figure*}

\begin{figure}[htbp]
  \begin{center}
        \includegraphics[width=0.48\textwidth]{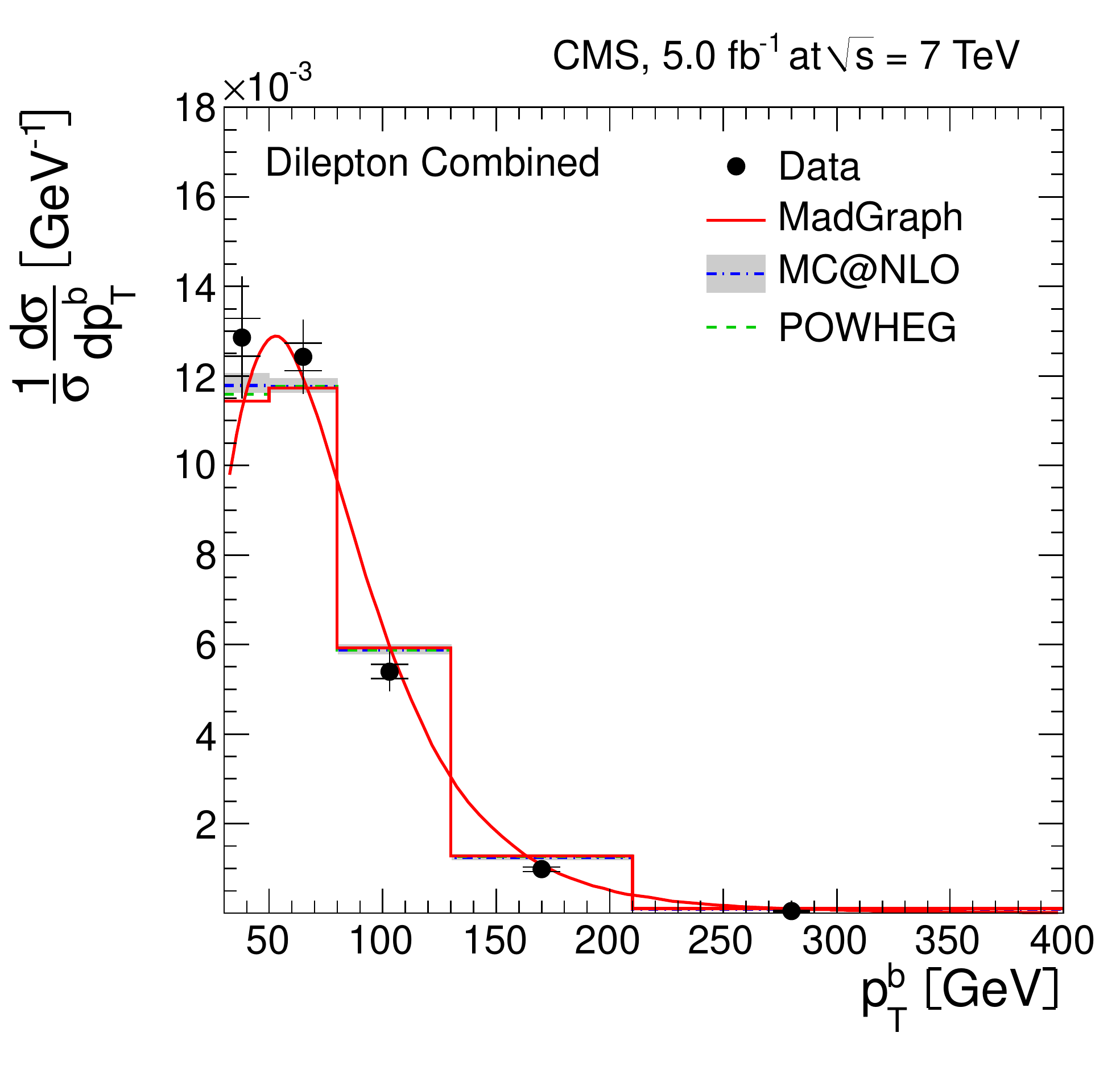}
	\includegraphics[width=0.48\textwidth]{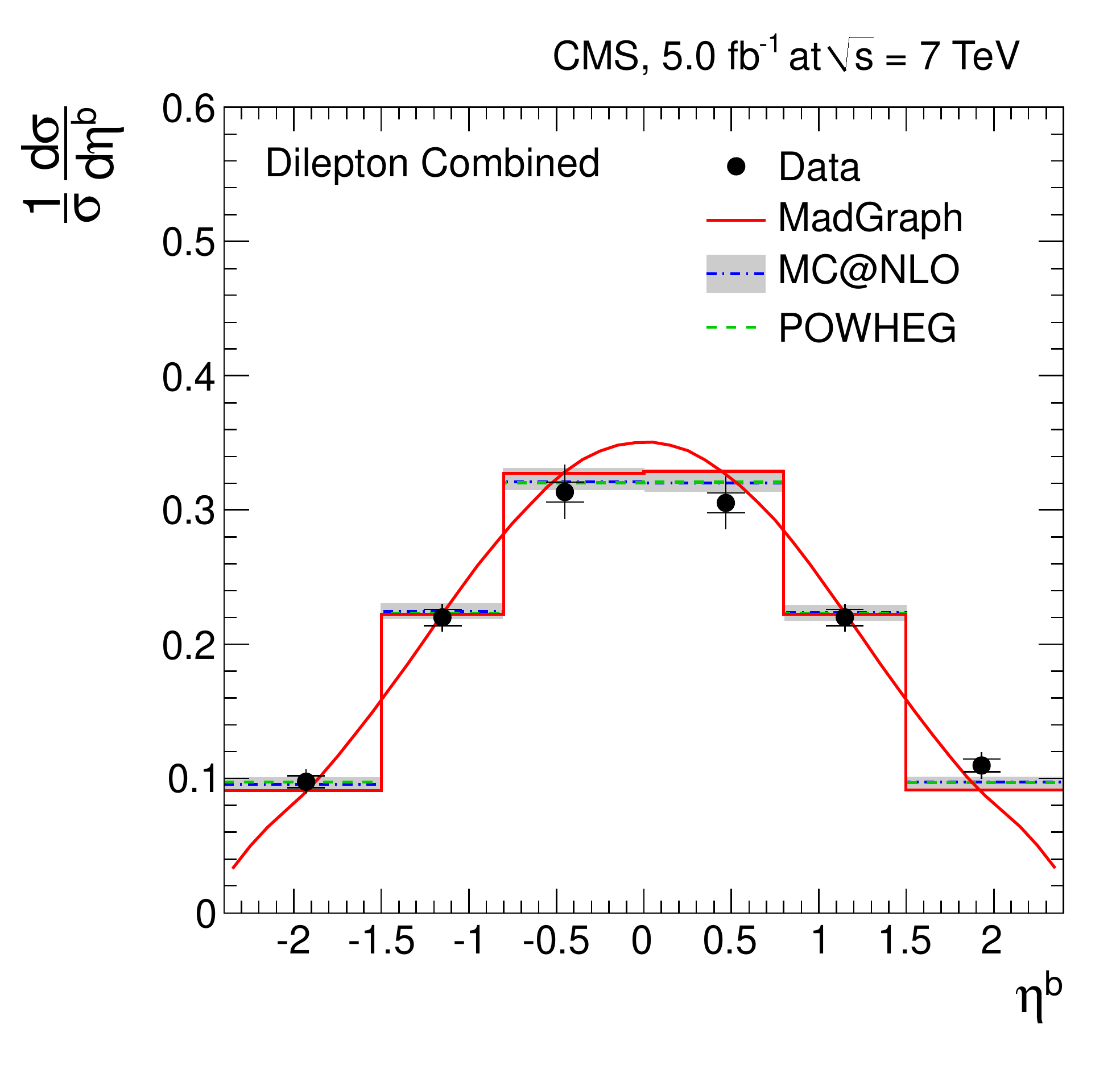}
    \caption{Normalised differential \ttbar\ production cross section in the dilepton channels as a function of the $\pt^{\text{b}}$ (\cmsLeft) and $\eta^{\text{b}}$ (\cmsRight) of the b jets. The superscript `b' refers to both b and $\bbbar$ jets. The inner (outer) error bars indicate the statistical (combined statistical and systematic) uncertainty. The measurements are compared to predictions from \MADGRAPH, \POWHEG, and \MCATNLO. The \MADGRAPH prediction is shown both as a curve and as a binned histogram.}
    \label{fig:diffXSec:bjets:dilepton}
  \end{center}
\end{figure}

In Fig.\,~\ref{fig:diffxsec:ll:dilepton}, the normalised differential cross section for the following lepton and lepton-pair observables are presented: the transverse momentum of the leptons $\pt^{\ell}$, the pseudorapidity $\eta^{\ell}$ of the leptons, the transverse momentum of the lepton pair $\pt^{\ell^{+}\ell^{-}}$, and the invariant mass of the lepton pair $m^{\ell^{+}\ell^{-}}$. The distributions for the transverse momentum of the b jets, $\pt^{\text{b}}$, and their pseudorapidity, $\eta^{\text{b}}$, are shown in Fig.\,~\ref{fig:diffXSec:bjets:dilepton}. Predictions from \MADGRAPH, \POWHEG, and \MCATNLO are also shown. Good agreement is observed between data and theoretical predictions within experimental uncertainties. The \MCATNLO and \POWHEG predictions, which take into account \ttbar spin correlations, suggest a better description of the lepton-pair observables in those bins in which there exists some discrepancy between the different generators.

\subsection{Top-Quark and \texorpdfstring{\ttbar}{Top-Quark-Pair} Differential Cross Sections}

The normalised differential \ttbar\ cross section as a function of the kinematic properties of the top quarks and the top-quark pair is presented at parton level and extrapolated to the full phase space using the \MADGRAPH prediction for both the \ljets and the dilepton channels.

\begin{figure*}[phtb]
  \begin{center}
        \includegraphics[width=\fiveup]{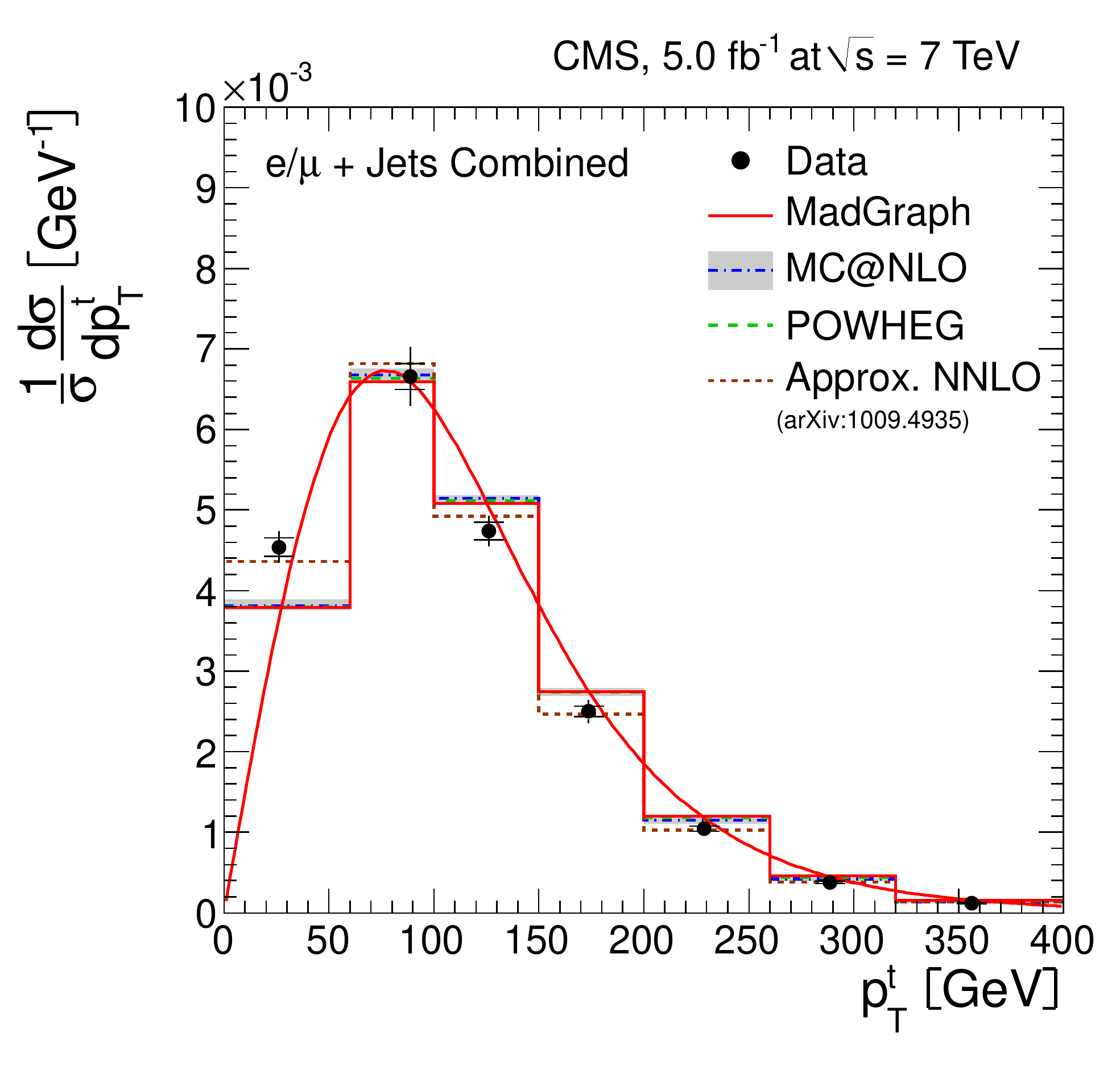}
	\includegraphics[width=\fiveup]{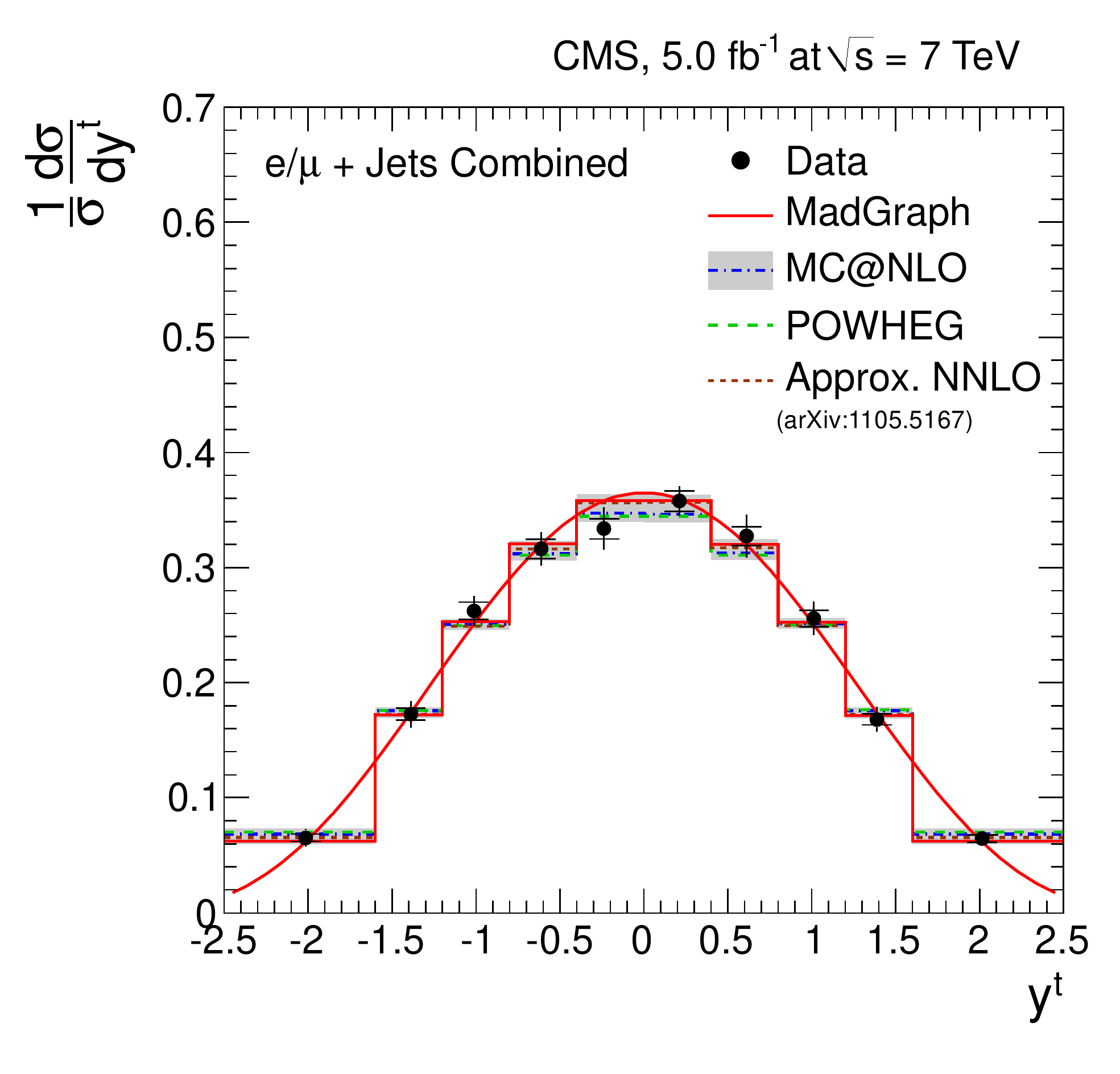}\\
	\includegraphics[width=\fiveup]{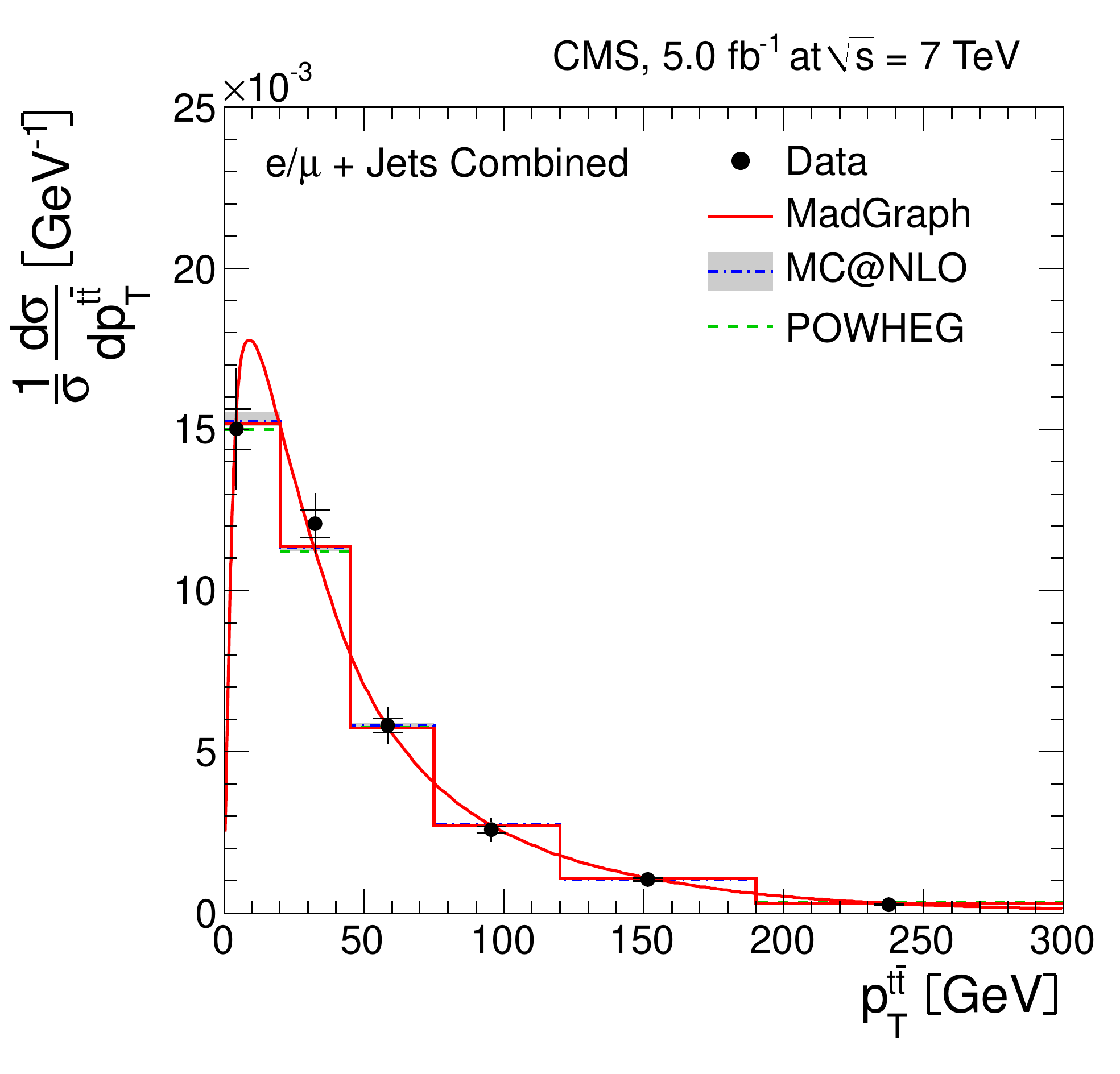}
	\includegraphics[width=\fiveup]{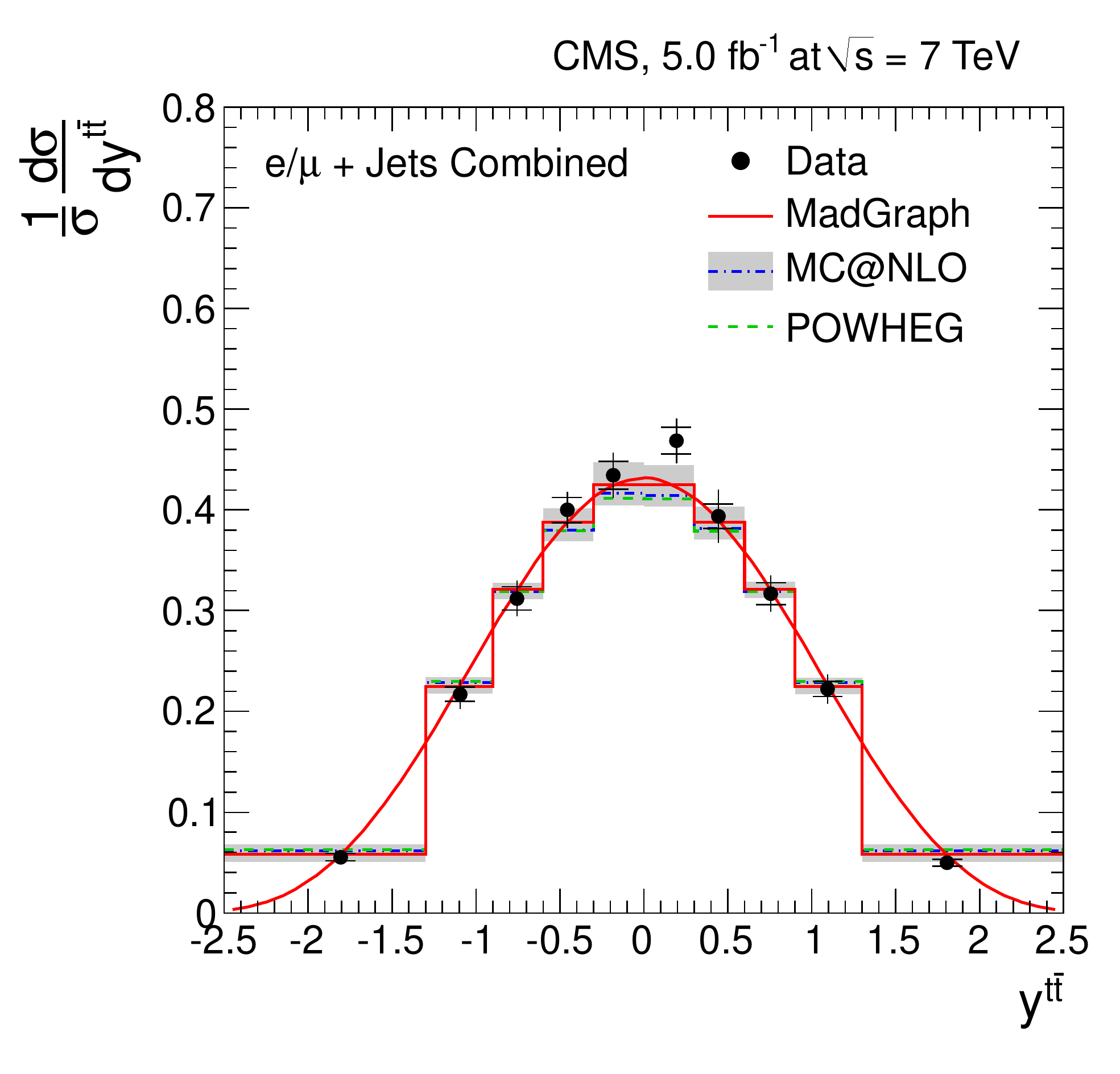}\\
        \includegraphics[width=\fiveup]{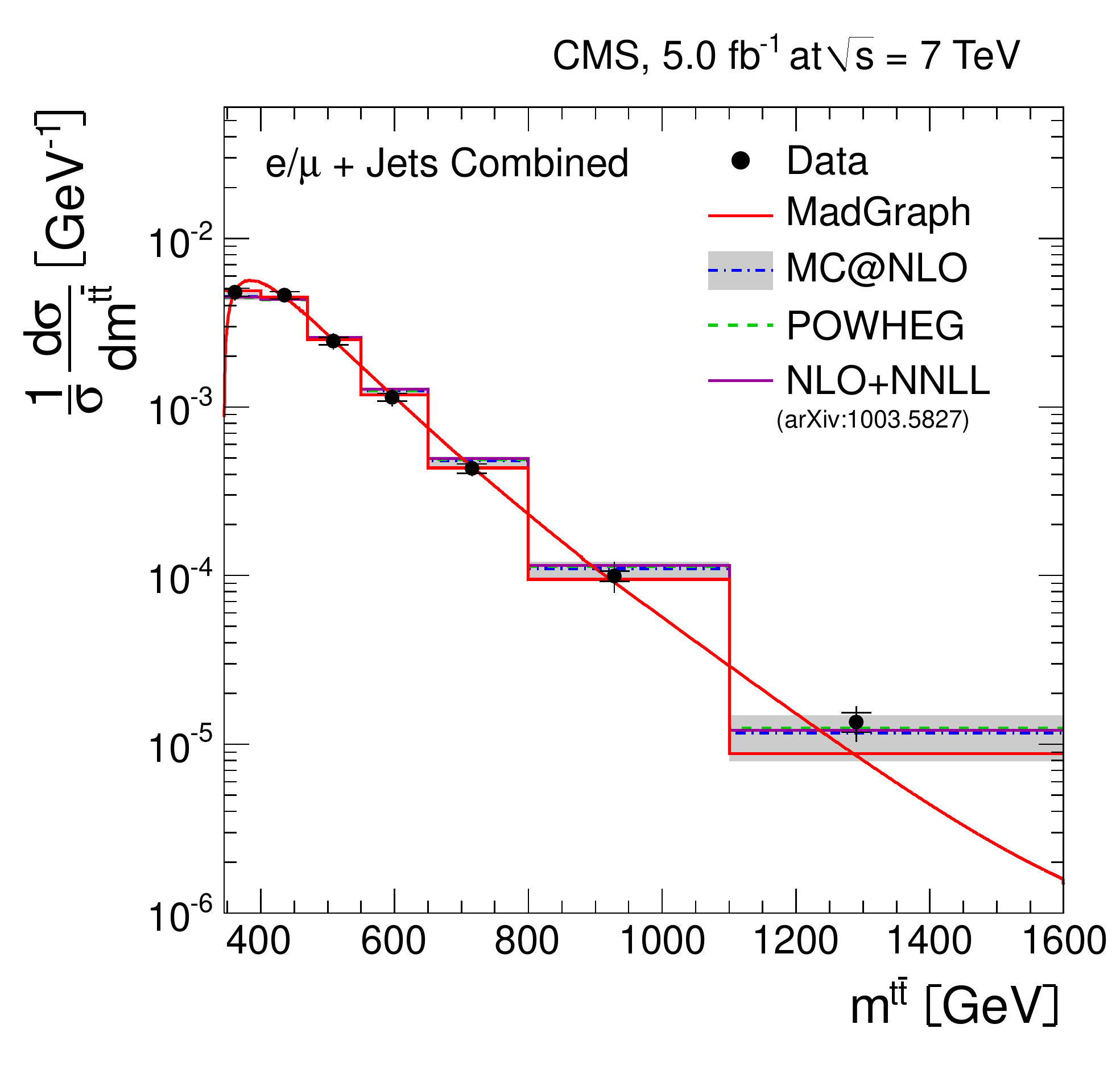}
    \caption{Normalised differential \ttbar\ production cross section in the \ljets channels as a function of the $\pt^{\text{t}}$ (top left) and $y^{\text{t}}$ (top right) of the top quarks, and the $\pt^{\ttbar}$ (middle left), $y^{\ttbar}$ (middle right), and $m^{\ttbar}$ (bottom) of the top-quark pairs. The superscript `t' refers to both top quarks and antiquarks. The inner (outer) error bars indicate the statistical (combined statistical and systematic) uncertainty. The measurements are compared to predictions from \MADGRAPH, \POWHEG, and \MCATNLO, and to NLO+NNLL~\cite{bib:ahrens_mttbar} and approximate NNLO~\cite{bib:kidonakis_pt, bib:kidonakis_y} calculations, when available. The \MADGRAPH prediction is shown both as a curve and as a binned histogram.}
\label{fig:diffXSec:tt:ljets}
  \end{center}
\end{figure*}

\begin{figure*}[phtb]
  \begin{center}
	\includegraphics[width=\fiveup]{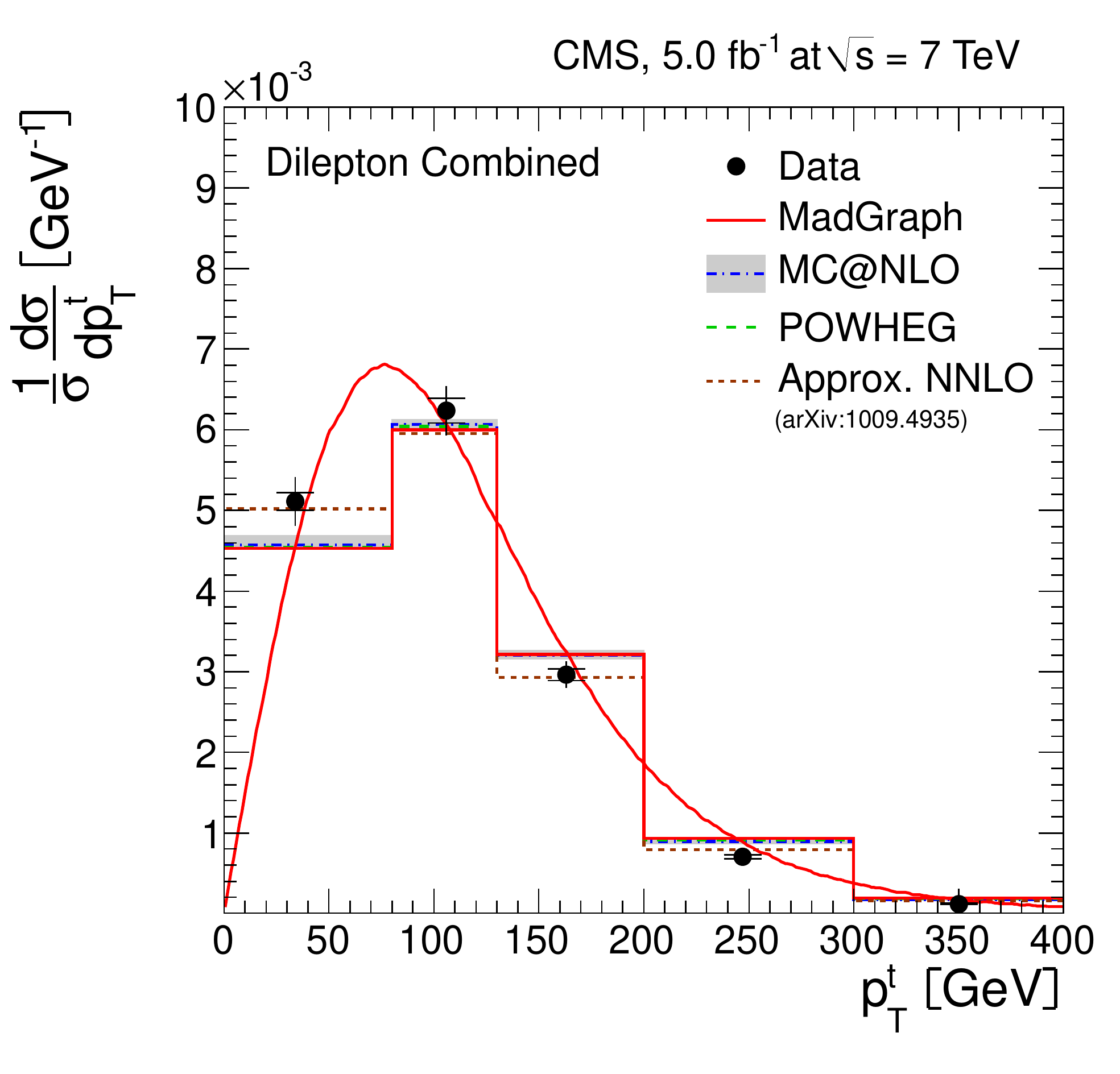}
	\includegraphics[width=\fiveup]{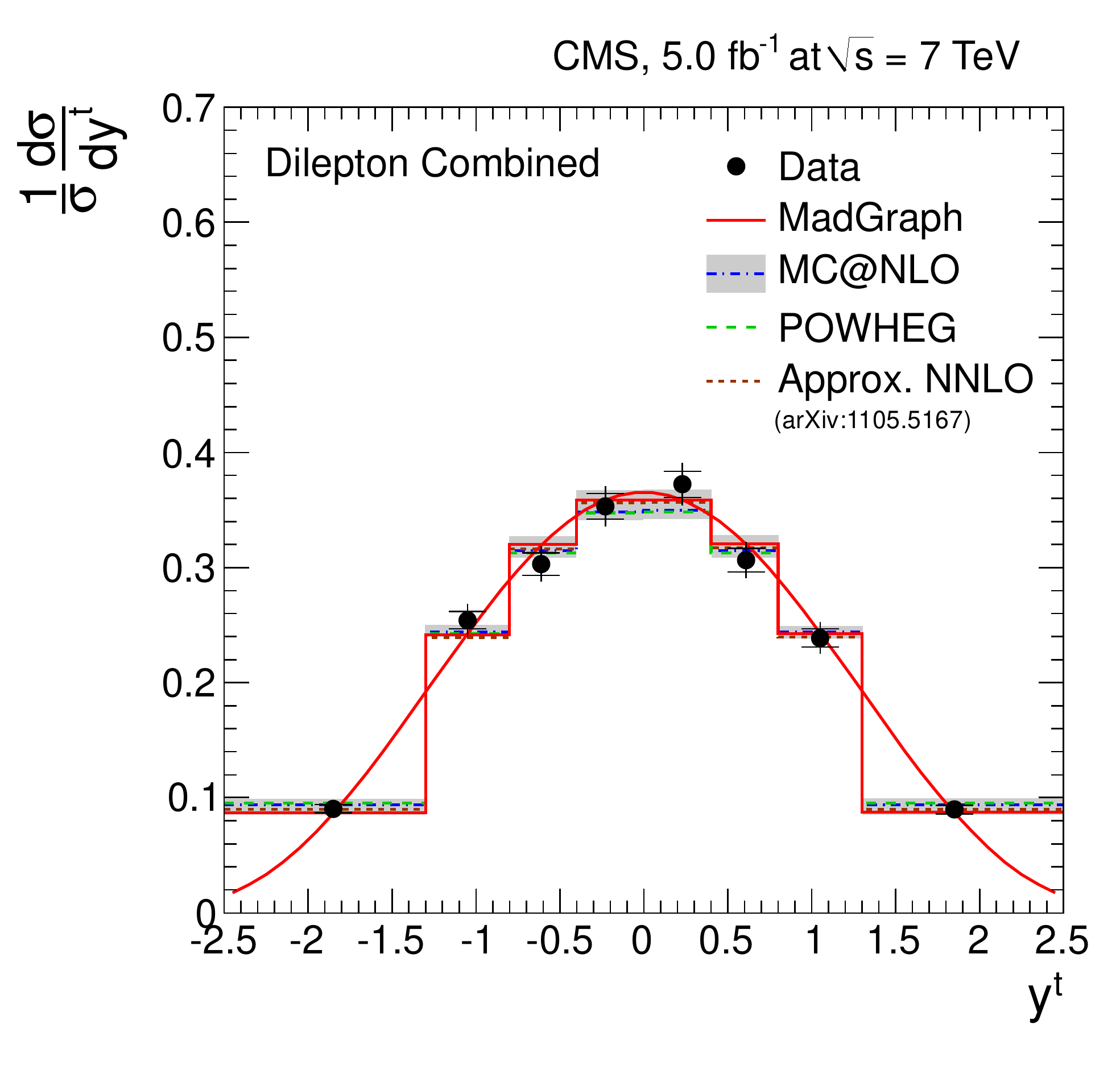}\\
	\includegraphics[width=\fiveup]{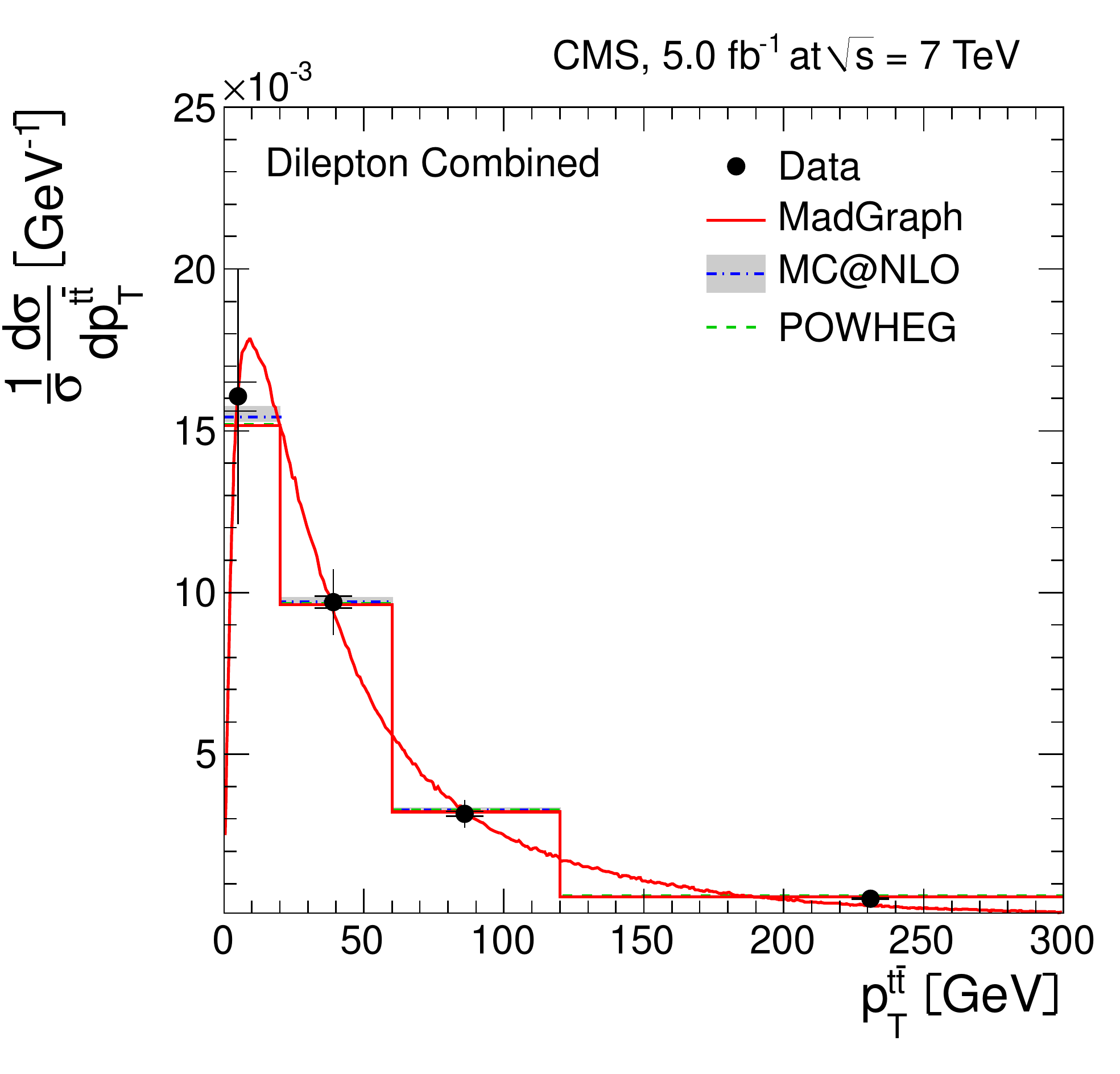}
	\includegraphics[width=\fiveup]{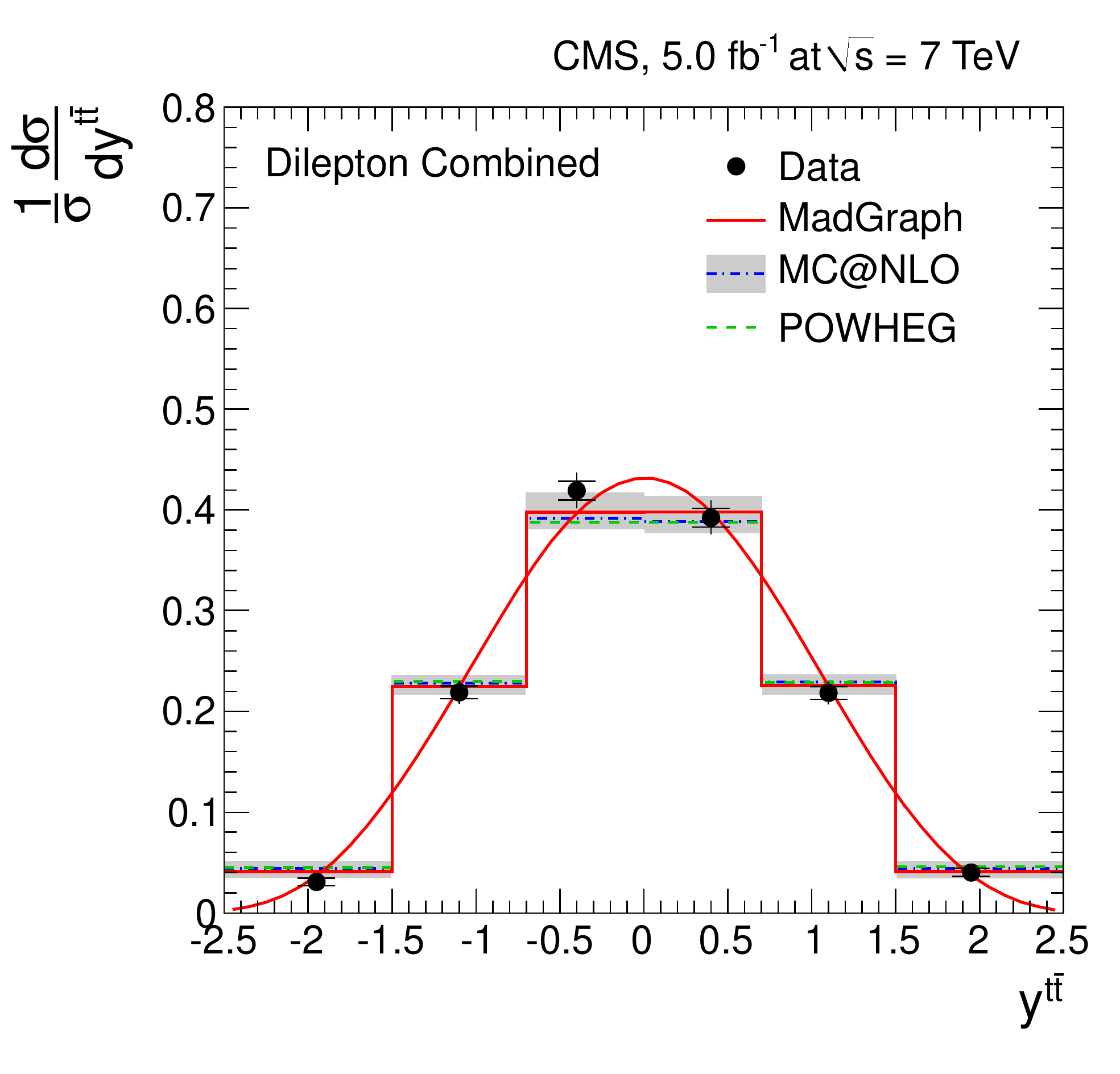}\\
        \includegraphics[width=\fiveup]{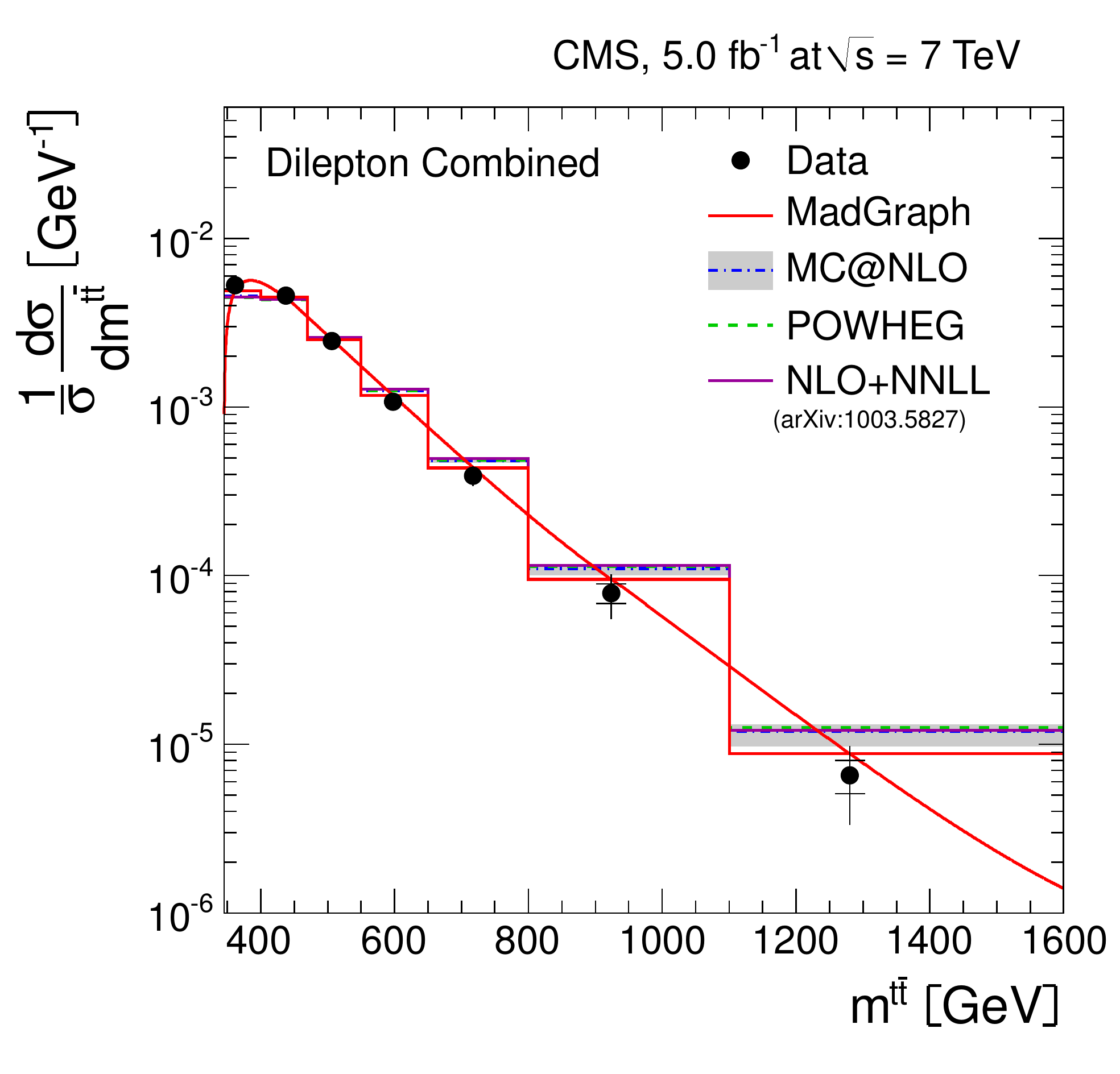}
    \caption{Normalised differential \ttbar\ production cross section in the dilepton channels as a function of the $\pt^{\text{t}}$ (top left) and $y^{\text{t}}$ (top right) of the top quarks, and the $\pt^{\ttbar}$ (middle left), $y^{\ttbar}$ (middle right), and $m^{\ttbar}$ (bottom) of the top-quark pairs. The superscript `t' refers to both top quarks and antiquarks. The inner (outer) error bars indicate the statistical (combined statistical and systematic) uncertainty. The measurements are compared to predictions from \MADGRAPH, \POWHEG, and \MCATNLO, and to NLO+NNLL~\cite{bib:ahrens_mttbar} and approximate NNLO~\cite{bib:kidonakis_pt, bib:kidonakis_y} calculations, when available. The \MADGRAPH prediction is shown both as a curve and as a binned histogram.}
    \label{fig:diffxsec:tt:dilepton}
  \end{center}
\end{figure*}

In Figs.~\ref{fig:diffXSec:tt:ljets} and~\ref{fig:diffxsec:tt:dilepton}, the distributions for the top-quark and the top-quark-pair observables in the \ljets channels and the dilepton channels are presented. Those are the transverse momentum $\pt^{\text{t}}$ and the rapidity $y^{\text{t}}$ of the top quarks and antiquarks, and the transverse momentum $\pt^{\ttbar}$, the rapidity $y^{\ttbar}$, and the invariant mass $m^{\ttbar}$ of the top-quark pair. Also shown are predictions from \MADGRAPH, \POWHEG, and \MCATNLO. In addition, the top-quark results are compared to the approximate NNLO calculations from Ref.~\cite{bib:kidonakis_pt, bib:kidonakis_y}, while the $m^{\ttbar}$ distribution is compared to the NLO+NNLL prediction in Ref.~\cite{bib:ahrens_mttbar}.

For both \ljets and dilepton channels, good agreement is observed between data and theoretical predictions within experimental uncertainties. Among the various predictions, the approximate NNLO calculation provides a better description of the data, as it predicts a slightly softer top-quark transverse momentum spectrum than the other three predictions.

\section{Summary}
\label{sec:concl}

First measurements of normalised differential top-quark-pair production cross sections in pp~collisions at $\sqrt{s}=7\TeV$ with the CMS detector are presented. The measurements are performed in the \ljets (\ejets and \mujets) and the dilepton (\ee, \mumu, and \mue) \ttbar decay channels. The normalised \ttbar cross section is measured as a function of the transverse momentum, (pseudo)rapidity, and invariant mass of the final-state leptons and b jets in the visible phase space, and of the top quarks and \ttbar system in the full phase space. The measurements among the different decay channels are in agreement with each other and with standard model predictions up to approximate next-to-next-to-leading-order precision. The prediction at approximate NNLO precision is found to give a particularly good description of the top-quark transverse momentum.

\section*{Acknowledgements}
{\tolerance=500
\hyphenation{Bundes-ministerium Forschungs-gemeinschaft Forschungs-zentren} We thank Nikolaos Kidonakis for providing the approximate NNLO calculations and for fruitful discussions, and Li Lin Yang for providing the NLO+NNLL calculation. We congratulate our colleagues in the CERN accelerator departments for the excellent performance of the LHC and thank the technical and administrative staffs at CERN and at other CMS institutes for their contributions to the success of the CMS effort. In addition, we gratefully acknowledge the computing centres and personnel of the Worldwide LHC Computing Grid for delivering so effectively the computing infrastructure essential to our analyses. Finally, we acknowledge the enduring support for the construction and operation of the LHC and the CMS detector provided by the following funding agencies: the Austrian Federal Ministry of Science and Research; the Belgian Fonds de la Recherche Scientifique, and Fonds voor Wetenschappelijk Onderzoek; the Brazilian Funding Agencies (CNPq, CAPES, FAPERJ, and FAPESP); the Bulgarian Ministry of Education, Youth and Science; CERN; the Chinese Academy of Sciences, Ministry of Science and Technology, and National Natural Science Foundation of China; the Colombian Funding Agency (COLCIENCIAS); the Croatian Ministry of Science, Education and Sport; the Research Promotion Foundation, Cyprus; the Ministry of Education and Research, Recurrent financing contract SF0690030s09 and European Regional Development Fund, Estonia; the Academy of Finland, Finnish Ministry of Education and Culture, and Helsinki Institute of Physics; the Institut National de Physique Nucl\'eaire et de Physique des Particules~/~CNRS, and Commissariat \`a l'\'Energie Atomique et aux \'Energies Alternatives~/~CEA, France; the Bundesministerium f\"ur Bildung und Forschung, Deutsche Forschungsgemeinschaft, and Helmholtz-Gemeinschaft Deutscher Forschungszentren, Germany; the General Secretariat for Research and Technology, Greece; the National Scientific Research Foundation, and National Office for Research and Technology, Hungary; the Department of Atomic Energy and the Department of Science and Technology, India; the Institute for Studies in Theoretical Physics and Mathematics, Iran; the Science Foundation, Ireland; the Istituto Nazionale di Fisica Nucleare, Italy; the Korean Ministry of Education, Science and Technology and the World Class University program of NRF, Korea; the Lithuanian Academy of Sciences; the Mexican Funding Agencies (CINVESTAV, CONACYT, SEP, and UASLP-FAI); the Ministry of Science and Innovation, New Zealand; the Pakistan Atomic Energy Commission; the Ministry of Science and Higher Education and the National Science Centre, Poland; the Funda\c{c}\~ao para a Ci\^encia e a Tecnologia, Portugal; JINR (Armenia, Belarus, Georgia, Ukraine, Uzbekistan); the Ministry of Education and Science of the Russian Federation, the Federal Agency of Atomic Energy of the Russian Federation, Russian Academy of Sciences, and the Russian Foundation for Basic Research; the Ministry of Science and Technological Development of Serbia; the Secretar\'{\i}a de Estado de Investigaci\'on, Desarrollo e Innovaci\'on and Programa Consolider-Ingenio 2010, Spain; the Swiss Funding Agencies (ETH Board, ETH Zurich, PSI, SNF, UniZH, Canton Zurich, and SER); the National Science Council, Taipei; the Thailand Center of Excellence in Physics, the Institute for the Promotion of Teaching Science and Technology and National Electronics and Computer Technology Center; the Scientific and Technical Research Council of Turkey, and Turkish Atomic Energy Authority; the Science and Technology Facilities Council, UK; the US Department of Energy, and the US National Science Foundation.

Individuals have received support from the Marie-Curie programme and the European Research Council (European Union); the Leventis Foundation; the A. P. Sloan Foundation; the Alexander von Humboldt Foundation; the Belgian Federal Science Policy Office; the Fonds pour la Formation \`a la Recherche dans l'Industrie et dans l'Agriculture (FRIA-Belgium); the Agentschap voor Innovatie door Wetenschap en Technologie (IWT-Belgium); the Ministry of Education, Youth and Sports (MEYS) of Czech Republic; the Council of Science and Industrial Research, India; the Compagnia di San Paolo (Torino); and the HOMING PLUS programme of Foundation for Polish Science, cofinanced from European Union, Regional Development Fund.
\par}

\bibliography{auto_generated}   
\ifthenelse{\boolean{cms@external}}{}{
\clearpage
\appendix
\section{ Normalised Differential Cross Section Values}
\label{app:suppMat}
\begin{table}[h]
	\begin{center}
		\caption{Normalised differential \ttbar\ cross section as a function of lepton observables in the \ljets channels: the transverse momentum of the leptons $\pt^{\ell}$ and the pseudorapidity of the leptons $\eta^{\ell}$.}
		\label{tab:diffxseclep:ljets}
		\begin{tabular}{c|c||c|c|c}
			\hline
			\hline          $\pt^{\ell}$ bin [GeV] & $ 1/\!\sigma\,\rd\sigma\!/\!\rd \pt^{\ell}$ & stat. [\%] & sys. [\%] & total [\%] \\
            		\hline
           $ 30$ to $ 35$ & $2.25 \cdot 10^{-2}$ &  2.4 &  6.1 &  6.6 \\
           $ 35$ to $ 40$ & $2.24 \cdot 10^{-2}$ &  2.5 &  3.8 &  4.6 \\
           $ 40$ to $ 45$ & $2.12 \cdot 10^{-2}$ &  2.5 &  6.0 &  6.5 \\
           $ 45$ to $ 50$ & $1.88 \cdot 10^{-2}$ &  2.6 &  3.7 &  4.6 \\
           $ 50$ to $ 60$ & $1.50 \cdot 10^{-2}$ &  2.1 &  2.2 &  3.0 \\
           $ 60$ to $ 70$ & $1.14 \cdot 10^{-2}$ &  2.3 &  3.6 &  4.3 \\
           $ 70$ to $ 80$ & $0.90 \cdot 10^{-2}$ &  2.5 &  4.3 &  5.0 \\
           $ 80$ to $100$ & $0.53 \cdot 10^{-2}$ &  2.4 &  3.7 &  4.4 \\
           $100$ to $120$ & $0.27 \cdot 10^{-2}$ &  3.3 &  5.6 &  6.5 \\
           $120$ to $150$ & $0.12 \cdot 10^{-2}$ &  3.9 &  5.6 &  6.8 \\
           $150$ to $200$ & $0.04 \cdot 10^{-2}$ &  5.8 &  8.5 & 10.3 \\
			\hline
			\hline          $\eta^{\ell}$ bin  & $ 1/\!\sigma\,\rd\sigma\!/\!\rd \eta^{\ell}$ & stat. [\%] & sys. [\%] & total [\%] \\
			\hline
           $-2.1$ to $-1.8$ & $0.83 \cdot 10^{-1}$ &  5.4 & 10.0 & 11.4 \\
           $-1.8$ to $-1.5$ & $1.35 \cdot 10^{-1}$ &  4.1 &  6.1 &  7.4 \\
           $-1.5$ to $-1.2$ & $1.74 \cdot 10^{-1}$ &  3.3 &  8.3 &  8.9 \\
           $-1.2$ to $-0.9$ & $2.54 \cdot 10^{-1}$ &  2.8 &  4.2 &  5.1 \\
           $-0.9$ to $-0.6$ & $3.03 \cdot 10^{-1}$ &  2.4 &  4.2 &  4.8 \\
           $-0.6$ to $-0.3$ & $3.49 \cdot 10^{-1}$ &  2.2 &  2.9 &  3.6 \\
           $-0.3$ to $ 0.0$ & $3.52 \cdot 10^{-1}$ &  2.3 &  3.8 &  4.4 \\
           $ 0.0$ to $ 0.3$ & $3.68 \cdot 10^{-1}$ &  2.3 &  3.2 &  3.9 \\
           $ 0.3$ to $ 0.6$ & $3.57 \cdot 10^{-1}$ &  2.2 &  3.4 &  4.0 \\
           $ 0.6$ to $ 0.9$ & $3.11 \cdot 10^{-1}$ &  2.3 &  2.2 &  3.2 \\
           $ 0.9$ to $ 1.2$ & $2.34 \cdot 10^{-1}$ &  2.7 &  4.3 &  5.1 \\
           $ 1.2$ to $ 1.5$ & $1.95 \cdot 10^{-1}$ &  3.2 &  6.8 &  7.5 \\
           $ 1.5$ to $ 1.8$ & $1.41 \cdot 10^{-1}$ &  4.1 &  5.1 &  6.5 \\
           $ 1.8$ to $ 2.1$ & $0.77 \cdot 10^{-1}$ &  5.4 & 12.8 & 13.9 \\
			\hline
			\hline
		\end{tabular}
	\end{center}
\end{table}

\begin{table}[h]
	\begin{center}
		\caption{Normalised differential \ttbar\ cross section as a function of b-jet observables in the \ljets channels: the transverse momentum of the b jets $\pt^{\rm{b}}$ and the pseudorapidity of the b jets $\eta^{\rm{b}}$.}
		\label{tab:diffxsecbjet:ljets}
		\begin{tabular}{c|c||c|c|c}
			\hline
			\hline            $\pt^{\rm{b}}$ bin [GeV] & $ 1/\!\sigma\,\rd\sigma\!/\!\rd \pt^{\rm{b}}$ & stat. [\%] & sys. [\%] & total [\%] \\
		        \hline
          $ 30$ to $ 60$ & $1.35 \cdot 10^{-2}$ &  1.0 &  4.5 &  4.6 \\
          $ 60$ to $ 95$ & $0.96 \cdot 10^{-2}$ &  1.3 &  2.6 &  2.9 \\
          $ 95$ to $140$ & $0.38 \cdot 10^{-2}$ &  1.8 &  3.6 &  4.0 \\
          $140$ to $200$ & $0.11 \cdot 10^{-2}$ &  2.6 &  9.7 & 10.0 \\
          $200$ to $400$ & $0.01 \cdot 10^{-2}$ &  5.3 & 16.3 & 17.1 \\
			\hline
			\hline          $\eta^{\rm{b}}$ bin & $ 1/\!\sigma\,\rd\sigma\!/\!\rd \eta^{\rm{b}}$ & stat. [\%] & sys. [\%] & total [\%] \\
			\hline
          $-2.4$ to $-1.5$ & $1.01 \cdot 10^{-1}$ & 2.1 & 4.2 & 4.7 \\
          $-1.5$ to $-1.0$ & $2.01 \cdot 10^{-1}$ & 1.7 & 2.8 & 3.3 \\
          $-1.0$ to $-0.5$ & $2.86 \cdot 10^{-1}$ & 1.4 & 2.2 & 2.6 \\
          $-0.5$ to $ 0.0$ & $3.39 \cdot 10^{-1}$ & 1.3 & 2.5 & 2.8 \\
          $ 0.0$ to $ 0.5$ & $3.30 \cdot 10^{-1}$ & 1.3 & 3.2 & 3.5 \\
          $ 0.5$ to $ 1.0$ & $2.84 \cdot 10^{-1}$ & 1.5 & 3.0 & 3.3 \\
          $ 1.5$ to $ 1.5$ & $2.06 \cdot 10^{-1}$ & 1.8 & 2.6 & 3.1 \\
          $ 1.6$ to $ 2.4$ & $0.95 \cdot 10^{-1}$ & 2.1 & 6.1 & 6.5 \\
  		\hline
                \hline
		\end{tabular}
	\end{center}
\end{table}

\begin{table}[h]
	\begin{center}
		\caption{Normalised differential \ttbar\ cross section as a function of top-quark observables in the \ljets channels: the transverse momentum of the top quarks $\pt^{\rm{t}}$, the rapidity of the top quarks $y^{\rm{t}}$, the transverse momentum of the top-quark pair $\pt^{\rm{\ttbar}}$, the rapidity of the top-quark pair $y^{\rm{\ttbar}}$, and the invariant mass of the top-quark pair $m^{\rm{\ttbar}}$.}
		\label{tab:diffxsectop:ljets}
		\begin{tabular}{c|c||c|c|c}
			\hline
			\hline          $\pt^{\rm{t}}$ bin [GeV] & $ 1/\!\sigma\,\rd\sigma\!/\!\rd \pt^{\rm{t}}$ & stat. [\%] & sys. [\%] & total [\%] \\
         		\hline
            $   0$ to $ 60$ & $4.54 \cdot 10^{-3}$ & 2.5 & 3.6 &  4.4 \\
            $  60$ to $100$ & $6.66 \cdot 10^{-3}$ & 2.4 & 4.9 &  5.5 \\
            $ 100$ to $150$ & $4.74 \cdot 10^{-3}$ & 2.4 & 3.2 &  4.0 \\
            $ 150$ to $200$ & $2.50 \cdot 10^{-3}$ & 2.6 & 5.1 &  5.8 \\
            $ 200$ to $260$ & $1.04 \cdot 10^{-3}$ & 2.9 & 5.5 &  6.2 \\
            $ 260$ to $320$ & $0.38 \cdot 10^{-3}$ & 3.7 & 8.2 &  9.0 \\
            $ 320$ to $400$ & $0.12 \cdot 10^{-3}$ & 5.8 & 9.5 & 11.1 \\
			\hline
			\hline          $y^{\rm{t}}$ bin & $ 1/\!\sigma\,\rd\sigma\!/\!\rd y^{\rm{t}}$ & stat. [\%] & sys. [\%] & total [\%] \\
			\hline
             $-2.$ to $-1.6$ & $0.65 \cdot 10^{-1}$ & 5.1 & 10.3 & 11.5 \\
             $-1.$ to $-1.2$ & $1.73 \cdot 10^{-1}$ & 2.9 &  5.9 &  6.6 \\
             $-1.$ to $-0.8$ & $2.62 \cdot 10^{-1}$ & 2.8 &  4.1 &  5.0 \\
             $-0.$ to $-0.4$ & $3.16 \cdot 10^{-1}$ & 2.6 &  3.8 &  4.6 \\
             $-0.$ to $ 0.0$ & $3.34 \cdot 10^{-1}$ & 2.7 &  4.8 &  5.5 \\
             $ 0.$ to $ 0.4$ & $3.58 \cdot 10^{-1}$ & 2.5 &  2.6 &  3.6 \\
             $ 0.$ to $ 0.8$ & $3.27 \cdot 10^{-1}$ & 2.5 &  5.2 &  5.8 \\
             $ 0.$ to $ 1.2$ & $2.56 \cdot 10^{-1}$ & 2.7 &  5.0 &  5.7 \\
             $ 1.$ to $ 1.6$ & $1.68 \cdot 10^{-1}$ & 3.0 &  5.7 &  6.4 \\
             $ 1.$ to $ 2.5$ & $0.64 \cdot 10^{-1}$ & 5.0 &  7.1 &  8.7 \\
			\hline
			\hline          $\pt^{\rm{\ttbar}}$ bin [GeV] & $ 1/\!\sigma\,\rd\sigma\!/\!\rd \pt^{\ttbar}$ & stat. [\%] & sys. [\%] & total [\%] \\
			\hline
             $  0 $ to $  20$ & $1.50 \cdot 10^{-2}$ & 4.1 & 11.8 & 12.5 \\
             $ 20 $ to $  45$ & $1.21 \cdot 10^{-2}$ & 3.5 &  7.0 &  7.8 \\
             $ 45 $ to $  75$ & $0.58 \cdot 10^{-2}$ & 3.8 &  9.2 & 10.0 \\
             $ 75 $ to $ 120$ & $0.26 \cdot 10^{-2}$ & 4.3 & 14.0 & 14.6 \\
             $120 $ to $ 190$ & $0.10 \cdot 10^{-2}$ & 4.5 &  7.8 &  8.9 \\
             $190 $ to $ 300$ & $0.02 \cdot 10^{-2}$ & 6.3 & 18.0 & 19.1 \\
			\hline
			\hline          $y^{\ttbar}$ bin & $ 1/\!\sigma\,\rd\sigma\!/\!\rd y^{\ttbar}$ & stat. [\%] & sys. [\%] & total [\%] \\
			\hline
             $-2.5 $ to $ -1.3$ & $0.55 \cdot 10^{-1}$ & 6.4 & 10.8 & 12.5 \\
             $-1.3 $ to $ -0.9$ & $2.17 \cdot 10^{-1}$ & 3.4 &  5.8 &  6.7 \\
             $-0.9 $ to $ -0.6$ & $3.12 \cdot 10^{-1}$ & 3.6 &  4.4 &  5.7 \\
             $-0.6 $ to $ -0.3$ & $4.00 \cdot 10^{-1}$ & 3.1 &  3.3 &  4.5 \\
             $-0.3 $ to $  0.0$ & $4.35 \cdot 10^{-1}$ & 3.1 &  4.1 &  5.1 \\
             $ 0.0 $ to $  0.3$ & $4.69 \cdot 10^{-1}$ & 2.8 &  3.8 &  4.8 \\
             $ 0.3 $ to $  0.6$ & $3.94 \cdot 10^{-1}$ & 3.1 &  5.9 &  6.7 \\
             $ 0.6 $ to $  0.9$ & $3.17 \cdot 10^{-1}$ & 3.4 &  4.7 &  5.8 \\
             $ 0.9 $ to $  1.3$ & $2.22 \cdot 10^{-1}$ & 3.3 &  5.8 &  6.6 \\
             $ 1.3 $ to $  2.5$ & $0.50 \cdot 10^{-1}$ & 6.8 &  9.7 & 11.9 \\
			\hline
			\hline          $m^{\ttbar}$ bin [GeV] & $ 1/\!\sigma\,\rd\sigma\!/\!\rd m^{\ttbar}$ & stat. [\%] & sys. [\%] & total [\%] \\
			\hline
             $   0 $ to $  345$ &  -                    &  -   &  -   & -    \\
             $ 345 $ to $  400$ & $ 4.81 \cdot 10^{-3}$ &  5.2 &  9.7 & 11.1 \\
             $ 400 $ to $  470$ & $ 4.60 \cdot 10^{-3}$ &  5.0 &  8.4 &  9.8 \\
             $ 470 $ to $  550$ & $ 2.46 \cdot 10^{-3}$ &  5.2 & 10.2 & 11.4 \\
             $ 550 $ to $  650$ & $ 1.14 \cdot 10^{-3}$ &  5.6 & 10.6 & 12.0 \\
             $ 650 $ to $  800$ & $ 0.43 \cdot 10^{-3}$ &  6.2 &  8.3 & 10.3 \\
             $ 800 $ to $ 1100$ & $ 0.99 \cdot 10^{-4}$ &  7.1 & 20.0 & 21.2 \\
             $1100 $ to $ 1600$ & $ 0.14 \cdot 10^{-4}$ & 13.5 & 19.4 & 23.7 \\
                        \hline
  		\hline
		\end{tabular}
	\end{center}
\end{table}

\begin{table}[h]
	\begin{center}
		\caption{Normalised differential \ttbar\ cross section as a function of lepton observables in the dilepton channels: the transverse momentum of the leptons $\pt^{\ell}$, the pseudorapidity of the leptons $\eta^{\ell}$, the transverse momentum of the lepton pair, $\pt^{\ell\ell}$, and the invariant mass of the lepton pair $m^{\ell\ell}$.}
		\label{tab:diffxseclep:dilepton}
		\begin{tabular}{c|c||c|c|c}
			\hline
			\hline          $\pt^{\ell}$ bin [GeV] & $ 1/\!\sigma\,\rd\sigma\!/\!\rd \pt^{\ell}$ & stat. [\%] & sys. [\%] & total [\%] \\
		\hline
$ 20 $ to $  40$  & 1.92 $\cdot 10^{-2}$ & 1.1 & 3.6 &  3.8\\
$ 40 $ to $  70$  & 1.27 $\cdot 10^{-2}$ & 1.2 & 3.1 &  3.4\\
$ 70 $ to $ 120$  & 0.38 $\cdot 10^{-2}$ & 1.7 & 3.8 &  4.2\\
$120 $ to $ 180$  & 0.07 $\cdot 10^{-2}$ & 3.5 & 6.7 &  7.6\\
$180 $ to $ 400$  & 0.32 $\cdot 10^{-4}$ & 9.5 & 7.4 & 12.0\\
			\hline
			\hline          $\eta^{\ell}$ bin  & $ 1/\!\sigma\,\rd\sigma\!/\!\rd \eta^{\ell}$ & stat. [\%] & sys. [\%] & total [\%] \\
			\hline
$-2.4 $ to $ -1.8$ & 0.07 $\cdot 10^{-1}$ & 3.8 & 4.4 & 5.8\\
$-1.8 $ to $ -1.2$ & 1.66 $\cdot 10^{-1}$ & 2.3 & 3.4 & 4.1\\
$-1.2 $ to $ -0.6$ & 2.65 $\cdot 10^{-1}$ & 1.7 & 3.4 & 3.8\\
$-0.6 $ to $  0.0$ & 3.37 $\cdot 10^{-1}$ & 1.6 & 3.2 & 3.5\\
$ 0.0 $ to $  0.6$ & 3.33 $\cdot 10^{-1}$ & 1.6 & 3.2 & 3.6\\
$ 0.6 $ to $  1.2$ & 2.62 $\cdot 10^{-1}$ & 1.8 & 3.4 & 3.8\\
$ 1.2 $ to $  1.8$ & 1.62 $\cdot 10^{-1}$ & 2.3 & 3.4 & 4.1\\
$ 1.8 $ to $  2.4$ & 0.71 $\cdot 10^{-1}$ & 3.6 & 4.4 & 5.7\\
			\hline
			\hline          $\pt^{\ell\ell}$ bin [GeV] & $ 1/\!\sigma\,\rd\sigma\!/\!\rd \pt^{\ell\ell}$ & stat. [\%] & sys. [\%] & total [\%] \\
			\hline
$   0 $ to $  10$ & 0.15 $\cdot 10^{-2}$ & 9.2 & 11.8 & 15.0\\
$  10 $ to $  20$ & 0.05 $\cdot 10^{-2}$ & 4.8 &  6.5 &  8.0\\
$  20 $ to $  40$ & 0.01 $\cdot 10^{-2}$ & 2.8 &  4.0 &  4.8\\
$  40 $ to $  60$ & 1.17 $\cdot 10^{-2}$ & 2.2 &  3.5 &  4.1\\
$  60 $ to $ 100$ & 0.96 $\cdot 10^{-2}$ & 1.7 &  3.4 &  3.8\\
$ 100 $ to $ 150$ & 0.29 $\cdot 10^{-2}$ & 2.8 &  5.2 &  5.9\\
$ 150 $ to $ 400$ & 0.92 $\cdot 10^{-4}$ & 7.0 & 10.3 & 12.5\\
			\hline
			\hline           $m^{\ell\ell}$ bin [GeV] & $ 1/\!\sigma\,\rd\sigma\!/\!\rd m^{\ell\ell}$ & stat. [\%] & sys. [\%] & total [\%] \\
			\hline
$ 12 $ to $  50$ & 4.36 $\cdot 10^{-3}$ & 2.3 & 4.4 & 4.9\\
$ 50 $ to $  76$ & 7.52 $\cdot 10^{-3}$ & 2.2 & 5.1 & 5.6\\
$ 76 $ to $ 106$ & 7.27 $\cdot 10^{-3}$ & 2.1 & 5.8 & 6.1\\
$106 $ to $ 200$ & 3.37 $\cdot 10^{-3}$ & 1.6 & 3.5 & 3.9\\
$200 $ to $ 400$ & 0.34 $\cdot 10^{-3}$ & 3.6 & 5.7 & 6.7\\
		\hline
  	\hline
		\end{tabular}
	\end{center}
\end{table}

\begin{table}[h]
	\begin{center}
		\caption{Normalised differential \ttbar\ cross section as a function of b-jet observables in the dilepton channels: the transverse momentum of the b jets $\pt^{\rm{b}}$ and the pseudorapidity of the b jets $\eta^{\rm{b}}$.}
		\label{tab:diffxsecbjet:dilepton}
		\begin{tabular}{c|c||c|c|c}
			\hline
			\hline           $\pt^{\rm{b}}$ bin [GeV] & $ 1/\!\sigma\,\rd\sigma\!/\!\rd \pt^{\rm{b}}$ & stat. [\%] & sys. [\%] & total [\%] \\
		\hline
$ 30 $ to $  50$ & 1.28 $\cdot 10^{-2}$ &  2.9 & 10.4 & 10.8\\
$ 50 $ to $  80$ & 1.25 $\cdot 10^{-2}$ &  3.0 &  6.3 &  7.0\\
$ 80 $ to $ 130$ & 0.54 $\cdot 10^{-2}$ &  3.2 &  7.9 &  8.5\\
$130 $ to $ 210$ & 0.10 $\cdot 10^{-2}$ &  5.0 &  7.1 &  8.7\\
$210 $ to $ 400$ & 0.49 $\cdot 10^{-4}$ & 19.1 & 19.5 & 27.3\\
			\hline
			\hline           $\eta^{\rm{b}}$ bin & $ 1/\!\sigma\,\rd\sigma\!/\!\rd \eta^{\rm{b}}$ & stat. [\%] & sys. [\%] & total [\%] \\
			\hline
 $-2.4 $ to $ -1.5$ & 0.98 $\cdot 10^{-1}$ & 4.0 & 8.5 & 9.4\\
 $-1.5 $ to $ -0.8$ & 2.20 $\cdot 10^{-1}$ & 3.0 & 4.0 & 5.0\\
 $-0.8 $ to $  0.0$ & 3.12 $\cdot 10^{-1}$ & 2.7 & 6.3 & 6.9\\
 $ 0.0 $ to $  0.8$ & 3.05 $\cdot 10^{-1}$ & 2.7 & 6.3 & 6.9\\
 $ 0.8 $ to $  1.5$ & 2.20 $\cdot 10^{-1}$ & 3.0 & 4.0 & 5.0\\
 $ 1.5 $ to $  2.4$ & 1.10 $\cdot 10^{-1}$ & 4.3 & 8.5 & 9.6\\
  		\hline
                \hline
		\end{tabular}
	\end{center}
\end{table}

\begin{table}[h]
	\begin{center}
		\caption{Normalised differential \ttbar\ cross section as a function of top-quark observables in the dilepton channels: the transverse momentum of the top quarks $\pt^{\rm{t}}$, the rapidity of the top quarks $y^{\rm{t}}$, the transverse momentum of the top-quark pair $\pt^{\rm{\ttbar}}$, the rapidity of the top-quark pair $y^{\rm{\ttbar}}$, and the invariant mass of the top-quark pair $m^{\rm{\ttbar}}$.}
		\label{tab:diffxsectop:dilepton}
		\begin{tabular}{c|c||c|c|c}
			\hline
			\hline           $\pt^{\rm{t}}$ bin [GeV] & $ 1/\!\sigma\,\rd\sigma\!/\!\rd \pt^{\rm{t}}$ & stat. [\%] & sys. [\%] & total [\%] \\
		\hline
$  0 $ to $  80$ & 5.10 $\cdot 10^{-3}$ & 2.2 & 5.6 & 6.0\\
$ 80 $ to $ 130$ & 6.26 $\cdot 10^{-3}$ & 2.6 & 3.9 & 4.7\\
$130 $ to $ 200$ & 2.96 $\cdot 10^{-3}$ & 2.6 & 4.9 & 5.6\\
$200 $ to $ 300$ & 0.70 $\cdot 10^{-3}$ & 3.5 & 6.2 & 7.1\\
$300 $ to $ 400$ & 0.12 $\cdot 10^{-3}$ & 7.5 & 5.4 & 9.3\\
			\hline
			\hline           $y^{\rm{t}}$ bin & $ 1/\!\sigma\,\rd\sigma\!/\!\rd y^{\rm{t}}$ & stat. [\%] & sys. [\%] & total [\%] \\
			\hline
$-2.5 $ to $ -1.3$ & 0.91 $\cdot 10^{-1}$ & 4.1 & 6.8 & 8.0\\
$-1.3 $ to $ -0.8$ & 2.55 $\cdot 10^{-1}$ & 3.1 & 4.9 & 5.8\\
$-0.8 $ to $ -0.4$ & 3.02 $\cdot 10^{-1}$ & 3.3 & 4.0 & 5.2\\
$-0.4 $ to $  0.0$ & 3.51 $\cdot 10^{-1}$ & 3.2 & 3.8 & 5.0\\
$ 0.0 $ to $  0.4$ & 3.71 $\cdot 10^{-1}$ & 3.2 & 3.8 & 4.9\\
$ 0.4 $ to $  0.8$ & 3.06 $\cdot 10^{-1}$ & 3.4 & 4.0 & 5.3\\
$ 0.8 $ to $  1.3$ & 2.41 $\cdot 10^{-1}$ & 3.3 & 4.9 & 5.9\\
$ 1.3 $ to $  2.5$ & 0.90 $\cdot 10^{-1}$ & 4.0 & 6.8 & 7.9\\
			\hline
			\hline            $\pt^{\rm{\ttbar}}$ bin [GeV] & $ 1/\!\sigma\,\rd\sigma\!/\!\rd \pt^{\ttbar}$ & stat. [\%] & sys. [\%] & total [\%] \\
			\hline
$  0 $ to $  20$ & 1.60 $\cdot 10^{-2}$ & 2.9 & 24.9 & 25.0\\
$ 20 $ to $  60$ & 0.97 $\cdot 10^{-2}$ & 2.1 & 10.7 & 10.9\\
$ 60 $ to $ 120$ & 0.32 $\cdot 10^{-2}$ & 2.5 & 13.2 & 13.5\\
$120 $ to $ 300$ & 0.05 $\cdot 10^{-2}$ & 3.7 &  6.9 &  7.9\\
			\hline
			\hline           $y^{\rm{\ttbar}}$ bin & $ 1/\!\sigma\,\rd\sigma\!/\!\rd y^{\ttbar}$ & stat. [\%] & sys. [\%] & total [\%] \\
			\hline
$-2.5 $ to $ -1.5$ & 0.30 $\cdot 10^{-1}$ & 13.7 & 14.7 & 20.1\\
$-1.5 $ to $ -0.7$ & 2.19 $\cdot 10^{-1}$ &  3.1 &  4.4 &  5.4\\
$-0.7 $ to $  0.0$ & 4.18 $\cdot 10^{-1}$ &  2.4 &  3.6 &  4.3\\
$ 0.0 $ to $  0.7$ & 3.93 $\cdot 10^{-1}$ &  2.5 &  3.6 &  4.4\\
$ 0.7 $ to $  1.5$ & 2.18 $\cdot 10^{-1}$ &  3.2 &  4.4 &  5.5\\
$ 1.5 $ to $  2.5$ & 0.40 $\cdot 10^{-1}$ & 10.4 & 14.7 & 18.0\\
			\hline
			\hline            $m^{\rm{\ttbar}}$ bin [GeV] & $ 1/\!\sigma\,\rd\sigma\!/\!\rd m^{\ttbar}$ & stat. [\%] & sys. [\%] & total [\%] \\
			\hline
$   0 $ to $   345$ &  -                  &  -    &  -    & -    \\
$ 345 $ to $   400$ & 5.26 $\cdot 10^{-3}$ &   5.4 &  10.4 &  11.7 \\
$ 400 $ to $   470$ & 4.58 $\cdot 10^{-3}$ &   3.8 &   4.1 &   5.6 \\
$ 470 $ to $   550$ & 2.46 $\cdot 10^{-3}$ &   4.9 &   7.6 &   9.0 \\
$ 550 $ to $   650$ & 1.07 $\cdot 10^{-3}$ &   6.1 &   3.9 &   7.2 \\
$ 650 $ to $   800$ & 0.39 $\cdot 10^{-3}$ &   6.9 &  11.4 &  13.4 \\
$ 800 $ to $  1100$ & 0.08 $\cdot 10^{-3}$ &  13.3 &  27.0 &  30.1 \\
$1100 $ to $  1600$ & 0.01 $\cdot 10^{-3}$ &  22.4 &  43.6 &  49.0 \\

                  \hline
  		\hline
		\end{tabular}
	\end{center}
\end{table}

}

\cleardoublepage \section{The CMS Collaboration \label{app:collab}}\begin{sloppypar}\hyphenpenalty=5000\widowpenalty=500\clubpenalty=5000\textbf{Yerevan Physics Institute,  Yerevan,  Armenia}\\*[0pt]
S.~Chatrchyan, V.~Khachatryan, A.M.~Sirunyan, A.~Tumasyan
\vskip\cmsinstskip
\textbf{Institut f\"{u}r Hochenergiephysik der OeAW,  Wien,  Austria}\\*[0pt]
W.~Adam, E.~Aguilo, T.~Bergauer, M.~Dragicevic, J.~Er\"{o}, C.~Fabjan\cmsAuthorMark{1}, M.~Friedl, R.~Fr\"{u}hwirth\cmsAuthorMark{1}, V.M.~Ghete, J.~Hammer, N.~H\"{o}rmann, J.~Hrubec, M.~Jeitler\cmsAuthorMark{1}, W.~Kiesenhofer, V.~Kn\"{u}nz, M.~Krammer\cmsAuthorMark{1}, I.~Kr\"{a}tschmer, D.~Liko, I.~Mikulec, M.~Pernicka$^{\textrm{\dag}}$, B.~Rahbaran, C.~Rohringer, H.~Rohringer, R.~Sch\"{o}fbeck, J.~Strauss, A.~Taurok, W.~Waltenberger, G.~Walzel, E.~Widl, C.-E.~Wulz\cmsAuthorMark{1}
% ...
\vskip\cmsinstskip
\dag:~Deceased\\
1:~~Also at Vienna University of Technology, Vienna, Austria\\
% ...

\end{sloppypar}
\end{document}